%% file: codedDS_main.tex
\documentclass[journal,onecolumn,10pt,twoside]{IEEEtran}

\usepackage{amsmath}
\usepackage{amsfonts}
\usepackage{amssymb}
\usepackage{amsthm}
\usepackage{amstext}
\usepackage{mathrsfs}
\usepackage{dsfont}
\usepackage{color}
\usepackage{cite}
\usepackage{times}
\usepackage{ifthen}
\usepackage{latexsym}
\usepackage{verbatim}
\usepackage{psfrag}
\usepackage{bbm}
\usepackage{float}
\usepackage{enumitem}
\usepackage{graphics}
\usepackage{graphicx}
\usepackage{epsfig}
\usepackage{epsf}
\usepackage{algorithm}
\usepackage{algpseudocode}

\ifCLASSOPTIONcompsoc
  \usepackage[caption=false,font=normalsize,labelfont=sf,textfont=sf]{subfig}
\else
  \usepackage[caption=false,font=footnotesize]{subfig}
\fi

\newtheorem{theorem}{Theorem}
\newtheorem{corollary}{Corollary}

\newtheorem{remark}{Remark}

\algnewcommand{\algorithmicand}{\textbf{and }}
\algnewcommand{\algorithmicor}{\textbf{or }}
\algnewcommand{\OR}{\algorithmicor}
\algnewcommand{\AND}{\algorithmicand}

\input{mySymbs.tex}
\newcommand{\lb}{\left\{ }
\newcommand{\rb}{\right\} }
\renewcommand{\F}[2]{ F^{#1}_{#2}}
\newcommand{\fu}[2]{\widetilde{F}^{#1}_{#2}}

\newcommand{\X}[1]{X_{#1}}
\newcommand{\set}{\Delta}
\newcommand{\mus}[2]{\mu^{#1}_{#2}}

\newcommand{\sett}{\widetilde{\Gamma}}
\renewcommand{\c}{^\complement}
\newcommand{\tF}{\widetilde{F}}
\newcommand{\proc}[1]{\cP^{#1}}
\renewcommand{\ex}[2]{\cE_{#1}^{#2}}

\newcommand{\partitionFiles}{\texttt{partitionFiles}}
\newcommand{\placeCache}{\texttt{placeCache}}

\newcommand{\encodeSubmessages}{\texttt{encodeSubmessages}}
\newcommand{\decodeSubfiles}{\texttt{decodeSubfiles}}
\newcommand{\updateCaches}{\texttt{updateCaches}}

\newcommand{\decomposeGraph}{\texttt{decomposeGraph}}
\newcommand{\Hungarian}{\texttt{Hungarian}}

\allowdisplaybreaks

\begin{document}
	
\title{On the Fundamental Limits of Coded Data Shuffling\\for Distributed Machine Learning}

\author{
Adel~Elmahdy, and Soheil~Mohajer
\thanks{This work was supported in part by the National Science Foundation under Grant CCF-1749981. A preliminary version of this work was presented at the 2018 IEEE International Symposium on Information Theory \cite{adel2018dataShuffle}.}  
\thanks{The authors are with the Department of Electrical and Computer Engineering, University of Minnesota, Minneapolis, MN, 55455 USA (e-mail: adel@umn.edu, soheil@umn.edu).}
}

\maketitle

\begin{abstract}
We consider the data shuffling problem in a distributed learning system, in which a master node is connected to a set of worker nodes, via a shared link, in order to communicate a set of files to the worker nodes.
The master node has access to a database of files. In every shuffling iteration, each worker node processes a new subset of files, and has excess storage to partially cache the remaining files, assuming the cached files are uncoded.
The caches of the worker nodes are updated every iteration, and they should be designed to satisfy any possible unknown permutation of the files in subsequent iterations.
For this problem, we characterize the exact load-memory trade-off for worst-case shuffling by deriving the minimum communication load for a given storage capacity per worker node.
As a byproduct, the exact load-memory trade-off for any shuffling is characterized when the number of files is equal to the number of worker nodes.
We propose a novel deterministic coded shuffling scheme, which improves the state of the art, by exploiting the cache memories to create coded functions that can be decoded by
several worker nodes.
Then, we prove the optimality of our proposed scheme by deriving a matching lower bound and showing that the placement phase of the proposed coded shuffling scheme is optimal over all shuffles.
\end{abstract}


\section{Introduction}
\IEEEPARstart{W}{ith} the emergence of big data analytics, distributed computing systems have attracted enormous attention in recent years.
The computational paradigm in the era of big data has shifted towards distributed systems, as an alternative to expensive supercomputers.
Distributed computing systems are networks that consist of a massive number of commodity computational nodes connected through fast communication links.
Examples of distributed computing applications span distributed machine learning, massively multilayer online games (MMOGs), wireless sensor networks, real-time process control, etc.
Prevalent distributed computing frameworks, such as Apache Spark \cite{zaharia2010spark}, and computational primitives, such as MapReduce \cite{dean2004mapreduce}, Dryad \cite{isard2007dryad}, and CIEL \cite{murray2011ciel}, are key enablers to process substantially large data-sets (in the order of terabytes), and execute production-scale data-intensive tasks.

Data shuffling is one of the core components in distributed learning algorithms. Broadly speaking, the data shuffling stage is introduced to prepare data partitions with desirable properties for parallel processing in future stages.
A prototypical iterative data processing procedure is outlined as follows: 
(i)~randomly shuffle the training data-set, 
(ii)~equally partition the data-set into non-overlapping batches, and assign each batch to a local worker\footnote{One may consider storing the entire training data-set in a massive shared storage system and let the workers directly access the new batches every learning epoch. Although this setting eliminates the communication overhead of the shuffling mechanism, it suffers from network and disk I/O bottlenecks, and hence, this approach is notoriously sluggish and cost-inefficient as well~\cite{chung2017UberSh}.}, 
(iii)~each local worker performs a local computational task to train a learning model, 
(iv)~reshuffle the training data-set to provide each worker with a new batch of data points at each learning model and continue the model training.
Data shuffling is known to enhance the learning model quality and lead to significant statistical gains in ubiquitous applications for machine learning and optimization. 
One prominent example is stochastic gradient descend (SGD) \cite{recht2012toward,bottou2012stochastic,gurbuzbalaban2015random,shamir2016without,ying2018stochastic,haochen2018random,meng2017convergence,safran2019good}.
Recht and R{\'e} \cite{recht2012toward} conjectured a non-commutative arithmetic-geometric mean inequality, and showed that the expected convergence rate of the random shuffling version of SGD is faster than that of the usual with-replacement version provided the inequality holds\footnote{It is a long-standing problem in the theory of SGD to prove this statement, and the correctness of the full conjecture is still an open problem.}.
In recent years, it has been demonstrated that shuffling the data before running SGD results in superior convergence performance \cite{bottou2012stochastic,gurbuzbalaban2015random,shamir2016without,ying2018stochastic,haochen2018random,meng2017convergence,safran2019good}.
For instance, Meng et al. \cite{meng2017convergence} have proposed an extensive analysis on the desirable convergence properties of distributed SGD with random shuffling, in both convex and non-convex cases.
In practice, however, the benefits of data shuffling come at a price. In every shuffling iteration, the entire data-set is communicated over the network of workers. Consequently, this leads to performance bottlenecks due to the communication overhead.

The idea of incorporating coding theory into the context of distributed machine learning has been introduced in a recent work by \cite{lee2017speeding}.
The authors posed an intriguing question as to how to use coding techniques to ensure robust speedups in distributed computing. 
To address this question, the work flow of distributed computation is abstracted into three main phases;  a storage phase, a communication phase, and a computation phase. 
Coding theory is utilized to alleviate the bottlenecks in the computation and communication phases of distributed learning algorithms. 
More specifically, the authors proposed novel algorithms for coded computation to speed up the performance of linear operations, and coded data shuffling to overcome the significant communication bottlenecks between the master node and worker nodes during data shuffling.

\subsection{Related Works}
The data shuffling problem has been extensively studied from various perspectives under different frameworks.
In what follows, we survey the literature and present the progress and the current status of the problem.

\subsubsection{Data Shuffling in Master-Worker Distributed Computing Framework}
In the master-worker distributed setup, the master node has access to the entire data-set that is randomly permuted and partitioned into batches at every iteration of the distributed algorithm. 
The data shuffling phase aims at communicating these batches to the worker nodes in order to locally perform their distributed tasks in parallel. Then, the master node aggregates the local results of the worker nodes to complete the computation and give the final result.
Lee et~al. \cite{lee2017speeding} proposed the first coded shuffling algorithm, based on random data placement, that leverages the excess storage of the local caches of the worker nodes to slash the communication bottlenecks.
The coded shuffling algorithm consists of three main strategies: a coded transmission strategy designed by the master node, and decoding and cache updating strategies executed by the worker nodes.
It is demonstrated, through extensive numerical experiments, the significant improvement in the achievable communication load\footnote{In the literature, the communication load is referred to as ``communication rate'', e.g. \cite{lee2017speeding, attia2018nearOpt}. However, the more accurate term should be communication (or delivery) load which we use throughout the manuscript.} and the average transmission time of coded shuffling framework, compared to uncoded shuffling. 
The theoretical guarantees of \cite{lee2017speeding} hold only when the number of data points approaches infinity. Moreover, the broadcast channel between the master node and the worker nodes in \cite{lee2017speeding} is assumed to be perfect.
In~pursuance of a practical shuffling algorithm, Chung et~al. \cite{chung2017UberSh} have recently proposed a novel coded shuffling algorithm, coined ``UberShuffle'', to individually address the practical considerations of the shuffling algorithm of \cite{lee2017speeding}.
However, it is not evident how far these coded shuffling algorithms are from the fundamental limits of communication load.

Attia and Tandon \cite{attia2016information,attia2016worst,attia2018nearOpt} investigated the coded data shuffling problem in a distributed computing system, consisting of a master node that communicates data points (or coded functions of them) to worker nodes with limited storage capacity.
An information-theoretic formulation of the data shuffling problem was proposed for data delivery and storage update phases. 
Furthermore, the worst-cast communication load is defined to be the maximum communication load from the master node to the worker nodes over all possible consecutive data shuffles for any achievable scheme characterized by the encoding, decoding, and cache updating functions.
Accordingly, the authors characterized the optimal trade-off between the storage capacity per worker node and the worst-case communication load for certain cases of the number of files $N$, the number of worker nodes $K$, and the available storage per worker node $S$.
More specifically, the communication load was characterized when the number of worker nodes is limited to $K \in \{2, 3\}$ in~\cite{attia2016information}. 
Furthermore, the special case of \emph{no-excess storage} (arbitrary $N$ and $K$, but $S = N/K$) was addressed in~\cite{attia2016worst}. 
However, the proposed schemes in these works do not generalize for arbitrary parameters.
Recently, the authors have proposed ``aligned coded shuffling scheme'' \cite{attia2018nearOpt} that is optimal for $K < 5$, and suboptimal for $K \geq 5$ with maximum multiplicative gap of $(K-\frac{1}{3}) / (K-1)$ from the lower bound on the load for the worst-case communication scenario.
The proposed placement strategy is similar to the one in coded caching literature \cite{maddah2014fundamental}, and the achievable scheme hinges on the concept of interference alignment \cite{jafar2011interference}. On the other hand, the proposed information-theoretic lower bound is based on a similar bounding technique introduced in \cite{yu2018exact}.

Under the same master-worker framework, Song et~al. \cite{song2017pliable} considered the data shuffling problem from the perspective of index coding \cite{bar2011index}, where the new data assigned by the master node at every iteration constitute the messages requested by the worker nodes, and the data cached at the worker nodes form the side information.
Motivated by the NP-hardness of the index coding problem \cite{bar2011index}, the authors proposed a pliable version of the index coding problem to enhance the communication efficiency for distributed data shuffling.
It is assumed that the worker nodes are pliable in such a way that they are only required to obtain new messages, that are randomly selected from original set of messages, at every iteration.
This degree of freedom enables the realization of semi-random data shuffling that yields more efficient coding and transmission schemes, as opposed to fully random data shuffling.

Recently, Wan et~al. \cite{wan2018fundamental} have considered a decentralized communication paradigm for the data shuffling problem, where worker nodes only communicate data points among each other, and the master node does not participate in the data communication, except for the initial placement.
The authors have proposed coded distributed data shuffling schemes that are within a factor of $2$ from the optimal trade-off, under the constraint of uncoded cache placement. Moreover, the exact trade-off is characterized for $K \leq 4$, and $\widehat{S} \in \left\{1, K-2, K-1\right\}$ where $\widehat{S}=S/(N/K)$.

\subsubsection{Data Shuffling in MapReduce Distributed Computing Framework}
MapReduce \cite{dean2004mapreduce} is a programming paradigm that allows for parallel processing of massive data-sets across large clusters of computational nodes.
More concretely, the overall computation is decomposed into computing a set of ``Map'' and ``Reduce'' functions in a distributed and parallel fashion.
Typically, a MapReduce job splits the input data-set into blocks, each of which is locally processed by a computing node that maps the input block into a set of intermediate key/value pairs.
Next, the intermediate pairs are transferred to a set of processors that reduce the set of intermediate values by merging those with the same intermediate key. The process of inter-server communication between the mappers and reducers is referred to as \textit{data shuffling}.
Li et~al. \cite{li2015coded} introduced a variant implementation of MapReduce, named ``Coded MapReduce'' that exploits coding to considerably reduce the communication load of the data shuffling phase. 
The key idea is to create coded multicast opportunities in the shuffling phase through an assignment strategy of repetitive mappings of the same input data block across different servers.
The fundamental trade-off between computation load and communication cost in Coded MapReduce is characterized in~\cite{li2016fundamental}. 
A unified coding framework for distributed computing in the presence of straggling servers was proposed in \cite{li2016unified}, where the trade-off between the computation latency and communication load is formalized for linear computation tasks.

We would like to highlight the subtle distinction between the coded caching problem and the coded shuffling problem.
Both problems share the property that the prefetching scheme is designed to minimize the communication load for any possible unknown demand (or permutation) of the data.
However, the coded shuffling algorithm is run over a number of iterations to store the data batches and compute some task across all worker nodes. 
In addition to that, the permutations of the data in subsequent iterations are not revealed in advance.  
Therefore, the caches of the worker nodes should be adequately updated after every iteration to maintain the structure of the data placement, guarantee the coded transmission opportunity, and achieve the minimum communication load for any undisclosed permutation of the data.
Another subtle distinction that we would like to emphasize is the difference between the concept of data shuffling in the master-worker setup and that in the MapReduce setup.
In the master-worker setup, a master node randomly shuffles data points among the computational worker nodes for a number of iterations to enhance the statistical efficiency of distributed computing systems.
A coded data shuffling algorithm enables coded transmission of batches of the data-set through exploiting the excess storage at the worker nodes. 
On the other hand, in the MapReduce setup, the whole data-set is divided among the computational nodes, and a data placement strategy is designed in order to create coding opportunities that can be utilized by the shuffling scheme to transfer locally computed results from the mappers to the reducers. 
In other words, coded MapReduce enables coded transmission of blocks of the data processed by the mappers in the shuffling phase through introducing redundancy in the computation of the Map stage.

Another distinction that should be highlighted is between index coding problem and coded shuffling problem.
As discussed in \cite{attia2018nearOpt} (Attia et al.), the data shuffling problem can also be considered as an index coding problem, in which the data cached at the worker nodes form the side information, and the new data assignments are the messages required by each worker node. 
It is worth noting that both the side information and requested files in index coding problem usually consist of one (of several) complete files, without sub-packetization, and this concern make the two problems different. Moreover, while the side information is given in index coding problem, here we can partially select the side information (the content of the excess storage) to reduce the communication load.
Perhaps the most related work in the context of index coding to data shuffling is \cite{song2017pliable} (Song et al.), where a pliable version of index coding is adopted for data shuffling. In the pliable  index coding it is assumed that the worker nodes are  only required to obtain new messages (not previously cached in their storage). This degree of freedom enables the realization of semi-random data shuffling that yields more
efficient coding and transmission schemes, as opposed to fully random data shuffling. It was demonstrated that a communication load of order $O(\log^2(K))$ can be achieved under this framework. However, it is worth mentioning that the constraints of the pliable index coding problem are very relaxed compared to the problem we study in this paper. Under the data shuffling model we study in this work, the worst case communication load is of order of $O(\frac{K-\widehat{S}}{\widehat{S}})$, where  $\widehat{S}$ is the normalized storage size of each worker. Therefore, the communication cost is still $O(K)$ when $\widehat{S}$ is small, but decreases to $O(1)$ if $\widehat{S}=\Theta(K)$. 

\subsection{Main Contributions}
In this paper, we consider a data shuffling problem in a master-worker distributed computing system, in which we have a master node and $K$~worker nodes. 
The master node has access to the entire data-set of $N$ files.
Each worker node has a limited cache memory that can store up to $S$ files.
In each iteration of the distributed algorithm, the master node randomly shuffles the data points among the worker nodes.
We summarize the main results of the paper as follows:
\begin{itemize}[leftmargin=*]
	\item We first study the data shuffling problem when $N=K$.
	We propose a novel linear coded shuffling algorithm that is based on interference alignment and elimination techniques. 
	It comprises the following phases: (i)~file partitioning and labeling, (ii)~cache placement, (iii)~encoding, (iv)~decoding, (v)~cache updating and subfile relabeling.
	We show how cache memories are leveraged in order to create coded functions that can be decoded by several worker nodes that process different files at every iteration of the distributed algorithm.
	The proposed scheme is generalized for arbitrary $K$ and~$S$.
	\item Next, we derive a matching information-theoretic lower bound on the communication load for the data shuffling problem when $N=K$, and we prove that among all possible placement and delivery strategies, our proposed coded shuffling scheme is universally optimal over all shuffling scenarios, and achieves the minimum communication load. Therefore, the optimal load-memory trade-off when $N=K$ is characterized for any shuffling.
	\item Finally, we extend the results obtained for the canonical setting of $N=K$ to investigate the general setting of the data shuffling problem when $N \geq K$. 
	Inspired by the concept of perfect matching in bipartite graphs, we develop a coded shuffling scheme by decomposing the file transition graph into $N/K$ subgraphs, each of which reduces to a canonical data shuffling problem with $K$ files, $K$ worker nodes, and storage capacity per worker node $S/(N/K)$. 
	Hence, we can apply our coded shuffling scheme for $N=K$ to each sub-problem and obtain a delivery scheme for the original shuffling problem with arbitrary parameters  $N$, $K$ and $S$. This leads to an achievable scheme whose delivery load can provides us with an upper bound for the optimum communication load of the general coded data shuffling problem.  
	Furthermore, we study an instance of the problem (the worst-case scenario), and show that the upper bound obtained by the achievable scheme is indeed tight to the proposed instance of the problem.
	As a result, the optimal load-memory trade-off is exactly characterized when $N \geq K$ for the worst-case shuffling.
\end{itemize}
We emphasize on the fact that the proposed coded shuffling algorithms works for any shuffling model.
Note that, while the proposed algorithm universally works for any shuffling, its required communication load is a function of the desired permutation. The worst-case shuffling is the one that requires the maximum communication load to permute the data points.
In other words, the communication load of random shuffling does not exceed that of the worst-case shuffling. Hence, we use this load as a benchmark to evaluate the performance of the proposed algorithm.

A brief report of the main results of this paper has been given in \cite{adel2018dataShuffle}. This paper presents complete proofs of all results, along with numerous illustrative examples to explain the essential concepts.

\subsection{Paper Outline}
The remainder of the paper is organized as follows.
We first present the formal definition of data shuffling problem as well as the main results of this work in Section~\ref{sec:probForm}.
The cache placement scheme is proposed in Section~\ref{sec:cache_placement}.
For the canonical setting of the shuffling problem, i.e. when $N=K$, two achievable coded shuffling schemes, along with illustrative examples, are delineated in Section~\ref{sec:achv_scheme_N=K_All}. 
Then, the optimality proof for our proposed delivery scheme is presented in Section~\ref{sec:converse_proof_N=K_All}.
Next, for the general and practical setting of the shuffling problem, i.e. when $N \geq K$, is studied in Section~\ref{sec:N>K_results}, where the achievable delivery scheme, an illustrative example, and the optimality of proposed delivery scheme for the worst-case shuffling are presented. 
Finally, the paper is concluded and directions for future research are discussed in Section~\ref{sec:conclusion}.

\section{Problem Formulation and Main Results}
\label{sec:probForm}
\subsection{Formulation of Data Shuffling Problem}
\begin{figure}
	\centering
	\includegraphics[width=.50\textwidth]{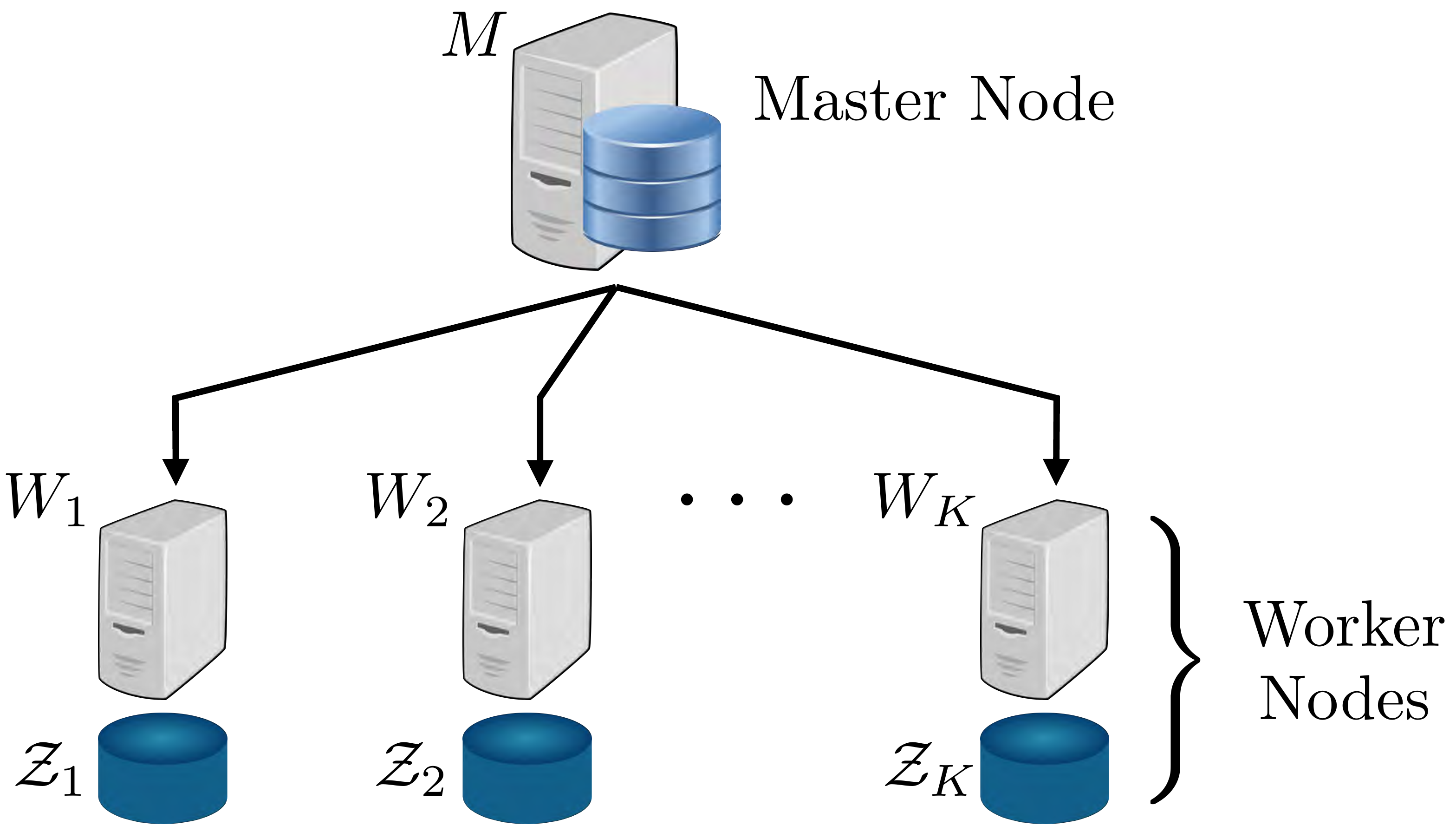}
	\caption{Data shuffling in a distributed computing system.}
	\label{fig:systemModel}
\end{figure}
For an integer $K$, let $\left[K\right]$ denote the set of integers $\left\{1, 2, \ldots, K\right\}$.
Fig.~\ref{fig:systemModel} depicts a distributed computing system with a master node, denoted by $M$, and a set of $K$ worker nodes, denoted by $\mathcal{W} = \{W_i: i \in [K]\}$.
The master node is assumed to have access to a data-set, including $N$ files, denoted by $\mathcal{F} = \{F^j: j \in [N]\}$, where the size of each file is normalized to $1$ unit. In practice, the number of files is remarkably larger than the number of worker nodes, and hence we study the data shuffling problem under the practical assumption of $N\geq K$.  At each iteration, each worker node should perform a local computational task on a subset of $N/K$ files\footnote{Unless otherwise stated, we assume that $N/K$ and $S/(N/K)$ are integers.}. The assignment of files to worker nodes is done by the master node, either randomly or according to some predefined mechanism. Each worker node $W_i$ has a cache~$\cZ_i$ that can store up to $S$ files, including those $N/K$ under-processing files. This imposes the constraint $S \geq N/K$ on the size of the cache at each worker node. Once the computation at the worker nodes is done, the result is sent back to the master node. A new batch of $N/K$ files will be assigned to each worker node for iteration $t+1$, and the cache contents of the worker nodes should be accordingly modified. The communication of files from the master node to the worker nodes occurs over a shared link, i.e., any information sent by the master node will be received by all of the worker nodes. 
Similar to \cite{lee2017speeding,attia2016information,attia2016worst,attia2018nearOpt}, the broadcast communication channel between the master node and worker nodes is assumed to be perfect.

For a given iteration $t$, we denote by $u(i)$ the set of indices of the files to be processed by $W_i$, and by $\proc{i}$ the portion of the cache of $W_i$  dedicated to the \emph{under-processing} files: $\proc{i}= \{F^j: j\in u(i)\}$. 
The subsets $\{u(i): i\in [K]\}$ provide a partitioning for the set of file indices, i.e., $u(i) \cap u(j) = \varnothing$ for $i \neq j$, and $\bigcup_{i=1}^{K} u(i) = [N]$. Similarly, $d(i)$ denotes the subset of indices of $N/K$ files to be processed by $W_i$ at iteration $t+1$, where $\{d(i): i\in [K]\}$ also forms a partitioning for $[N]$. 
When $S > N/K$, each worker node has an excess storage to cache (parts of) the other files in $\mathcal{F}$, in addition to the $N/K$ files in $\proc{i}$.  
We denote by $\ex{i}{} = \mathcal{Z}_i \setminus \proc{i}$ the contents of the remaining space of the cache of $W_i$, which is called the \emph{excess} storage.   
Therefore, $\cZ_i = \proc{i} \cup \ex{i}{}$.
Let $\proc{i}(t)$, $\ex{i}{}(t)$ and $\cZ_i(t)$ denote the realizations of $\proc{i}$, $\ex{i}{}$ and $\cZ_i$ at iteration $t$, respectively.
For the sake of brevity, we may drop the iteration index $t$ whenever it is clear from the context.

Filling the excess part of the caches of worker nodes is performed \emph{independently} of the new assigned subsets $\{d(i): i\in [K]\}$. 
Between iterations $t$ and $t+1$, the master node should compute and broadcast a message (a function of all files in $\mathcal{F}$), such that each worker node $W_i$ can retrieve all files in $d(i)$ from its cached data $\mathcal{Z}_i$ and the broadcast message $\cX$. 
The communication load $R=R(K,N,S)$ is defined as the size of the broadcast message $\cX$ for the parameters introduced above.  
We interchangeably refer to $R$ as \emph{delivery load} and \emph{communication load} of the underlying data-shuffling system. 
The goal is to develop a cache placement strategy and design a broadcast message $\cX$ to minimize $R$ for any $\{d(i): i\in [K]\}$. 
For $S \geq N$, we have $R=0$ since each worker node can store all the files in its cache and no communication is needed between the master node and worker nodes for any shuffling.
Thus, we can focus on the regime of $N/K \leq S \leq N$. We define $\widehat{S}=S/(N/K)$ to be the cache size normalized by the size of data to be processed by each worker node. Accordingly, we have $1\leq \widehat{S} \leq K$.

\begin{figure}
	\centering
	\subfloat[]{
		\includegraphics[width=3.3in]{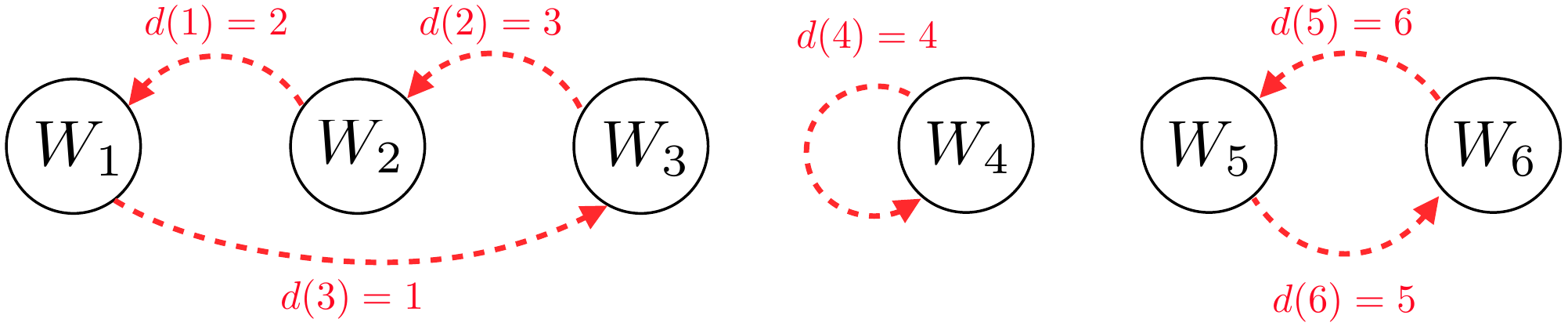}
		\label{fig:fileTransGraph_instance1}
	}
	\hfill
	\subfloat[]{
	\includegraphics[width=3.3in]{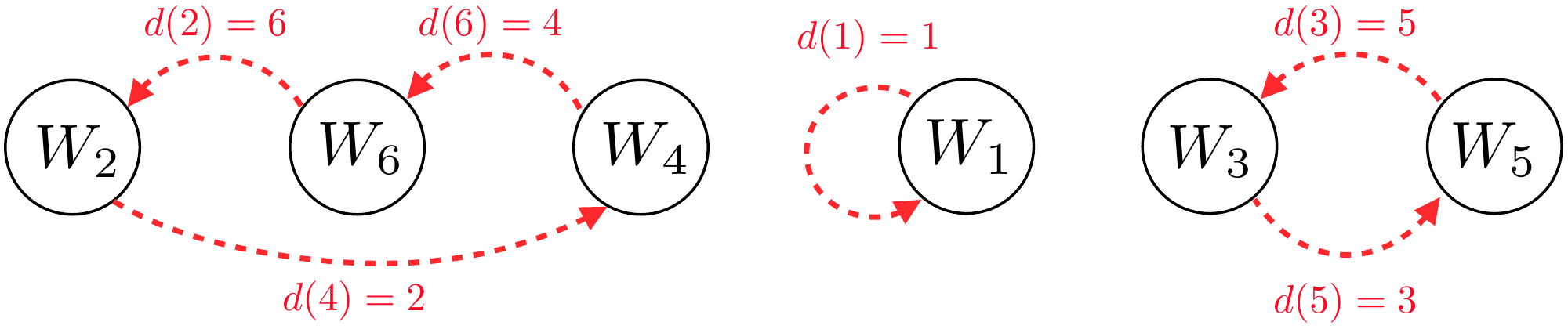}
	\label{fig:fileTransGraph_instance2}
	}
	\caption{
		The file transition graphs for two instances of a data shuffling system with $K =6$ worker nodes and $N = 6$ files. Assume $u(i)=i$ for $i\in [6]$, and consider two different assignment functions; $d_1(\cdot)$ and $d_2(\cdot)$. The shown graphs are isomorphic. 
		(a) $d_1(1)=2, d_1(2)=3, d_1(3)=1, d_1(4)=4, d_1(5)=6$ and $d_1(6)=5$.
		(b) $d_2(1)=1, d_2(2)=6, d_2(3)=5, d_2(4)=2, d_2(5)=3$ and $d_2(6)=4$.
	}
	\label{fig:fileTransGraph_instances}
\end{figure}

\subsection{File Transition Graph}
A file transition graph is defined as a directed graph $\mathcal{G} (V,E)$, where $V$, with $|V|=K$, denotes the set of vertices each corresponding to a worker node, and $E$, with $|E|=N$, is the set of directed edges, each associated to one file (see Fig.~\ref{fig:fileTransGraph_instances}).  An edge $e_j=(i,\ell)\in E$ with $j\in[N]$ and $i,\ell\in [K]$ indicates that $j\in u(i) \cap d(\ell)$, i.e., file $F^j$ is being processed by worker node $W_i$ at iteration $t$, and assigned to worker node $W_\ell$ to be processed at iteration $t+1$. Note that in general $\mathcal{G} (V,E)$ is a multigraph, since there might be multiple files in $u(i) \cap d(\ell)$, and we include one edge from $W_i$ to $W_\ell$ for each of such files.

Without loss of generality, let us assume a fixed assignment function $u(\cdot)$ at iteration $t$, for example, $u(i)=\{(i-1)N/K + \ell: \ell \in [N/K]\}$ for $i\in [K]$, otherwise we can relabel the files. Hence, the problem and its file transition graph are fully determined by the assignment function $d(\cdot)$ at iteration $t+1$. Let $\cD_\cG$ be the set of assignment functions $d(\cdot)$ whose corresponding file transition graphs are \emph{isomorphic} to $\cG$. Fig.~\ref{fig:fileTransGraph_instances} captures two instances of a shuffling problem with isomorphic file transition graphs. For a given graph $\cG(V,E)$, we define the average delivery load over all assignment functions in $\cD_\cG$ as
\begin{IEEEeqnarray}{l}
	R(\cG) = \frac{1}{|\cD_\cG|}\sum_{d\in \cD_\cG} R(N,K,S; d).
\end{IEEEeqnarray}
Our ultimate goal in this paper is to characterize $R(\cG)$ for given parameter $(N,K,S)$ and for all feasible file transition graphs~$\cG$.

\subsection{Main Results}
\label{sec:main_results_all}
First, we present our main results to characterize the exact load-memory trade-off for the canonical setting of data shuffling problem, when $N=K$, for any shuffling. 
Since $N = K$, then $\widehat{S} = S$, each worker node processes one file at each iteration. 
Without loss of generality, we assume that $W_i$ processes file $\F{i}{}$ at every iteration, i.e., $u(i) = i$, for $i \in [K]$, otherwise we can relabel the files. 
The following theorems summarize our main results.
\begin{theorem} 
	\label{thrm1}
	For a data shuffling problem with a master node, $K$ worker nodes, each with a cache of size $S$ files with $S \in [N]$, the communication load $R=R(N=K,K,S)$ required to shuffle $N=K$ files among the worker nodes for any file transition graph is upper bounded~by\footnote{Note that $\binom{n}{k} = 0$ when $n < k$.}
	\begin{IEEEeqnarray}{lCl}
		\label{eq:R_achv1}
		R & \leq & \frac{\binom{K-1}{S}}{\binom{K-1}{S-1}}.
	\end{IEEEeqnarray}
	For non-integer values of $S$, where $1 \leq S \leq N$, the lower convex envelope of the $N$ corner points, characterized by \eqref{eq:R_achv1}, is achievable by memory-sharing.
\end{theorem}
An achievability argument consists of a cache placement strategy and a delivery scheme. We propose a cache placement in Section~\ref{sec:cache_placement} which will be used for all achievable schemes discussed in this paper. The delivery scheme, along with the memory-sharing argument for non-integer values of $S$, is presented in Section~\ref{sec:achv_scheme_N=K}.
Illustrative examples are then given in Section~\ref{sec:achv_scheme_N=K_ex}.

The next theorem provides an achievable delivery load (depending on the file transition graph) by an opportunistic coding scheme.
We will show later that the underlying file transition graph of any data shuffling problem, $\cG(V,E)$, comprises a number of directed cycles. 
We denote the number of cycles in the file transition graph by~$\gamma$, and denote the cycle lengths by $(\ell_1,\ell_2,\dots, \ell_\gamma)$ where $\sum_{i=1}^{\gamma} \ell_i = K$. 
\begin{theorem} 
	\label{thrm2}
	For a data shuffling system with a master node and $K$ worker nodes, each with a cache of size $S$ files, for $S \in [N]$, the shuffling of $N=K$ files among the worker nodes for a given file transition graph that comprises $\gamma$ cycles can be performed by broadcasting a message of size $R$, where
	\begin{IEEEeqnarray}{lCl}
		\label{eq:R_achv2}
		R & \leq & 
		\frac{\binom{K-1}{S} - \binom{\gamma-1}{S}}{\binom{K-1}{S-1}}.
	\end{IEEEeqnarray}
	For non-integer values of $S$, where $1 \leq S \leq N$, the lower convex envelope of the $N$ corner points, characterized by \eqref{eq:R_achv2}, is achievable by memory-sharing.
\end{theorem}
The proposed delivery scheme and achievability proof for Theorem~\ref{thrm2}  are presented in Section~\ref{sec:achv_scheme_optimal_N=K}.
The memory-sharing argument for non-integer values of $S$ follows a similar reasoning as the one in Theorem~\ref{thrm1}.
We provide an illustrative example in Section~\ref{sec:achv_scheme_optimal_N=K_ex}.

\begin{theorem} 
	\label{thrm3}
	For the data shuffling system introduced in Theorem~\ref{thrm2}, the communication load $R$ required to shuffle $N=K$~files among the worker nodes for a given assignment with a file transition graph that comprises $\gamma$ cycles is lower bounded~by
	\begin{IEEEeqnarray}{lCl}
		\label{eq:R_lowerBound}
		R & \geq & 
		\frac{\binom{K-1}{S} - \binom{\gamma-1}{S}}{\binom{K-1}{S-1}}.
	\end{IEEEeqnarray}
\end{theorem}
The proof of optimality (converse) is presented in Section~\ref{sec:converse_proof_N=K_All}, where we also provide an illustrative example to describe the proof technique. 

\begin{corollary}
	\label{remrk_opt_NeqK}
	Theorems \ref{thrm2} and \ref{thrm3} prove the optimality of the proposed coded shuffling scheme for an arbitrary number of worker nodes $K$, storage capacity per worker node $S$, and file transition graph with $\gamma$ cycles, when $N=K$. 
	Therefore, the optimal delivery load $R^\star$ is characterized as
	\begin{IEEEeqnarray}{lCl}
		\label{eq:opt_R_N=K}
		R^\star(N=K,K,S) & = &
		\frac{\binom{K-1}{S} - \binom{\gamma-1}{S}}{\binom{K-1}{S-1}},
		\quad S \in [N].
	\end{IEEEeqnarray}
	For non-integer values of $S$, where $1 \leq S \leq N$, the optimal delivery load $R^\star$ is equal to the lower convex envelope of the $N$ corner points given in \eqref{eq:opt_R_N=K}. 
	Furthermore, when $\gamma-1 < S$, the achievable delivery load of Theorem~\ref{thrm2} is equal to that of Theorem~\ref{thrm1}, and takes its maximum. This characterizes the optimal worst-case delivery load $R^\star_{\text{worst-case}}$ which is given by
	\begin{IEEEeqnarray}{lCl}
		R^\star_{\text{worst-case}} & = & \frac{\binom{K-1}{S}}{\binom{K-1}{S-1}}.
	\end{IEEEeqnarray}
	This indicates that the upper bound of Theorem~\ref{thrm1} is the best universal (assignment independent) bound that holds for all instances of the data shuffling problem. 
\end{corollary}

\begin{remark}
	The essence of the information-theoretic lower bound on the communication load for the worst-case shuffling proposed in this paper for $N=K$ is equivalent to that of \cite{attia2018nearOpt} (Attia et al.) for general $N$ and $K$. However, in more details, the proof of \cite{attia2018nearOpt} involves solving a linear program, while we use set theoretic arguments to obtain a closed-form expression. Moreover, for general parameters $N$ and $K$, we provide a simple reduction argument to re-use the bound proved for the canonical setting.
\end{remark}

\begin{figure}
	\centering
	\includegraphics[width=0.45\textwidth]{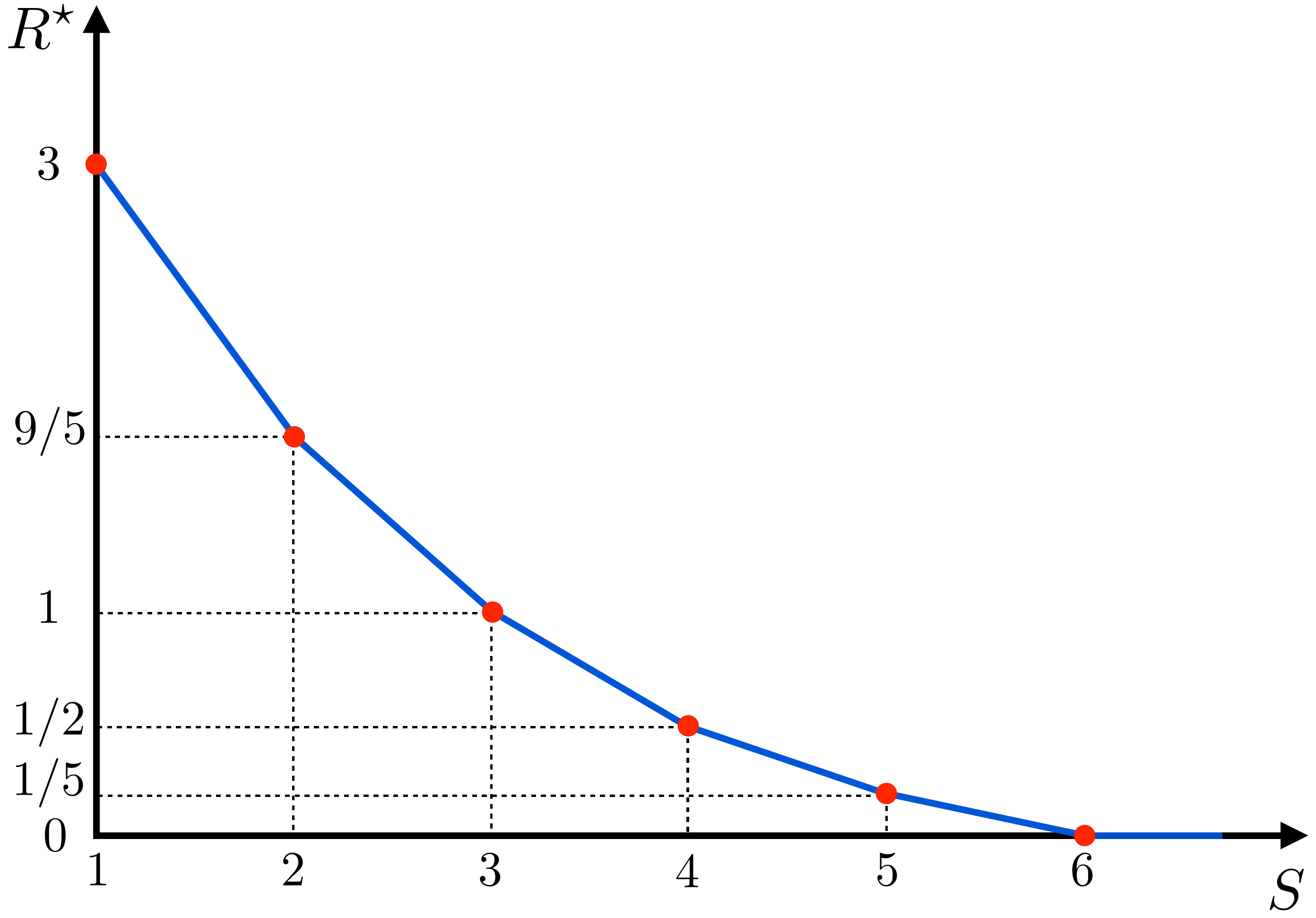}
	\caption{
		The optimum trade-off curve between the delivery load $R^\star$ and the storage capacity per worker node $S$, when $N = K = 6$ and $\gamma = 3$.
	}
	\label{fig:tradeoffCurve}
\end{figure}

Fig.~\ref{fig:tradeoffCurve} captures the optimum trade-off curve between $R^\star(K,K,S)$ as a function of $S$ for $N=K=6$ and a file transition graph with $\gamma = 3$ cycles. 

Next, based on the results obtained for the canonical setting of data shuffling problem when $N = K$, we present our main results in Theorem~\ref{thrm4} to characterize an upper bound on the load-memory trade-off for the general setting of data shuffling problem when $N \geq K$. This upper bound turns out to be optimum for the worst-case shuffling, as stated in Theorem~\ref{thrm5}.
\begin{theorem} 
	\label{thrm4}
	For a data shuffling system that processes $N$ files, and consists of a master node and $K$ worker nodes, each with a normalized storage capacity of $\widehat{S} = S/(N/K)$ files, the achievable delivery load $R=R(N,K,S)$ required to shuffle $N$~files among the worker nodes for any file transition graph is upper bounded~by
	\begin{IEEEeqnarray}{lCl}
		\label{eq:R_achv_gen}
		\displaystyle
		R & \leq & \frac{N}{K} \frac{\binom{K-1}{\widehat{S}}}{\binom{K-1}{\widehat{S}-1}}, \quad \widehat{S} \in [K].
	\end{IEEEeqnarray}
	For non-integer values of $\widehat{S}$, where $1 \leq \widehat{S} \leq K$, the lower convex envelope of the $N$ corner points, characterized by \eqref{eq:R_achv_gen} is achievable by memory-sharing.
\end{theorem}
The delivery scheme and achievability proof are presented in Section~\ref{sec:N_larger_K_system}. 
Note that the proposed achievable scheme is an extension to the one developed for the canonical setting of $N=K$ in Theorem~\ref{thrm2}.
The memory-sharing argument for non-integer values of $\widehat{S}$ follows a similar reasoning as the one in Theorem~\ref{thrm1}.
We also present an illustrative example in Section~\ref{sec:N_larger_K_system_ex}.

\begin{theorem} 
	\label{thrm5}
	For the data shuffling system introduced in Theorem~\ref{thrm4}, the  communication load $R_{\textrm{worst-case}}$ required to shuffle $N$~files among the worker nodes according to the worst-case shuffling is given by
	\begin{IEEEeqnarray}{lCl}
		\label{eq:R_lowerBound_N_large_K}
		\displaystyle
		R_{\textrm{worst-case}} & = & \frac{N}{K} \frac{\binom{K-1}{\widehat{S}}}{\binom{K-1}{\widehat{S}-1}}. 
	\end{IEEEeqnarray}
\end{theorem}
The proof of Theorem~\ref{thrm5} is presented in Section~\ref{sec:converse_proof_NlargerK}.

Before we start discussing the results of the paper, we present an example to explain the general idea of the proposed coded shuffling scheme.

\subsection{Illustrative Example}
\noindent\textbf{Example 1 (Single-Cycle File Transition Graph)}:  
{\it 
Consider a shuffling system with a master node and $K= 4$ worker nodes. The size of the cache at each worker node is $S = 2$ files. There are $N = 4$ files, denoted by $\{F^1, F^2, F^3, F^4\}$.
For notational simplicity, we rename the files as $\{A,B,C,D\}$.
Without loss of generality, we assume that worker nodes $W_1$, $W_2$, $W_3$ and $W_4$ are processing files $A$, $B$, $C$, and $D$, respectively, at iteration $t$, that is $u(1)=A$, $u(2)=B$, $u(3)=C$, and $u(4)=D$. 
The file transition graph is $d(1) = B$, $d(2) = C$, $d(3) = D$ and $d(4) = A$, as depicted by Fig.~\ref{fig:ex1_fileTransGraph}.
\begin{figure}
	\centering
	\subfloat[]{\includegraphics[width=3in]{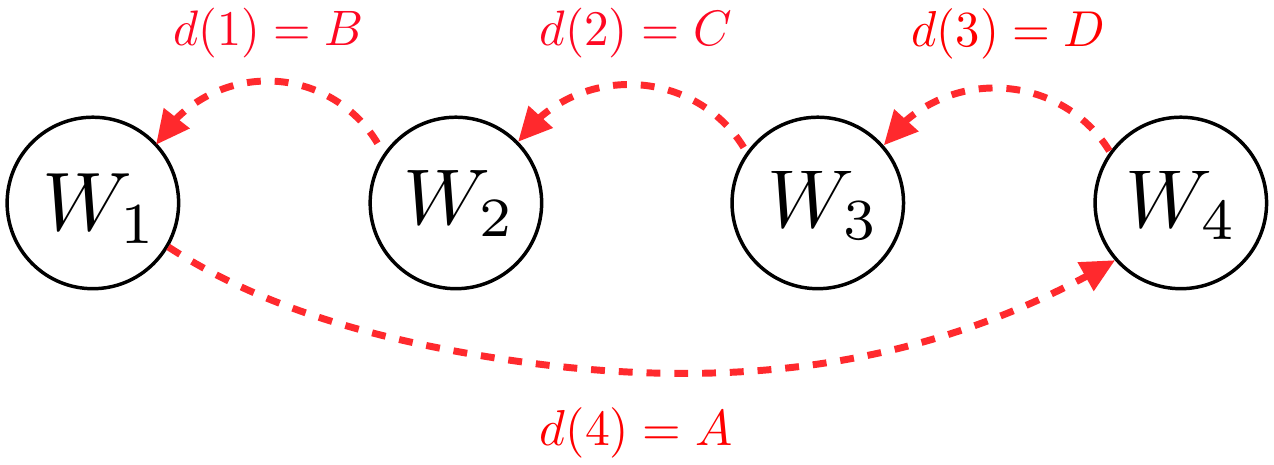}%
	\label{fig:ex1_fileTransGraph}}
	\\
	\centering
	\subfloat[]{\includegraphics[width=2.8in]{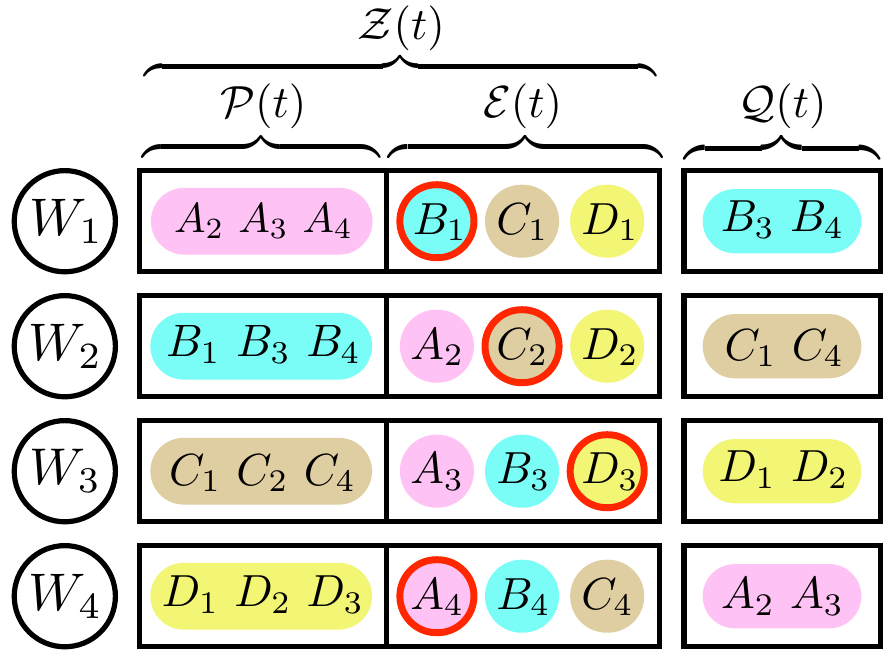}%
	\label{fig:ex1_cacheOrg}}
	\hspace{10mm}
	\subfloat[]{\includegraphics[width=1.71in]{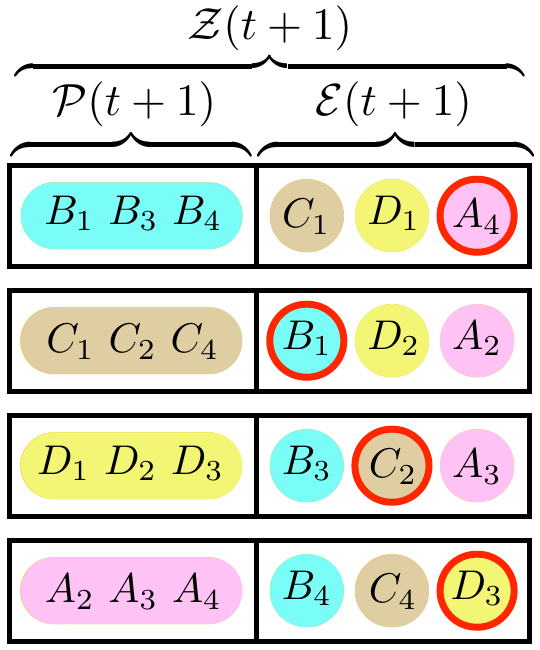}%
	\label{fig:ex1_cacheUpdate}}
	\hspace{10mm}
	\subfloat[]{\includegraphics[width=1.726in]{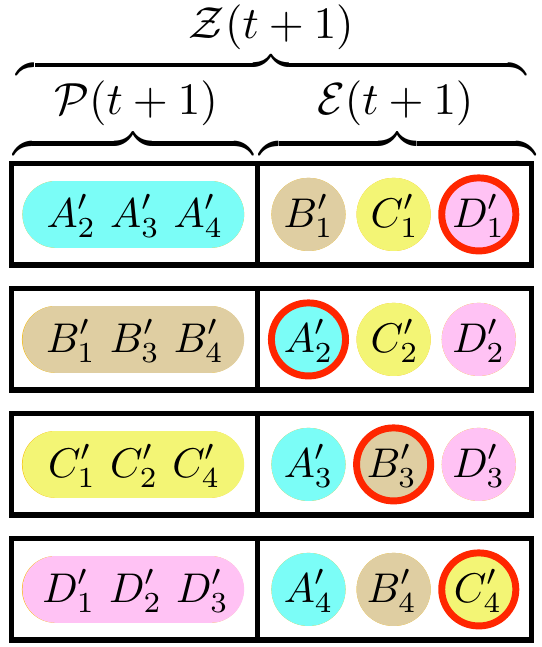}%
	\label{fig:ex1_cacheUpdate_subfileRelabel}}
	\\
	\centering
	\subfloat[]{\includegraphics[width=4.24in]{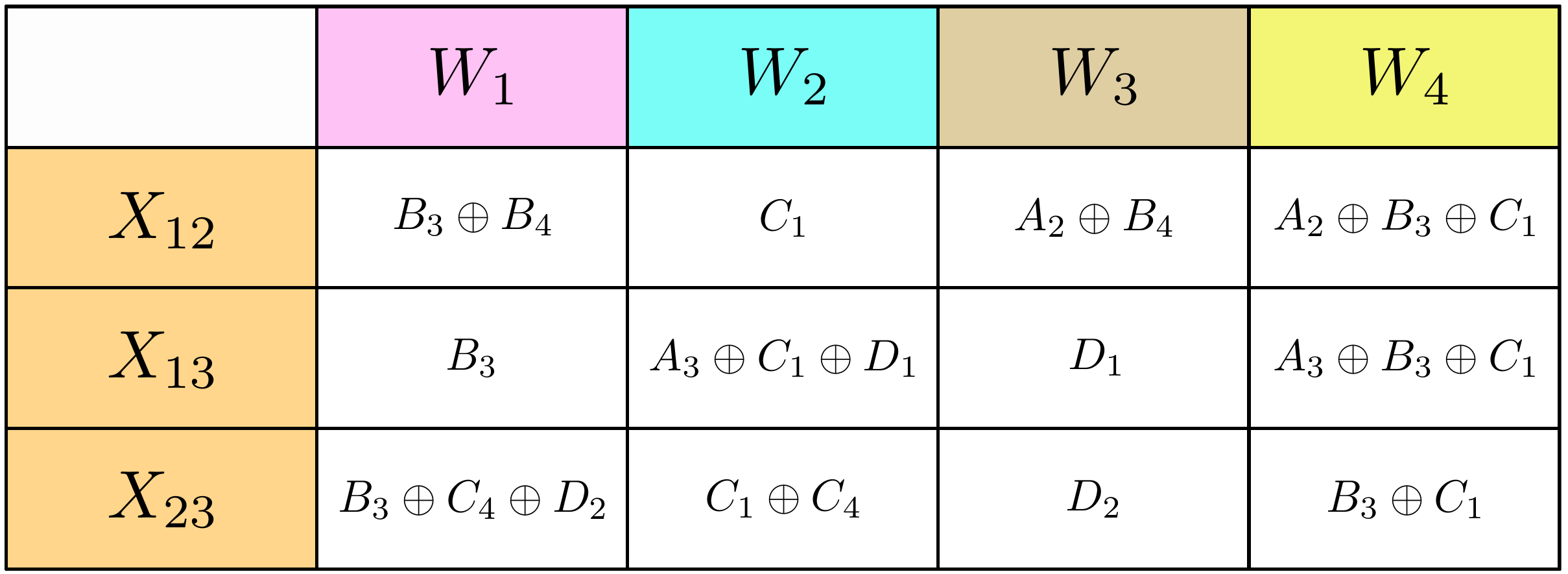}%
	\label{fig:ex1_functions}}
	\caption{
		Data shuffling system with $N = K = 4$, $S = 2$ and $\gamma = 1$.
		(a) The file transition graph for a data shuffling system with $N = K = 4$, $S = 2$ and $\gamma = 1$. Worker nodes $W_1$, $W_2$, $W_3$ and $W_4$ are processing files $A$, $B$, $C$, and $D$, respectively.
		(b) Cache organization of worker nodes at iteration $t$, along with the set of subfiles which are not available in the caches at iteration $t$ and need to be processed at iteration $t+1$.
		(c) Cache organization of worker nodes at iteration $t+1$ after updating the caches.
		Subfiles $A_4$, $B_1$, $C_2$ and $D_3$ in $\ex{4}{}(t)$, $\ex{1}{}(t)$, $\ex{2}{}(t)$ and $\ex{3}{}(t)$ are moved to $\ex{1}{}(t+1)$, $\ex{2}{}(t+1)$, $\ex{3}{}(t+1)$ and $\ex{4}{}(t+1)$, respectively.
		(d) Cache organization of worker nodes at iteration $t+1$ after updating the caches and relabeling the subfiles of Fig.~\ref{fig:ex1_cacheOrg}.
		Subfiles $A_4$, $B_1$, $C_2$ and $D_3$ in $\ex{4}{}(t)$, $\ex{1}{}(t)$, $\ex{2}{}(t)$ and $\ex{3}{}(t)$ are moved to $\ex{1}{}(t+1)$, $\ex{2}{}(t+1)$, $\ex{3}{}(t+1)$ and $\ex{4}{}(t+1)$ and relabeled to $D_1'$, $A_2'$, $B_3'$ and $C_4'$, respectively.
		(e) Received functions by worker nodes after removing the cached subfiles. The complete received functions at worker nodes are expressed in~\eqref{eqn:recvFunction_Ex0}.
	}
	\label{fig:ex1_fTG_cache_updateRelabel_fn}
\end{figure}
The proposed placement strategy partitions each file into $\binom{K-1}{S-1}=3$ subfiles of equal sizes. The subfiles are labeled with sets $\Gamma\subseteq [4]$, where $|\Gamma| = \widehat{S}-1=S/(N/K)-1=1$. For instance, file $A$ being processed by worker node $W_1$ is partitioned into $A_2$, $A_3$, and $A_4$. Accordingly, the cache $\cZ_i$ of $W_i$ is divided into two parts; $\proc{i}$ that is dedicated to the under-processing file $\F{i}{}$, and $\ex{i}{}$ that is dedicated to store parts of other files. Fig.~\ref{fig:ex1_cacheOrg} captures the cache organization of worker nodes, along with the missing subfiles (i.e., the ones in $\cQ_i$, as defined in \eqref{eq:def:demand}) that need to be processed at iteration $t+1$.
The broadcast message $\cX$ transmitted from the master node to the worker nodes is formed by the concatenation of $3$ sub-messages $\cX=(X_{12}, X_{13}, X_{23})$, where
\begin{align}
	\begin{split}
		X_{12} &= A_2 \oplus B_3 \oplus B_4 \oplus C_1,
		\\
		X_{13} &= A_3 \oplus B_3 \oplus C_1 \oplus D_1,
		\\
		X_{23} &= B_3 \oplus C_1 \oplus C_4 \oplus D_2.
	\end{split}
	\label{eqn:recvFunction_Ex0}
\end{align}
We need to show that each worker node $W_i$ can decode all the missing subfiles in $\cQ_i$ from the broadcast message. 
Fig.~\ref{fig:ex1_functions} shows the received sub-messages by each worker node after removing the subfiles that exist in its cache. For instance, $W_1$ can decode $B_3$ from $X_{13}$ and $B_4$ from $X_{13} \oplus X_{12}$, respectively. These, together with $B_1$ that is already cached in $\cZ_1$, enable $W_1$ to fully recover $B$, which is the file to be processed at iteration $t+1$.  

Decoding file $A$ at $W_4$, which is the ignored worker node, is more involved. Worker node $W_4$ can decode $A_2$ and $A_3$ from $X_{12} \oplus X_{23}$ and $X_{13} \oplus X_{23}$, respectively, and $A_4$ already exists in its cache $\cZ_4$.  
Consequently, the proposed scheme can achieve a delivery load of $R_{\text{coded}} = 1$, due to sending $3$ sub-messages, each of size $1/3$. On the other hand, the delivery load achieved by the uncoded shuffling scheme, under the same placement strategy, involves sending $8$ sub-messages, each of size $1/3$, resulting in $R_{\text{uncoded}} = 8/3$.
Hence, the proposed coded shuffling scheme can save around $62\%$ of the communication load, compared to the uncoded shuffling scheme.

After the decoding phase, each worker node has access to (at least) $8$ subfiles, including $6$ subfiles pre-stored in its cache $\cZ_i$, and $2$ subfiles in $\cQ_i$. However, only $6$ subfiles can be stored in the cache, and the remaining ones should be discarded. 
It remains to show that the caches of the worker nodes can be updated using the broadcast message $\cX$ to maintain a similar arrangement in preparation for the next data shuffle from iteration $t+1$ to $t+2$. 
This is done in two phases, namely cache updating and subfile relabeling. Fig.~\ref{fig:ex1_cacheUpdate} depicts the cache organization of the worker nodes after updating the caches, while Fig.~\ref{fig:ex1_cacheUpdate_subfileRelabel} captures the cache organization of the worker nodes after updating the caches and relabeling the subfiles.
For example, as shown in Fig.~\ref{fig:ex1_cacheUpdate}, $W_1$ needs to keep a full copy of $B=\{B_1,B_2,B_3\}$ at iteration $t+1$ because $d(1) = B$. 
Moreover, subfiles $C_1$ and $D_1$ already exist in $\ex{1}{}(t)$, and hence they will remain in $\ex{1}{}(t+1)$. Among the remaining subfiles $\{A_2, A_3, A_4\}$, only one of them can be kept in the cache. According to \eqref{eq:cache_update_add}, the subfile to be kept is $A_\Gamma$ such that $d^{-1}(1) \in \Gamma$, that is $A_4$. 
Finally, in order to maintain a cache configuration consistent with the original one in \eqref{eq:Z-def}, all the subfiles in the excess storage of $W_i$ should have a subscript of $\{i\}$, and $W_i$ should process $F_i$, for $i \in [K]$, at every iteration. 
For example, subfiles $B_1$, $B_3$ and $B_4$, that are processed in $\proc{1}{}(t+1)$ by $W_1$, in Fig.~\ref{fig:ex1_cacheUpdate} are relabeled to $A_2'$, $A_3'$ and $A_4'$ in Fig.~\ref{fig:ex1_cacheUpdate_subfileRelabel}, respectively.
Similarly, subfiles $C_1$, $D_1$, $A_4$, that are cached in $\ex{1}{}(t+1)$ by $W_1$, in Fig.~\ref{fig:ex1_cacheUpdate} are relabeled to $B_1'$, $C_1'$ and $D_1'$ in Fig.~\ref{fig:ex1_cacheUpdate_subfileRelabel}, respectively.~$\blacklozenge$
}

\section{Cache Placement}
\label{sec:cache_placement}
In this section we introduce our proposed cache placement, in which the contents of each worker node's cache at iteration $t$ are known. Note that the cache placement does not depend on the files to be processed by the worker nodes at iteration $t+1$, i.e., it does not depend on $\{d(i): i\in [K]\}$. 

\subsection{File Partitioning and Labeling}
Throughout this work, we assume that $N/K$ and $S/(N/K)$ are integer numbers, unless it is specified otherwise. Let $\widehat{S} = S/(N/K)$.
Let $\F{j}{}$ be a file being processed by worker node $W_i$ at iteration $t$, i.e., $j\in u(i)$. We partition $\F{j}{}$ into $\binom{K-1}{\widehat{S}-1}$~equal-size subfiles, and label the subfiles with a subscript~as
\begin{IEEEeqnarray}{l}
	\F{j}{} = \{
	\F{j}{\Gamma}: \Gamma\subseteq [K] \setminus \{i\}, |\Gamma|=\widehat{S}-1
	\},\:\:
	\forall i \in [K],\: j \in [N], \: j\in u(i).
\end{IEEEeqnarray}
Since the size of each file is normalized to $1$,  the size of each subfile will be $1 / \binom{K-1}{\widehat{S}-1}$. 
For the sake of completeness, we also define dummy subfiles
$\F{j}{\Gamma} = 0$ (with size $0$)  for every $\Gamma\subseteq [K]$ with $j\in \Gamma$ or $|\Gamma|\neq \widehat{S}-1$.

\subsection{Cache Placement}
The cache $\mathcal{Z}_i$ of $W_i$ consists of two parts:
(i) the  \emph{under-processing} part $\proc{i}$, in which \emph{all subfiles} of files to be processed at iteration $t$ are stored;  
(ii) the  \emph{excess} storage part $\ex{i}{}$, which is equally distributed among all other files.  We denote by $\ex{i}{\ell}$ the portion of~$\ex{i}{}$ dedicated to the file $\F{\ell}{}$, in which all subfiles $\F{\ell}{\Gamma}$ with $i\in\Gamma$ are cached. 
Hence, we have 
\begin{IEEEeqnarray}{lCl}
	\cZ_i 
	&=& \proc{i} \cup \ex{i}{} = \proc{i} \cup \left(\bigcup\nolimits_{\ell \in [N] \setminus u(i)} \ex{i}{\ell} \right),
	\:\: \forall i \in [K], 
	\label{eq:Z-def}
\end{IEEEeqnarray}
where
\begin{IEEEeqnarray}{lCl}
	\proc{i}
	& = & \left\{\F{j}{\Gamma}: j \in u(i),\: \Gamma \subseteq [K] \setminus \{i\},\: |\Gamma| = \widehat{S}-1\right\},
	\label{eq:P-def}\\
	\ex{i}{\ell} 
	&=& \left\{\F{\ell}{\Gamma}: \ell \notin u(i), \: i \in \Gamma \subseteq [K], \: |\Gamma| = \widehat{S} - 1\right\}.
	\label{eq:E-def}
\end{IEEEeqnarray} 
For any worker node $W_i$, there are $N/K$ complete files in $\proc{i}$. 
Moreover, for each of the remaining $N - N/K$ files, there are $\binom{K-2}{\widehat{S}-2}$ subfiles, out of a total of $\binom{K-1}{\widehat{S}-1}$ subfiles, that are cached in the excess storage part.
Thus, we~have
\begin{IEEEeqnarray}{l}
	\label{eqn:SanityCh_CacheSize} 
	\displaystyle
	|\cZ_i| = \left|\proc{i}\right| + \sum_{\ell \in [N] \setminus u(i)} \left|\ex{i}{\ell}\right|
	=
	\frac{N}{K}  +   \left(N - \frac{N}{K}\right) \frac{\binom{K - 2}{\widehat{S} - 2}}{\binom{K - 1}{\widehat{S} - 1}} 
	= S,
	\nonumber
\end{IEEEeqnarray}
which satisfies the memory constraints.

\begin{remark}
	The proposed cache placement is different from the one in [14] (Lee et al.). The placement in \cite{lee2017speeding} follows a random sampling of the files independently across the users. Moreover, file splitting is not allowed in \cite{lee2017speeding}, and a file (data point) is either fully stored in the storage of a worker node, or no bit of that is cached by the worker node. The proposed placement strategy, however, is deterministic and fully characterized at every shuffling iteration.
	On the other hand, the so-called structural invariant placement strategy introduced in \cite{attia2018nearOpt} is deterministic, but  different from the one proposed in this paper. In particular, the structural invariant placement method works for values of $\widehat{S}$ satisfying $\widehat{S} = \left(1+i\left(\frac{K-1}{K}\right)\right)$ for some integer $i=0,1,\dots, K$, while we  need $\widehat{S}$ to be some integer in $[K]$. Consequently, the communication load achieved in \cite{attia2018nearOpt} is sub-optimum for general $K$. A variation of the  structural invariant storage placement is proposed in \cite[Appendix D]{attia2018nearOpt} only for $\widehat{S}\in \{1,K-2,K-1\}$ which is identical the placement strategy proposed in our work for any integer value of $\widehat{S}$. It is worth noting that in spite of similarities between the placement strategies, the proposed encoding and decoding methods in~\cite{attia2018nearOpt} (referred to as ``aligned coded shuffling'') are completely different from those proposed in this paper.
\end{remark}

Recall that the worker node $W_i$ should be able to recover files $\{F^{\ell}: \ell\in d(i)\}$ from its cache $\cZ_i$ and the broadcast message $\cX$. Communicating files in $d(i)$ from the master node to a worker node $W_i$ can be limited to sending only the desired subfiles that do not exist in the cache of $W_i$. For a worker node $i\in [K]$, let $\cQ_{i}$  denote the set of subfiles to be processed by~$W_i$ at iteration $t+1$, which are not available in its cache $\cZ_i$ at iteration $t$, that is,
\begin{IEEEeqnarray}{l}
	\cQ_{i} = \lb \F{\ell}{\Gamma} : \ell \in d(i), \: 
	\ell \notin u(i), \: i  \notin \Gamma, \:
	\Gamma \subseteq [K], \:
	|\Gamma| = \widehat{S} - 1 \rb. 
	\label{eq:def:demand}
\end{IEEEeqnarray}
It is evident that each worker node needs to decode at most $\binom{K-1}{\widehat{S}-1} - \binom{K-2}{\widehat{S}-2} = \binom{K-2}{\widehat{S}-1}$ subfiles for each of the $N/K$ files in $d(i)$, in order to process them at iteration $t+1$.

The pseudocodes of the proposed file partitioning and labeling, and cache placement are given in Algorithm~\ref{AchvScheme1_1} and Algorithm~\ref{AchvScheme1_2}, respectively, in Appendix~\ref{app:psuedocodes_p1}.

\section{Coded Shuffling for the Canonical Setting $(N=K)$}
\label{sec:achv_scheme_N=K_All}
We describe two delivery strategies in this section.  The first delivery scheme is universal, in the sense that it does not exploit the properties of the underlying file transition graph. By analyzing this scheme in Section~\ref{sec:achv_scheme_N=K}, we show that the delivery load in Theorem~\ref{thrm1} is achievable. Two  illustrative examples are presented in Section~\ref{sec:achv_scheme_N=K_ex} to better describe the coding and decoding strategies. We then demonstrate that the size of the broadcast message can be reduced by exploiting the cycles in the file transition graph. A graph-based delivery strategy is proposed in Section~\ref{sec:achv_scheme_optimal_N=K}. This new scheme can achieve the reduced delivery load proposed in Theorem~\ref{thrm2}. Finally, we conclude this section by presenting an illustrative example for the graph-based delivery scheme in Section~\ref{sec:achv_scheme_optimal_N=K_ex}. 

\subsection{A Universal Delivery Scheme for Any Shuffling: Proof of Theorem~\ref{thrm1}}
\label{sec:achv_scheme_N=K}
Recall that for $N=K$ we have $\widehat{S} = S/(N/K)=S$. In order to prove Theorem~\ref{thrm1}, we propose a coded shuffling scheme to show that a delivery load of $R = \binom{K-1}{S} / \binom{K-1}{S-1}$ is achievable for the canonical setting ($N=K$) for any integer $1 \leq S \leq N$. 
We assume, without loss of generality, that $W_i$ processes file $\F{i}{}$ at  iteration $t$, i.e., $u(i) = i$ for $i \in [K]$, otherwise we can relabel the files.

\subsection*{Encoding}
Given all cache contents $\{\cZ_i: i\in [K]\}$, characterized by \eqref{eq:Z-def}, and $\{d(i): i\in [K]\}$, the  
broadcast message $\cX$ sent from the master node to the worker nodes is obtained  by the concatenation of a number of sub-messages $\X{\set}$, each specified for a group of worker nodes $\set$, that is,
\begin{IEEEeqnarray}{lCl}
	\label{eq:X_all}
	\cX &=& \{X_\set : \set  \subseteq [K-1], |\set|=S \}, 
\end{IEEEeqnarray}
where
\begin{IEEEeqnarray}{l}
	\label{eq:X_Delta}
	X_{\Delta}  \triangleq 
	\bigoplus_{i \in \Delta} \left( \F{i}{\Delta \setminus \{i\}} 
	\oplus
	\F{d(i)}{\Delta \setminus \{d(i)\}}
	\oplus	
	\bigoplus_{ j \in [K] \setminus \set } 
	\F{d(i)}{(\{j\} \cup \set) \setminus \{i, d(i)\}} 
	\right).
\end{IEEEeqnarray}
The encoding design hinges on $K-1$ worker nodes. 
Without loss of generality, we consider $W_1, W_2, \ldots, W_{K-1}$ for whom the broadcast sub-messages are designed, and designate $W_K$ as the \emph{ignored} worker node. 
We will later show how $W_K$ is served for free using the sub-messages designed for other worker nodes. 

\begin{remark}
	It is true that subfile $F^{d(i)}_{\Delta\setminus\{d(i)\}}$ does exist in the cache of worker node $W_i$, and is not needed to be broadcast, and indeed the encoded sub-message does not include this subfile. 
	This is due to the fact that such a subfile $F^{d(i)}_{\Delta\setminus\{d(i)\}}$ appears twice in $X_\Delta$, and hence will be canceled by the XOR operation. 
	Note that, this subfile is not dummy only if $d(i)\in \Delta$. Therefore, one copy is attained in the second term of the summand  as $F^{d(i)}_{\Delta\setminus\{d(i)\}}$, and the second copy is attained as the first term in the summand as $F^{j}_{\Delta\setminus\{j\}}$ for  $j=d(i)\in \Delta$.
\end{remark}

According to the proposed encoding scheme, there are a total of $\binom{K-1}{S}$ encoded sub-messages, each corresponds to one subset $\Delta$, and the size of each sub-message is $1 / \binom{K-1}{S-1}$. 
Hence, the overall broadcast communication load is upper bounded~by 
\begin{IEEEeqnarray}{lCl}
	R & \leq & \frac{\binom{K-1}{S}}{\binom{K-1}{S-1}},
	\nonumber
\end{IEEEeqnarray}
as claimed in Theorem~\ref{thrm1}.

\subsection*{Decoding}
\label{sec:achv_scheme_decode}
The following lemmas demonstrate how each worker node decodes the missing subfiles, that constitute the file to be processed at iteration $t+1$, from the broadcast sub-messages and its cache contents.

\begin{lm}
	\label{lm:decode-all-OFC}
	For a worker node $W_\ell$, where $\ell \in [K-1]$, a~missing subfile $\F{d(\ell)}{\Gamma}\in \cQ_{\ell}$ can be decoded  
	\begin{itemize}[leftmargin=*]
		\item from $\cZ_\ell$ and the broadcast sub-message $X_{\{\ell\} \cup \Gamma}$, if $K\notin \Gamma$; and
		\item from $\cZ_\ell$, the broadcast sub-message $X_{(\Gamma\setminus\{K\}) \cup \{\ell,d(\ell)\}}$, and other subfiles previously decoded by~$W_\ell$, if $K\in \Gamma$. 
	\end{itemize}
\end{lm}
We refer to Appendix~\ref{app:lm:dec-all} for the proof of lemma~\ref{lm:decode-all-OFC}.

\begin{remark}
	Here, we provide an intuitive justification for Lemma~\ref{lm:decode-all-OFC}. Consider a worker node $W_i$ and a set of worker nodes $\set$ of size $S$ that includes $i$. One can show that every subfile appearing in $X_{\set}$ belongs to either $\cZ_i$ or $\cQ_i$. Therefore, worker node $W_i$ can recover a linear equation in the subfiles in $\cQ_i$ by removing the subfiles in its cache $\cZ_i$ from $X_{\set}$. It turns out that all such equations are linearly independent. The number of such equations is $\binom{K-2}{S-1}$ (because $\set\subseteq [K-1]$ and $i\in \set$). On the other hand, the number of subfiles in $\cQ_i$ is (at most) $\binom{K-1}{S-1}-\binom{K-2}{S-2} = \binom{K-2}{S-1}$, since out of a total of $\binom{K-1}{S-1}$ subfiles of $F^{d(i)}$, $\binom{K-2}{S-2}$ of them are cached in $\ex{i}{d(i)}$, characterized by \eqref{eq:E-def}. Therefore, the obtained set of linearly independent equations suffices to recover all the subfiles in~$\cQ_i$. 
\end{remark}

\begin{lm}
	\label{lm:decode-K-OFC}
	For the worker node $W_K$, any missing subfile $\F{d(K)}{\Gamma} \in \cQ_{K}$  can be decoded from the cache contents $\cZ_K$ and the summation of the broadcast sub-messages $\displaystyle \bigoplus_{\ell\in [K-1]\setminus \Gamma} X_{\{\ell\} \cup \Gamma }$. 
\end{lm}
We refer to Appendix~\ref{app:lm:decode-K-OFC} for the proof of lemma~\ref{lm:decode-K-OFC}.

\subsection*{Cache Updating and Subfile Relabeling}
\label{sec:achv_scheme_cacheUpdateRelabel}
After worker nodes decode the missing subfiles, characterized by \eqref{eq:def:demand}, the caches of worker nodes need to be updated and the subfiles need to be relabeled before processing the files at iteration $t+1$. The goal of cache updating and subfile relabeling is to maintain a similar cache configuration for the worker nodes for shuffling iteration $t+2$.
First, the caches are updated as follows:
\begin{itemize}[leftmargin=*]
	\item For $i \in [K]$, all the subfiles of $\F{d(i)}{}$ are placed in $\proc{i}$ at iteration $t+1$, i.e.,
	\begin{IEEEeqnarray}{lCl}
		\proc{i}(t+1)&=& \left\{\F{d(i)}{\Gamma}: \Gamma \subseteq [K]\setminus\{d(i)\},\: |\Gamma|=S-1\right\}.
	\end{IEEEeqnarray}
	\item For $i \in [K]$, the excess storage is updated by removing all the subfiles of $\F{d(i)}{}$, and replacing them by the subfiles of $\F{i}{}$ that were cached  at $W_{d^{-1}(i)}$, i.e., 
	\begin{IEEEeqnarray}{lCl}
		\ex{i}{}(t+1) & = & \left( \ex{i}{}(t) \setminus \cS_i \right) \cup \cA_i,
	\end{IEEEeqnarray}
	where
	\begin{IEEEeqnarray}{lCl}
		\cS_i &=& \left\{\F{d(i)}{\Gamma}: i \in \Gamma,\: \Gamma \subseteq [K]\setminus\{d(i)\},\: |\Gamma|=S-1\right\},\label{eq:cache_update_remove}
		\\
		\cA_i &=& \left\{\F{i}{\Gamma}: d^{-1}(i) \in \Gamma,\: \Gamma \subseteq [K]\setminus\{i\},\: |\Gamma|=S-1\right\}. \label{eq:cache_update_add}
	\end{IEEEeqnarray}
\end{itemize}
Consequently, we have
$\cZ_i(t+1) = \proc{i}(t+1) \cup \ex{i}{}(t+1)$ by definition. 
Note that the cache updating procedure is feasible, since the subfiles needed for $\cZ_i(t+1)$ either exist in $\cZ_i(t)$ or appear in the set of missing subfiles $\cQ_i(t)$ to be decoded after the broadcast message delivery. In particular, all the subfiles of $F^{i}$ already exist in $\proc{i}(t)$, and hence those in~$\cA_i$ will be simply moved from the under-processing part to the excess storage part of the cache. 

Next, the subfiles are \emph{relabeled} as follows: 
\begin{enumerate}
	\item For every subfile $\F{i}{\Gamma}$, where $i \in [K]$, $d^{-1}(i) \in \Gamma$, $\Gamma \subseteq [K] \setminus \{i\}$ and $|\Gamma|=S-1$, relabel the subfile's subscript to $\F{i}{\Lambda}$, where $\Lambda = \left(\Gamma \setminus \left\{d^{-1}(i)\right\}\right) \cup \left\{i\right\}$.
	\item For every subfile $\F{i}{\Gamma}$, where $i \in [K]$, $ \Gamma \subseteq [K] \setminus \{i\}$ and $|\Gamma|=S-1$, relabel the subfile's superscript to $\F{d^{-1}(i)}{\Gamma}$. 
\end{enumerate}

At the end of cache updating and subfile relabeling phase, the cache configuration of each worker node at iteration $t+1$ maintains a similar arrangement to that introduced initially at iteration $t$ and characterized by~\eqref{eq:Z-def}. 
More specifically, the cache updating step ensures that the under-processing part of the cache includes all subfiles of the file to be processed at iteration $t+1$. 
It also guarantees that the excess storage part of the cache stores an equal share of subfiles of all other files that are not in the under-processing part.
The subfile relabeling step, however, ensures that two properties are satisfied at any shuffling iteration; (i) the index of a worker node appears in the new subscript of all subfiles in the excess storage part of its cache, (ii)~the name of the file to be processed by worker node $W_i$ is changed to $F^i$ (from $F^{d(i)}$). 
Therefore, the proposed scheme can be systematically applied at the following shuffling iterations. This completes the proof of Theorem~\ref{thrm1}.
\hfill $\blacksquare$

The pseudocodes of the proposed encoding at the master node, decoding at the worker nodes, and cache updating and subfile relabeling are given in Algorithm~\ref{AchvScheme1_3}, Algorithm~\ref{AchvScheme1_4}, and Algorithm~\ref{AchvScheme1_5}, respectively, in Appendix~\ref{app:psuedocodes_p1}.

\begin{remark}
	Let $S$ be a non-integer cache size with $1\leq S \leq N$. We can always write $S = \alpha \lfloor S \rfloor + \left( 1-\alpha \right) \lceil S \rceil$, for some  $\alpha \in (0,1)$. 
	The data shuffling problem for non-integer cache size $S$ can be addressed by a memory-sharing argument, similar to \cite{attia2016information}.  More precisely, we can show that the pairs
	$\left(\lfloor S \rfloor, R\left(\lfloor S \rfloor\right)\right)$ and $\left(\lceil S \rceil, R\left(\lceil S \rceil\right)\right)$ are achievable, and conclude that, for $S = \alpha \lfloor S \rfloor + \left( 1-\alpha \right) \lceil S \rceil$, a communication load of $R\left(S \right) = \alpha R\left(\lfloor S \rfloor\right) + (a-\alpha) R\left(\lceil S \rceil\right)$ can be achieved. 
	Recall that the size of each file is normalized to $1$~unit. For the memory-sharing argument, each file will be partitioned into two parts of sizes $\alpha$~and $1-\alpha$. The cache of each worker node is also divided into two parts of sizes $\alpha S$ and $(1-\alpha)S$. Then, the files of size $\alpha$ will be cached and shuffled within the parts of the caches of size $\alpha S$. Similarly, the files of size $1-\alpha$, together with the parts of the caches of size $(1-\alpha)S$, form another isolated instance of the problem. Summing the delivery loads of the two instances, we get 
	\begin{IEEEeqnarray}{lCl}
		R(\alpha \lfloor S \rfloor + \left( 1-\alpha \right) \lceil S \rceil)
		&=&
		\alpha R (\lfloor S \rfloor) +  \left( 1-\alpha \right) R (\lceil S \rceil)
		\nonumber\\
		&\leq&
		\alpha \frac{\binom{K-1}{\lfloor S \rfloor} - \binom{\gamma-1}{\lfloor S \rfloor}}{\binom{K-1}{\lfloor S \rfloor-1}} 
		+ \left( 1-\alpha \right) \frac{\binom{K-1}{\lceil S \rceil} - \binom{\gamma-1}{\lceil S \rceil}}{\binom{K-1}{\lceil S \rceil-1}}.
	\end{IEEEeqnarray} 
	This shows that  the convex hull of the pairs $\{(S,R): S \in [N]\}$ is achievable. 
\end{remark}

\begin{figure}
	\centering
	\includegraphics[width=0.6\textwidth]{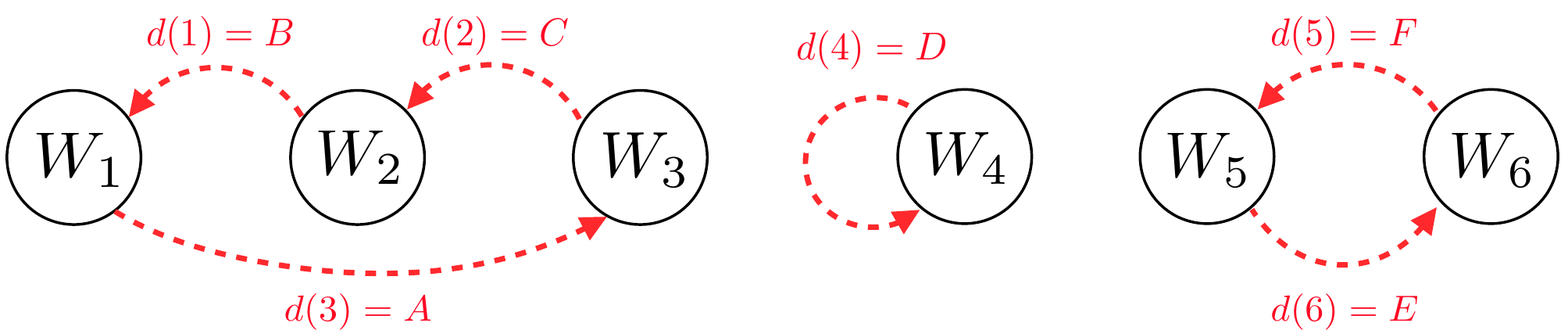}
	\caption{
		The file transition graph for a data shuffling system with $N = K = 6$ and $\gamma = 3$. Worker nodes $W_1$, $W_2$, $W_3$, $W_4$, $W_5$ and $W_6$ process files $A$, $B$, $C$, $D$, $E$, and $F$ at iteration $t$, respectively.
	}
	\label{fig:ex2_ex3_fileTransGraph}
\end{figure}

\subsection{Illustrative Examples}
\label{sec:achv_scheme_N=K_ex}
\noindent\textbf{Example 2 (Multiple-Cycle File Transition Graph)}:  
\label{sec:illus_ex}
{\it 
Consider a shuffling problem with parameters $K = 6$, $S = 3$, and $N = 6$.  
For simplicity, we rename the files $\{F^1, F^2, F^3, F^4, F^5, F^5\}$ to $\{A,B,C,D,E,F\}$.
Assume worker nodes $W_1$, $W_2$, $W_3$, $W_4$, $W_5$ and $W_6$ are processing files $A$, $B$, $C$, $D$, $E$, and $F$, respectively.
The file transition graph is depicted by Fig.~\ref{fig:ex2_ex3_fileTransGraph}. 
It comprises $\gamma = 3$ cycles, with cycle lengths $(\ell_1, \ell_2, \ell_3)=(3,1,2)$. That is, $d(1)=B$, $d(2)=C$, $d(3)=A$ in the first cycle,  $d(4)=D$ in the second cycle, and $d(5)=F$, $d(6)=E$ in the third cycle. 
Fig.~\ref{fig:ex2_cacheOrg} captures the cache organization of worker nodes, along with the missing subfiles that need to be processed at the next iteration.
Note that $\proc{i}$ and $\ex{i}{}$, for $i \in [K]$, are designed according to \eqref{eq:P-def} and \eqref{eq:E-def}, respectively.
We use \eqref{eq:X_all} and \eqref{eq:X_Delta} to design the broadcast message $\mathcal{X}$, which is constructed by concatenating a number of $X_\Delta$, each intended for $S=3$ worker nodes. For example, $X_{123}$ is expressed~as
\begin{IEEEeqnarray}{lCl}
	X_{123} 
	&=&
	\bigoplus_{i \in \{1,2,3\}} \left( 
	\F{i}{\{1,2,3\} \setminus \{i\}} 
	\oplus
	\F{d(i)}{\{1,2,3\} \setminus \{d(i)\}}
	\oplus	
	\bigoplus_{ j \in [K] \setminus \{1,2,3\}} 
	\F{d(i)}{(\{j\} \cup \{1,2,3\})\setminus \{i, d(i)\}} 
	\right)
	\nonumber\\
	&=&
	\left(F^1_{23} \oplus F^2_{13} \oplus \left(F^2_{34} \oplus F^2_{35} \oplus F^2_{36}\right)\right)
	\oplus
	\left(F^2_{13} \oplus F^3_{12} 
	\oplus \left(F^3_{14}
	\oplus F^3_{15} \oplus F^3_{16}\right)\right)
	\oplus
	\left(F^3_{12} \oplus F^1_{23} \oplus \left(F^1_{24} \oplus F^1_{25} \oplus F^1_{26}\right)\right)
	\nonumber\\
	&=&
	F^1_{24} \oplus F^1_{25} \oplus F^1_{26}
	\oplus
	F^2_{34} \oplus F^2_{35} \oplus F^2_{36}
	\oplus
	F^3_{14} \oplus F^3_{15} \oplus F^3_{16}
	\nonumber\\
	&=&
	A_{24} \oplus A_{25} \oplus A_{26} \oplus B_{34} \oplus B_{35} \oplus B_{36} \oplus C_{14} \oplus C_{15} \oplus C_{16}.
	\label{eqn:recvFunctionEx1_part0} 
\end{IEEEeqnarray}
Similarly, the set of other broadcast sub-messages is expressed~as
\begin{align}
	\begin{split}
		X_{124} &= A_{24} \oplus B_{34} \oplus B_{45} \oplus B_{46} \oplus C_{14},
		\\
		X_{125} &= A_{25} \oplus B_{35} \oplus B_{45} \oplus B_{56} \oplus C_{15} \oplus E_{12} \oplus F_{12},
		\\
		X_{134} &= A_{24} \oplus A_{45} \oplus A_{46} \oplus B_{34} \oplus C_{14},
		\\
		X_{135} &= A_{25} \oplus A_{45} \oplus A_{56} \oplus B_{35} \oplus C_{15} \oplus E_{13} \oplus F_{13},
		\\
		X_{145} &= A_{45} \oplus B_{45} \oplus E_{14} \oplus F_{14},
		\\
		X_{234} &= A_{24} \oplus B_{34} \oplus C_{14} \oplus C_{45} \oplus C_{46},
		\\
		X_{235} &= A_{25} \oplus B_{35} \oplus C_{15} \oplus C_{45} \oplus C_{56} \oplus E_{23} \oplus F_{23},
		\\
		X_{245} &= B_{45} \oplus C_{45} \oplus E_{24} \oplus F_{24},
		\\
		X_{345} &= A_{45} \oplus C_{45} \oplus E_{34} \oplus F_{34}.
	\end{split}
	\label{eqn:recvFunctionEx1} 
\end{align}
It should be noted that no subfiles of file $D$ appears in any of the sub-messages defined \eqref{eqn:recvFunctionEx1} since $u(4)=d(4)=D$, i.e., $D$ is processed by $W_4$ at iterations $t$ and $t+1$,  and hence does not need to be  transmitted by the master node for this data shuffle. Moreover, note that index $6$ does not appear in the subscript of the broadcast sub-messages, since $W_6$ is the ignored worker node. However, the subfiles assigned to $W_6$ can be recovered by linear combination of other transmitted sub-messages. 

\begin{figure}
	\centering
	\subfloat[]{\includegraphics[width=3.75in]{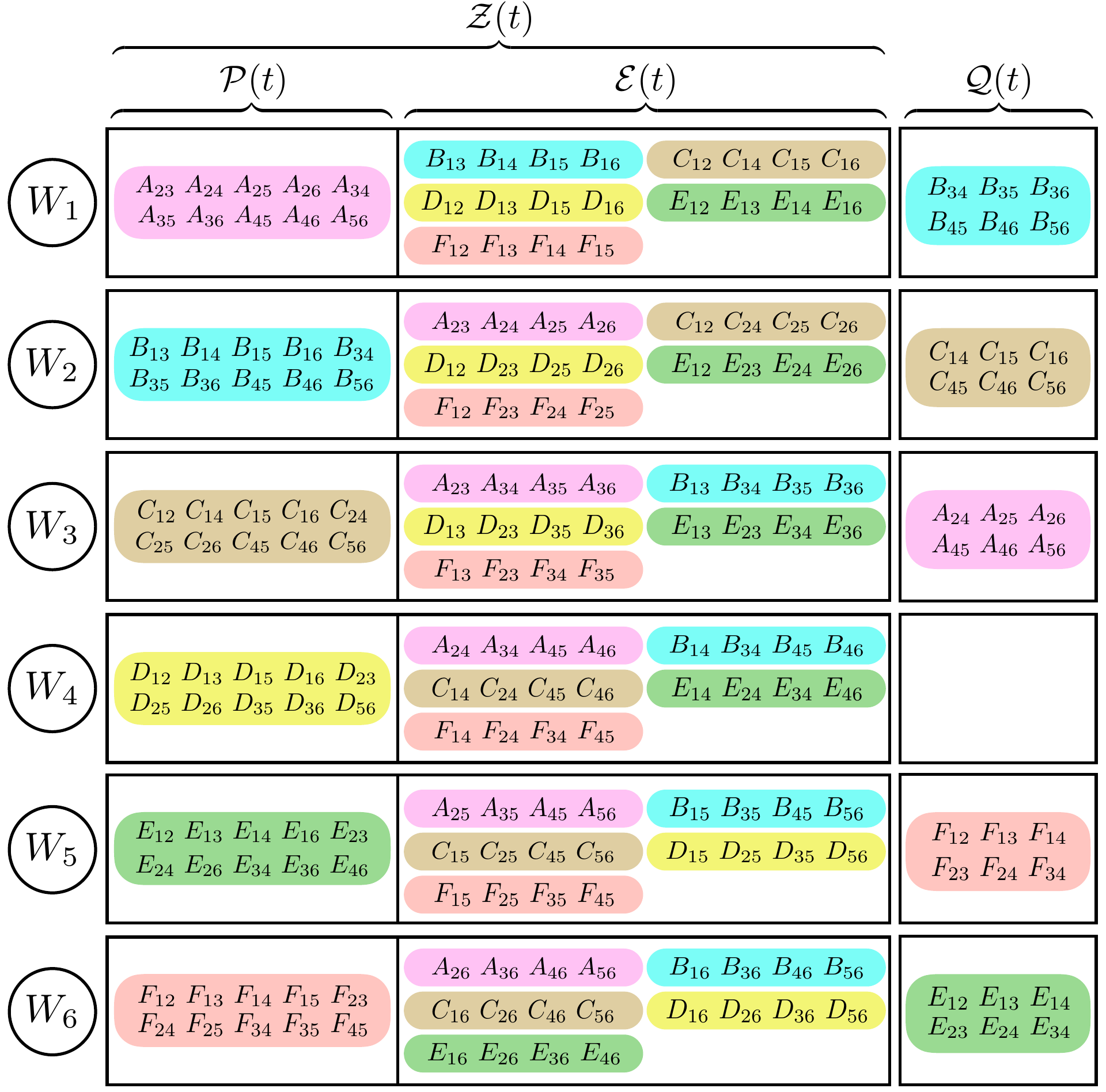}%
	\label{fig:ex2_cacheOrg}}
	\hspace{15mm}
	\subfloat[]{\includegraphics[width=2.72in]{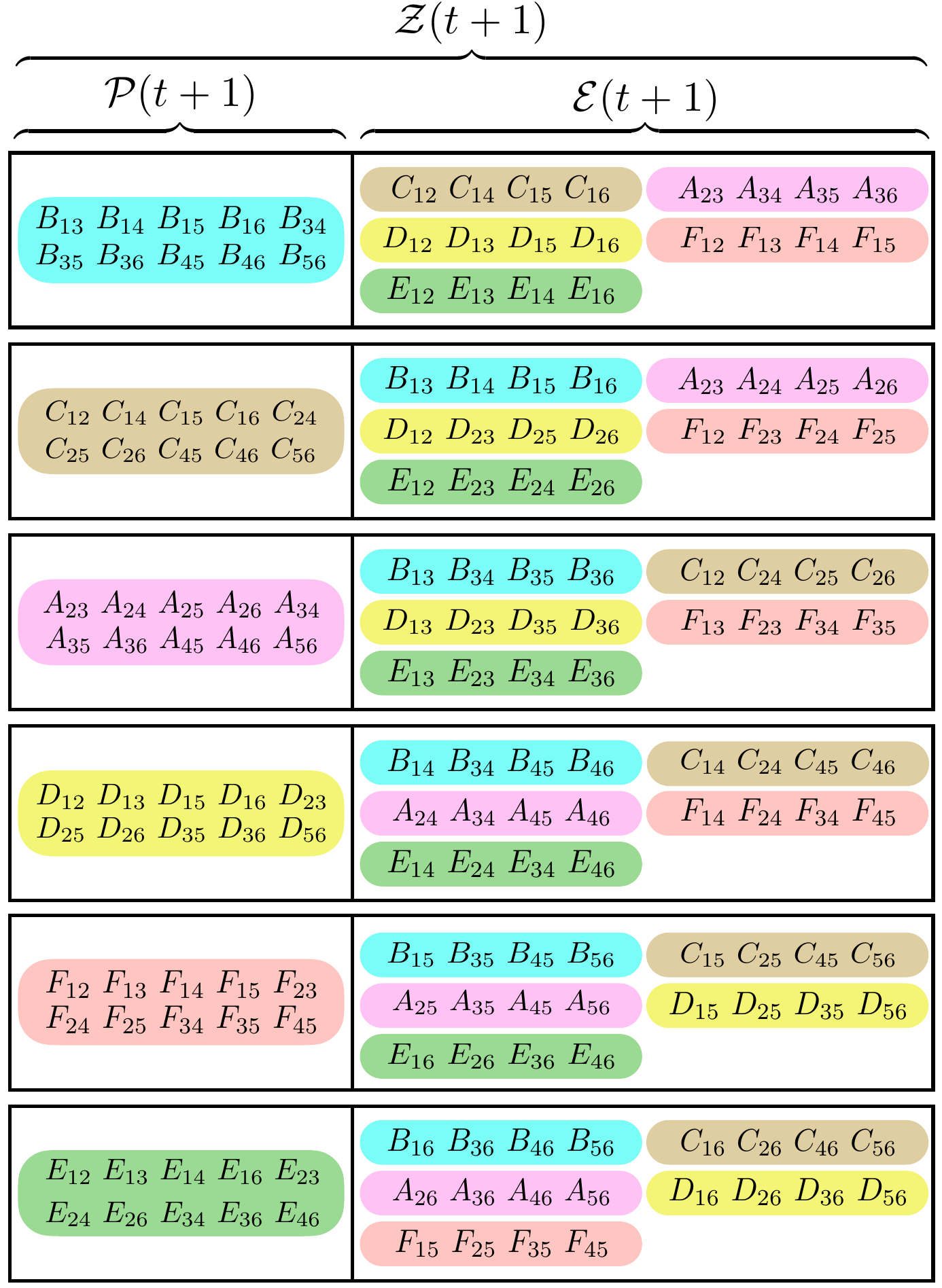}%
	\label{fig:ex2_cacheUpdate}}
	\\
	\centering
	\subfloat[]{\includegraphics[width=7in]{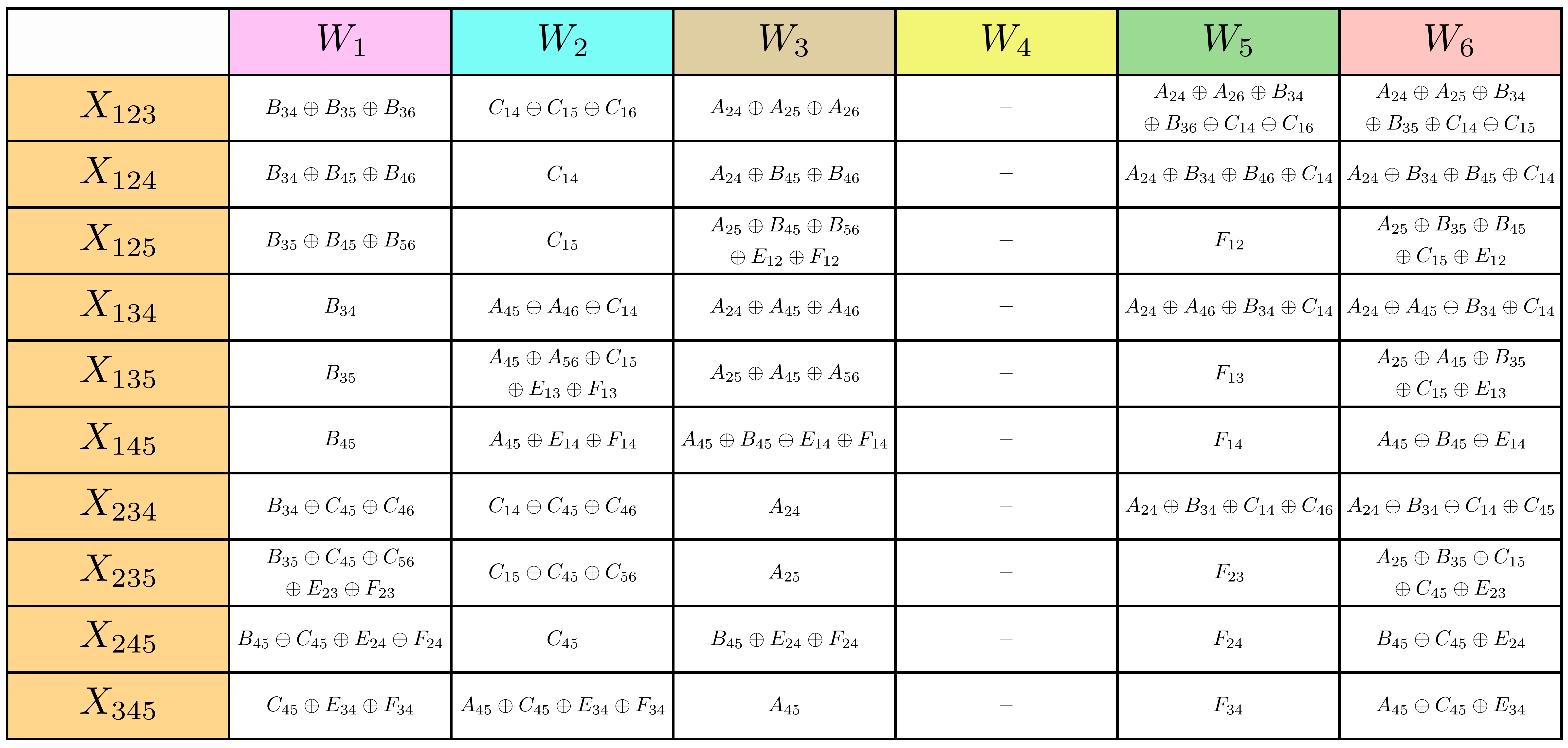}%
	\label{fig:ex2_functions}}
	\caption{
		Data shuffling system with $N = K = 6$, $S = 3$ and $\gamma = 3$. The file transition graph is depicted in Fig.~\ref{fig:ex2_ex3_fileTransGraph}.
		(a)~Cache organization of worker nodes at iteration $t$, along with the set of subfiles which are not available in the caches at iteration $t$ and need to be processed at iteration $t+1$.
		(b) Cache organization of worker nodes at iteration $t+1$ after updating the caches.
		For instance, subfiles $\left\{B_{13}, B_{14}, B_{15}, B_{16}\right\}$ in $\ex{1}{}(t)$ are moved to $\ex{2}{}(t+1)$.
		(c) Received functions by worker nodes after removing the cached subfiles. The complete received functions at worker nodes are expressed in \eqref{eqn:recvFunctionEx1_part0} and \eqref{eqn:recvFunctionEx1}.
	}
	\label{fig:ex2_cache_update_fn}
\end{figure}

For each worker node, Fig.~\ref{fig:ex2_functions} shows the received sub-messages  from the master node after removing the subfiles that already exist in its cache.
The decoding procedure of the proposed coded shuffling scheme is analogous to interference mitigation techniques in wireless communications. To present this analogy, we focus on three different cases of the decoding procedure.

(i) Decoding $C_{14}$ from $X_{124}$ by $W_2$: 
\begin{IEEEeqnarray}{lCl}
	X_{124} & = & 
	\underbrace{C_{14}}_{\text{Desired subfile}} \oplus 
	\underbrace{A_{24} \oplus B_{34} \oplus B_{45} \oplus B_{46}}_{\text{Cached subfiles in $\mathcal{Z}_2$}}.
	\nonumber
\end{IEEEeqnarray}
The decoding procedure is analogous to interference suppression technique. 
$W_2$ decodes $C_{14}$ by canceling the interfering subfiles using its cache contents~$\mathcal{Z}_2$.

(ii) Decoding $C_{16}$ from $X_{123}$ by $W_2$: 
\begin{IEEEeqnarray} {lCl}
	X_{123} & = & 
	\underbrace{C_{16}}_{\text{Desired subfile}} \oplus 
	\underbrace{C_{14}}_{\substack{\text{Decoded subfile}\\ \text{from $X_{124}$}}} \oplus
	\underbrace{C_{15}}_{\substack{\text{Decoded subfile} \\ \text{from $X_{125}$}}}
	\oplus
	\underbrace{A_{24} \oplus A_{25} \oplus A_{26} \oplus B_{34} \oplus B_{35} \oplus B_{36}}_{\text{Cached subfiles in $\mathcal{Z}_2$}}.
	\nonumber
\end{IEEEeqnarray}	
The decoding procedure is analogous to successive interference cancellation (SIC) technique.
$W_2$ decodes $C_{16}$ by first canceling the subfiles that exist in~$\mathcal{Z}_2$. Next, it exploits the subfiles decoded from $X_{124}$ and $X_{125}$ to successively cancel the remaining interfering subfiles.

(iii) Decoding $E_{23}$ from $X_{235}$ by $W_6$:
\begin{IEEEeqnarray} {lCl}
	X_{235} & = & 
	\underbrace{E_{23}}_{\text{Desired subfile}} \oplus
	\underbrace{\left(C_{56} \oplus  F_{23}\right)}_{\substack{\text{Cached subfiles} \\ \text{in $\mathcal{Z}_6$}}} \oplus
	\underbrace{\left(A_{25} \oplus B_{35} \oplus C_{15} \oplus C_{45} \right)}_{\text{$X_{123} \oplus X_{234}$}}.
	\nonumber
\end{IEEEeqnarray}
The decoding procedure is analogous to aligned interference suppression (interference alignment) technique. 
$W_6$ decodes $E_{23}$ by first canceling the subfiles cached in~$\mathcal{Z}_6$.
Then, the remaining interfering subfiles are the result of XORing some other received sub-messages, i.e., $X_{123} \oplus X_{234}$, and hence, they can be canceled accordingly.

As a result, the achieved delivery load is 
$R_{\text{coded}} = \binom{5}{3} / \binom{5}{2} = 1$.
On the other hand, the delivery load achieved by the uncoded shuffling scheme, under the same placement strategy, is 
$R_{\text{uncoded}} = (5 \times 6) / \binom{5}{2} = 3$.
That is, the proposed coded shuffling scheme can save around $66\%$ of the communication load, and thus, it speeds up the overall run-time of the data shuffling process. 
When each worker node decodes all missing subfiles at iteration $t$, Fig.~\ref{fig:ex2_cacheUpdate} depicts the cache organization of each worker node after updating the cache in preparation for the following data shuffle at iteration $t+1$.
Note that the subfiles can be relabeled in a similar way as in Example~$1$. \hfill $\blacklozenge$
}

\subsection{A Graph-Based Delivery Scheme for Any Shuffling: Proof of Theorem~\ref{thrm2}}
\label{sec:achv_scheme_optimal_N=K}
The coded shuffling scheme proposed in Section~\ref{sec:achv_scheme_N=K} provides the worker nodes with the missing parts of their assigned files, using the broadcast message and cached subfiles.
However, depending on the file transition graph, the delivery load obtained by that scheme may be sub-optimum. As an extreme and hypothetical example, consider a file transition graph where each worker node $W_i$ is assigned the same file to process at iterations $t$ and $t+1$, i.e., $d(i)=u(i)$.  
Clearly, no communication between the master node and worker nodes is needed in this case, and hence, $R=0$ is achievable. This implies the scheme in Section~\ref{sec:achv_scheme_N=K} is sub-optimal for this instance of the shuffling problem.

It turns out that the \emph{number of cycles} in the file transition graph is the main characteristic to determine the optimum delivery scheme. More concretely, for a file transition graph with $\gamma$ cycles  where $\gamma - 1 \geq S$, we show that there are precisely $\binom{\gamma-1}{S}$~sub-messages in \eqref{eq:X_all} that are 
linearly dependent on the other sub-messages. Thus, by refraining from broadcasting these sub-messages, we can reduce the delivery load, and achieve the one given in~\eqref{eq:R_achv2}. In the decoding phase, worker nodes can first recover all the redundant sub-messages that have not been transmitted by the master node by computing linear combinations of appropriate sub-messages that have been received.
Then, each worker node follows the same decoding rules, discussed in Section~\ref{sec:achv_scheme_decode}, to decode the assigned file at iteration $t+1$. This results in an opportunistic coded shuffling scheme based on the scheme proposed in  Section~\ref{sec:achv_scheme_N=K}. We will later show in Section~\ref{sec:converse_proof_N=K_All} that this scheme is indeed optimum, and achieves the minimum possible delivery load.  

The following lemma characterizes the linearly dependent sub-messages, and quantifies the reduced delivery load:
\begin{lm}
	\label{lm:OptAchvScheme}
	Consider a data shuffling  system with a master node, $K$ worker nodes with storage capacity per worker node $S$, and $N=K$ files, and a given file transition graph that comprises $\gamma$ cycles. Consider the placement strategy given in Section~\ref{sec:cache_placement} and the delivery scheme provided in Section~\ref{sec:achv_scheme_N=K}. Then, there are a total of $\binom{\gamma-1}{S}$ redundant (linearly dependent) sub-messages among the $\binom{K-1}{S}$ broadcasting sub-messages. 
\end{lm}
We refer to Appendix~\ref{app:opt_achvScheme} for the proof of lemma~\ref{lm:OptAchvScheme}.

In fact, we can explicitly characterize the set of redundant sub-messages that are not broadcast in the delivery phase. To this end, 
\begin{itemize}[leftmargin=*]
	\item consider the first  $\gamma -1$ cycles out of the total of $\gamma$ cycles formed by the file transition graph (generally, we have to consider all cycles except the one that includes the ignored worker node);
	\item from these $\gamma -1$ cycles, consider all possible combinations of sets of cycles that have $S$ distinct cycles. There are $\binom{\gamma-1}{S}$~such sets;
	\item consider all sub-messages $X_{\Delta}$  (defined in \eqref{eq:X_Delta}) where $\Delta$ has exactly one worker node from each of the chosen $S$ cycles. 
\end{itemize}
In the proof of Lemma~\ref{lm:OptAchvScheme}, we show that the sum of all such sub-messages is zero. 
Thus, the master node can remove one of the sub-messages from each group and transmit the rest of them to the worker nodes.
Therefore, the resulting broadcast communication load $R$ is upper bounded~by
\begin{IEEEeqnarray}{lCl}
	R & \leq & \frac{\binom{K-1}{S} - \binom{\gamma-1}{S}}{\binom{K-1}{S-1}}.
	\nonumber
\end{IEEEeqnarray}
In the decoding phase, each worker node first reconstructs the missing, redundant sub-messages by adding the other sub-messages in the group, and then follows the same decoding scheme, and cache updating and subfile relabeling strategies introduced in Section~\ref{sec:achv_scheme_N=K}.
This completes the proof of Theorem~\ref{thrm2}. 
In the next example, we illustrate the concept of redundancy within the the set of broadcast sub-messages. 

\subsection{Illustrative Example}
\label{sec:achv_scheme_optimal_N=K_ex}
\noindent\textbf{Example 3 (Multiple-Cycle File Transition Graph with Opportunistic Transmission)}: 
{\it 
\begin{figure}   
	\centering
	\includegraphics[height=0.38\textwidth]{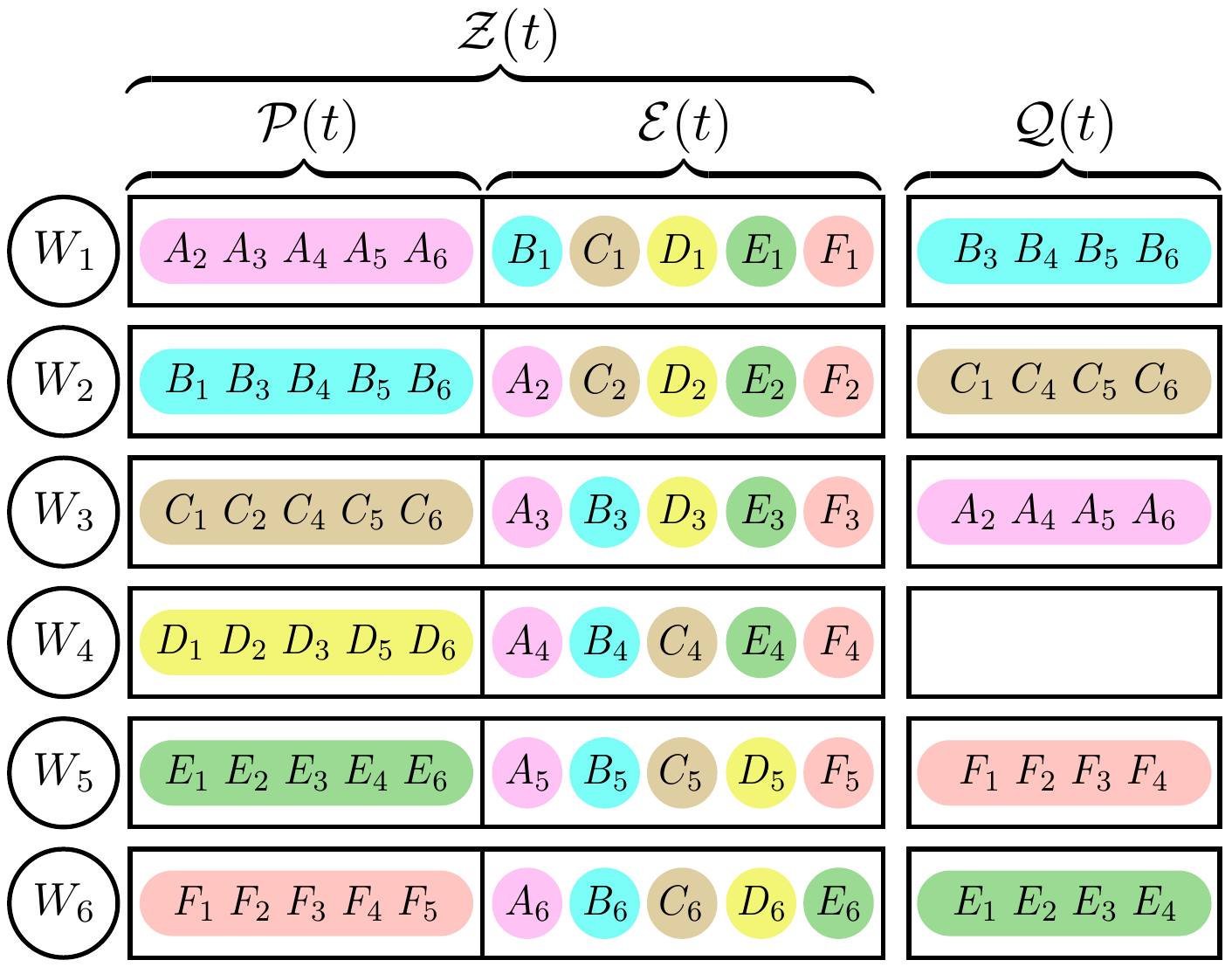}
	\caption{
	Cache organization of worker nodes at iteration $t$, along with the set of subfiles which are not available in the caches at iteration $t$ and need to be processed at iteration $t+1$, for a data shuffling system with $N = K = 6$, $S = 2$ and $\gamma = 3$. 
	The file transition graph for this system is depicted in Fig.~\ref{fig:ex2_ex3_fileTransGraph}.
	}
	\label{fig:ex3_cacheOrg}
\end{figure}
We consider the same system parameters of Example~2 in Section~\ref{sec:illus_ex} with the same file transition graph given in Fig.~\ref{fig:ex2_ex3_fileTransGraph}, except that here the cache size of each worker node is $S = 2$ files.
Fig.~\ref{fig:ex3_cacheOrg} captures the cache organization of worker nodes, along with the missing subfiles that need to be processed at iteration $t+1$. 
Following the achievable scheme proposed in \ref{sec:achv_scheme_N=K}, the set of sub-messages transmitted by the master node to the worker nodes is expressed~as
\begin{align}
	\begin{split}
		X_{12} & =  A_2 \oplus B_3 \oplus B_4 \oplus B_5 \oplus B_6 \oplus C_1,
		\\
		X_{13} & =  A_2 \oplus A_4 \oplus A_5 \oplus A_6 \oplus B_3 \oplus C_1,
		\\
		X_{14} & =  A_4 \oplus B_4,
		\\
		X_{15} & =  A_5 \oplus B_5 \oplus E_1 \oplus F_1,
		\\
		X_{23} & =  A_2 \oplus B_3 \oplus C_1 \oplus C_4 \oplus C_5 \oplus C_6,
		\\
		X_{24} & =  B_4 \oplus C_4,
		\\
		X_{25} & =  B_5 \oplus C_5 \oplus E_2 \oplus F_2,
		\\
		X_{34} & =  A_4 \oplus C_4,
		\\
		X_{35} & =  A_5 \oplus C_5 \oplus E_3 \oplus F_3,
		\\
		X_{45} & =  E_4 \oplus F_4.
	\end{split}
	\label{eqn:recvFunction_Ex2}
\end{align}
Note that the file transition graph consists of $\gamma=3$ cycles, namely, $\{W_1, W_2, W_3\}$, $\{W_4\}$, and $\{W_5, W_6\}$, where the ignored worker node $W_6$ appears in the third cycle. Ignoring the third cycle, we have two remaining ones. The family of all $\set$'s with exactly one entry from each of the first and second cycle is given by $\{\{1,4\}, \{2,4\}, \{3,4\}\}$. It is evident that $X_{14}$, $X_{24}$ and $X_{34}$ are linearly dependent, since $X_{34}\oplus X_{14} \oplus X_{24} = 0$. Therefore, we can safely omit~$X_{34}$ from the set of the transmitted sub-messages. 
For decoding purposes, worker nodes can recover $X_{34}$ from $X_{14}$ and $X_{24}$. Therefore, each one of them is able to decode the assigned file at iteration $t+1$ by following the decoding procedure presented in Section~\ref{sec:achv_scheme_decode}. \hfill $\blacklozenge$
}

\section{Converse Proof for Data Shuffling with Canonical Setting $(N=K)$}
\label{sec:converse_proof_N=K_All}
We prove the optimality of the coded shuffling scheme proposed in Section~\ref{sec:achv_scheme_optimal_N=K}, as stated in Theorem~\ref{thrm3}. The instance of the problem is fully determined by the indices of files to be processed by the worker nodes at iterations $t$ and $t+1$, i.e., $u(\cdot)$~and $d(\cdot)$, respectively. Without loss of generality, we may assume $u(i)=i$ for every $i\in [K]$, otherwise we can relabel the files. These assignment functions induce a directed file transition graph $\cG(V,E)$, with node set $V=\{W_1,W_2,\dots, W_K\}$, in which there is an edge from $W_i$ to  $W_j$  if and only if $ d(W_j) = F^i$. 

Since the in-degrees and out-degrees of each vertex are equal to $1$, such a graph consists of a number of directed cycles. Let  $\gamma$ denote the number of cycles in this graph with cycle lengths given by $\{\ell_1,\ell_2,\dots, \ell_\gamma\}$. 
Then each node in the graph can be represented by a pair $(c,p)$, where $c\in \{1,2,\dots, \gamma\}$ denoted the cycle number, and $p\in \{1,2,\dots, \ell_c\}$ denotes the position within cycle $c$. With this notation\footnote{The selection of the order of cycles, as well as the starting position within each cycle, is arbitrary.}, we have $d(c,p) = u(c,p+1) = (c,p+1)$, i.e., the worker node at position $(c,p)$ will process file $F^{(c,p)}$ and $F^{(c,p+1)}$ at iterations $t$ and $t+1$, respectively\footnote{We define $(c, \ell_c+1) = (c, 1)$ for the sake of consistency.}.
Note that the \emph{positional label} $(c,p)$ essentially induces an \emph{order} on the nodes and edges of the graph. For two pairs $(c',p')$ and $(c,p)$, we say $(c',p')$ appears before $(c,p)$ and denote it by $(c',p') \prec (c,p)$ if either $(c',p')$ appears in a cycle with smaller index ($c'<c$) or in the same cycle but at an smaller position ($c'=c$ and $p'<p$). Similarly, $(c',p') \preceq (c,p)$ indicates that either $(c',p') \prec (c,p)$  or $(c',p') = (c,p)$. 

Consider an arbitrary \emph{uncoded placement} of the files in worker nodes' cache, and  denote by $\tF^i_j$ the bits of file $F^i$ cached at $W_j$. Note that we do not make any assumptions on the size or symmetry of $\fu{i}{j}$'s. Then, the cache contents of worker node $W_j$ is given by 
$\cZ_j = F^{j} \cup \bigcup_{i\neq j} \fu{i}{j}$,
which is equivalent to 
\begin{IEEEeqnarray}{lCl}
	\cZ_{(c,p)} = F^{(c,p)} \cup \left(\bigcup_{(c',p') \neq (c,p)} \fu{(c',p')}{(c,p)}\right),
	\label{eq:Z_positional}
\end{IEEEeqnarray}
using the positional labeling, where $j$ is the worker node in the $p$th position of the $c$th cycle. 

For a given file transition graph $\cG(V,E)$, we introduce a virtual worker node $W_\star$ equipped with a cache $\cZ_\star$, in which we store
\begin{IEEEeqnarray}{lCl}
	\cZ_\star &= & \left( \bigcup_{c\in \left[\gamma\right]} F^{(c,1)} \right) \cup 
	\left( \bigcup_{\substack{c \in \left[\gamma\right] \\ p < \ell_{c} }}  \left( \bigcup_{\substack{(c',p') \succ (c,p)\\ p'>1}}   \fu{(c',p')}{(c,p)}
	\right)  \right) 
	\label{eq:Z_star_0}
	\\
	& = & \left( \bigcup_{c\in \left[\gamma\right]} F^{(c,1)} \right) \cup 
	\left( \bigcup_{\substack{c \in \left[\gamma\right] \\ p >1}} 
	\left(\bigcup_{\substack{(c',p') \prec (c,p) \\p'<\ell_{c'} }} \fu{(c,p)}{(c',p')} \right)\right),
	\label{eq:Z_star}
\end{IEEEeqnarray}
where
\begin{itemize}[leftmargin=*]
	\item in \eqref{eq:Z_star_0}, 
	the first equality reads as follows. the cache $\mathcal{Z}_\star$ of the virtual worker node $W_\star$ is the union of two sets. 
	The first set is the union of files being processed by the first worker node in each cycle. The second set only consists the sub-files of all other files, and only includes sub-files that are stored at worker nodes that (i) whose rank is the circular order is less that that of the processing worker, and (ii) do not appear at the last position of their cycle.
	\item in \eqref{eq:Z_star}, 
	the second equality in (25) is just a re-arrangement of the sub-files selected in (24). That will  follows form two steps applied simultaneously, i.e., (i)  swapping the labels used for the subscript and superscripts of $\widetilde{F}$, and (ii)  changing the order of the unions.
\end{itemize}

In order to prove the optimality of the proposed coded shuffling scheme, we first show that the broadcast message~$\cX$ and the cache contents of the virtual worker node suffice to decode all the files (Lemma~\ref{lemma_virtualNode}). Then, we lower bound the size of the broadcast message by upper bounding the size of the data cached at the virtual worker node (Lemma~\ref{lemma_R} and Lemma~\ref{lemma_opt_mu}).

\begin{lm}
	\label{lemma_virtualNode}
	For any file transition graph $G$, given the cache contents $\cZ_\star$ of the virtual worker node~$W_\star$ and the broadcast message $\cX$, all files in the data shuffling system can be decoded, that is 
	\begin{IEEEeqnarray}{lCl}
		H\left(\{F^{i}\}_{i=1}^{K}| \cX, \cZ_\star\right) & = & 0.
		\label{eq:induction}
	\end{IEEEeqnarray}
\end{lm}
We refer to Appendix~\ref{app:lemma_virtualNode} for the proof of Lemma~\ref{lemma_virtualNode}. 

For a file $F^i$ and a subset of worker nodes $\cJ\subseteq [K] \setminus \{i\}$, we define the size of a union of bits of $F^i$ that are cached at worker nodes in $\cJ$ as $\mu^i_{\cJ} = \left| \bigcup\nolimits_{j\in \cJ} \tF^i_j \right|$. For a given integer $\alpha$, let $\mu_{\alpha}$ denote the average (over files and worker nodes) size of a set of union of bits that are cached in the excess storage of $\alpha$ worker nodes, that is,
\begin{IEEEeqnarray}{lCl}
	\label{eqn:mu_eta}
	\mu_{\alpha} & = &
	\frac{1}{K \binom{K-1}{\alpha}} \sum_{i \in [K]} \sum_{\substack{\cJ \subseteq [K] \setminus i \\ |\cJ| = \alpha}} \mu^i_\cJ \nonumber\\
	&=& \frac{1}{K \binom{K-1}{\alpha}} \sum_{i \in [K]} \sum_{\substack{\cJ \subseteq [K] \setminus i \\ |\cJ| = \alpha}}
	\left| \bigcup\nolimits_{j\in \cJ} \tF^i_j \right|.
\end{IEEEeqnarray}

Next, the communication load for an instance of the shuffling problem, determined by a graph $\cG(V,E)$, can be lower bounded in terms of $\mu_i$'s as~follows.
\begin{lm}
	\label{lemma_R}
	For a data shuffling problem determined by a file transition graph $\cG(V,E)$,  the communication load $R$ is lower bounded~by
	\begin{IEEEeqnarray}{lCl}
		\label{eq:R}
		R(\cG)
		& \geq &
		K - \gamma - \sum_{i=1}^{K - \gamma}\mus{}{i}.	
	\end{IEEEeqnarray}
\end{lm}
We refer to Appendix~\ref{app:lemma_R} for the proof of Lemma~\ref{lemma_R}.

\begin{figure}
	\centering
	\subfloat[]{\includegraphics[width=5.3in]{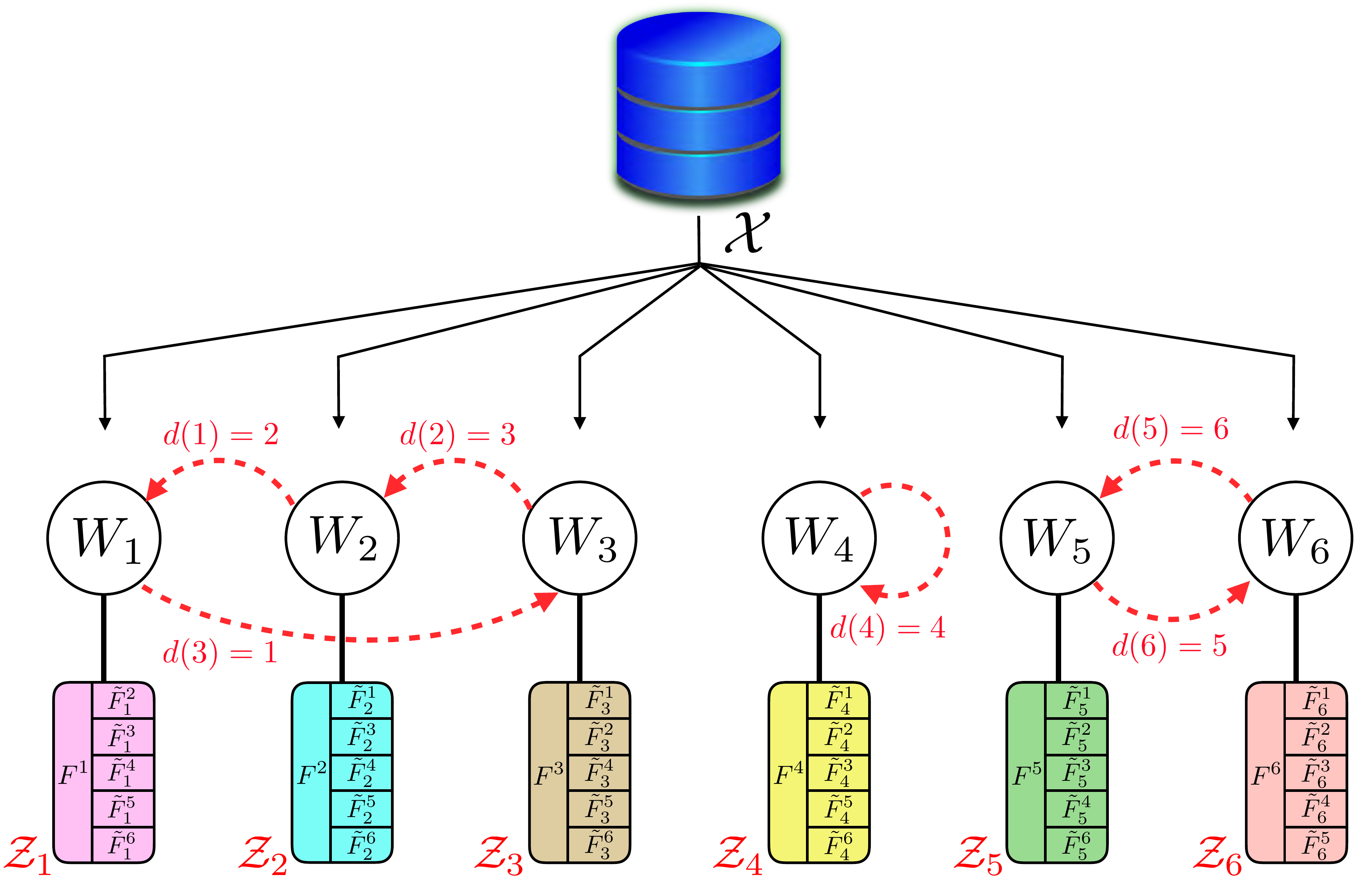}%
	}
	\caption{
		The file transition graph, along with the cache organization of worker nodes, for a data shuffling system with $N = K = 6$, and $\gamma = 3$.
	}
	\label{fig:ex4_converse}
\end{figure}

Before we continue with the formal proof, we present the defined notation and the main steps of our argument through an illustrative example. 

\noindent \textbf{Example 4}: 
{\it 
Consider a data shuffling system with $K=6$ worker nodes, and $N = 6$ files, namely $\left\{F^1, F^2, F^3, F^4, F^5, F^6\right\}$. We assume worker node $i$ is processing $F^i$ at time $t$. Moreover, the file assignments for iteration $t+1$ are given by $d(1)=2$, $d(2)=3$, $d(3)=1$, $d(4)=4$, $d(5)=6$, and $d(6)=5$. The file transition graph of the problem is depicted by Fig.~\ref{fig:ex4_converse}, that consists of $\gamma=3$ cycles, with cycle lengths $(\ell_1, \ell_2, \ell_3)=(3,1,2)$. Hence, the nodes are labeled using the $(c,p)$ notation as follows: $1 \leftrightarrow (1,1)$, $2 \leftrightarrow (1,2)$, $3 \leftrightarrow (1,3)$, $4 \leftrightarrow (2,1)$, $5 \leftrightarrow (3,1)$, and $6 \leftrightarrow (3,2)$. 
The cache contents of the virtual worker node will be
\begin{IEEEeqnarray*}{lCl}
	\cZ_\star &=& \left\{F^{(1,1)}, F^{(2,1)}, F^{(3,1)}\right\} 
	\cup 
	\left\{\fu{(1,2)}{(1,1)}, \fu{(1,3)}{(1,1)}, \fu{(1,3)}{(1,2)}, \fu{(3,2)}{(1,1)}, \fu{(3,2)}{(1,2)}, \fu{(3,2)}{(3,1)}\right\},
\end{IEEEeqnarray*}

Next, according to Lemma~\ref{lemma_virtualNode}, we argue that given $\cZ_\star$ and $\cX$,  the virtual worker node is able to decode all files as follows:
\begin{itemize}[leftmargin=*]
	\item $W_\star$ has the entire files $F^{(1,1)}$, $F^{(2,1)}$, and  $F^{(3,1)}$,  i.e.,    $H\left(F^{(1,1)}, F^{(2,1)}, F^{(3,1)} \Big|Z_\star\right)=0$. 
	\item It also has the entire data cached at $W_1$, that is, 
	\begin{IEEEeqnarray*}{l}
		\cZ_1=\left\{F^{(1,1)}, \fu{(1,2)}{(1,1)}, \fu{(1,3)}{(1,1)}, \fu{(2,1)}{(1,1)}, \fu{(3,1)}{(1,1)}, \fu{(3,2)}{(1,1)}\right\}.
	\end{IEEEeqnarray*}    
	Hence, it can decode $d(1)=F^{(1,2)}$ from $(\cZ_1,\cX)$, which implies $H\left(F^{(1,2)}\Big|\cZ_\star, \cX\right)=0$. 
	\item Then, having $F^{(1,2)}$ decoded, the virtual worker node has the entire content of $\cZ_2$ which, together with~$\cX$, suffices to decode the file assigned to $W_2$, which is $d(2)=3 \leftrightarrow (1,3)$. That is, $H\left(F^{(1,3)}\Big|\cZ_\star, \cX, F^{(1,2)}\right)=0$.
	\item Finally,  since files $F^{(1,1)}$, $F^{(2,1)}$,  $F^{(3,1)}$ and $\fu{(3,2)}{(3,1)}$ are in $\cZ_\star$, and files $F^{(1,2)}$, $F^{(1,3)}$ are decoded, the virtual worker node has the entire $\cZ_5$. Hence, the only remaining file $F^{(3,2)}$ can be recovered from $(\cZ_5,\cX)$ since $d(5)=6 \leftrightarrow (3,2)$. This implies $H\left(F^{(3,2)}\Big|\cZ_\star, \cX, F^{(1,2)}, F^{(1,3)}\right)=0$. 
\end{itemize}
This argument shows that 
\begin{IEEEeqnarray}{lCl}
	6 &=&  H\left(F^{1}, F^{2}, F^{3}, F^{4}, F^{5}, F^{6}\right) 
	\nonumber\\
	& \leq & H(\cX, \cZ_\star) 
	\nonumber\\
	& \leq & H(\cX) +  H\left(F^{1}\right) + H\left(F^{4}\right) + H\left(F^{5}\right) + H\left(\fu{2}{1}\right) 
	+ H\left(\fu{3}{1}, \fu{3}{2}\right) + H\left(\fu{6}{1},\fu{6}{2}, \fu{6}{5}\right), \nonumber
\end{IEEEeqnarray}
which implies 
	\begin{IEEEeqnarray}{lCl}
		R &=& H(\cX) \nonumber \\
		& \geq & 
		6 - \left(H\left(F^{1}\right) + H\left(F^{4}\right) + H\left(F^{5}\right)\right) 
		- \left(H\left(\fu{2}{1}\right) + H\left(\fu{3}{1}, \fu{3}{2}\right) + H\left(\fu{6}{1},\fu{6}{2}, \fu{6}{5}\right) \right)
		\nonumber\\
		& = & 3 - \left(H\left(\fu{2}{1}\right) + H\left(\fu{3}{1}, \fu{3}{2}\right) + H\left(\fu{6}{1},\fu{6}{2}, \fu{6}{5}\right) \right)
		\nonumber\\
		& \geq & 3 - \left(\mus{}{1} + \mus{}{2} + \mus{}{3}\right), 
		\nonumber	
	\end{IEEEeqnarray}
	where the last inequality follows from a similar argument on several versions of the same problem with re-labeled files, using set-theoretic operations. To avoid any repetition, we omit these steps here, and refer to the proof of Lemma~\ref{lemma_R} in \ref{app:lemma_R}. Finally, we need to bound $\mu_1$, $\mu_2$, and $\mu_3$, which is elaborated by lemma~\ref{lemma_opt_mu}. 
\hfill $\blacklozenge$
} 

Finally, we seek an upper bound on $\mu_i$'s using a set theoretic argument, as stated in the following lemma:
\begin{lm}
	\label{lemma_opt_mu}
	The variables $\mu_{i}$'s introduced in \eqref{eqn:mu_eta} satisfy 
	\begin{IEEEeqnarray}{lCl}
		\mu_{i} & \leq & 1 - \frac{\binom{K-i-1}{S-1}}{\binom{K-1}{S-1}},
		\label{eq:opt-sol}
	\end{IEEEeqnarray}
	for $\alpha \in \{0,1,2,\dots, K-1\}$. 
\end{lm}
The proof of this lemma is presented in Appendix~\ref{app:lemma_lemma_opt_mu}. 
Having the lemmas above, we are ready to prove the optimality of the proposed coded shuffling scheme. 

\begin{IEEEproof}[Proof of Theorem~\ref{thrm3}]
	We  start with Lemma~\ref{lemma_R}, and use Lemma~\ref{lemma_opt_mu} to upper bound $\mu_i$'s. We have
	\begin{IEEEeqnarray}{lCl}
		R(\cG)
		& \geq &
		K - \gamma - \sum_{i=1}^{K - \gamma}\mu_i
		\nonumber\\
		& \geq &
		K - \gamma -  \left[K-\gamma - \sum_{i=1}^{K - \gamma} \frac{\binom{K-i-1}{S-1}}{\binom{K-1}{S-1}}\right].
		\nonumber\\
		&=&
		\frac{1}{\binom{K-1}{S-1}}
		\sum_{i=1}^{K - \gamma}  \binom{K-i-1}{S-1}
		\nonumber\\
		&=&
		\frac{1}{\binom{K-1}{S-1}}
		\sum_{j=\gamma-1}^{K - 2}  \binom{j}{S-1}
		\nonumber\\
		&=&
		\frac{1}{\binom{K-1}{S-1}}
		\left[
		\sum_{j=0}^{K-2}  \binom{j}{S-1} - \sum_{j=0}^{\gamma-2}  \binom{j}{S-1}
		\right]
		\nonumber\\
		&=&
		\frac{1}{\binom{K-1}{S-1}}
		\left[
		\binom{K-1}{S} - \binom{\gamma-1}{S}
		\right].
	\end{IEEEeqnarray}
	 This completes the proof of Theorem~\ref{thrm3}.
\end{IEEEproof}

\section{Coded Shuffling Scheme for General Setting $(N \geq K)$}
\label{sec:N>K_results}
We shift our attention to the general and practical setting of the data shuffling problem when $N \geq K$. We assume $N/K$, the number of files to be processed by each worker node, is integer. Furthermore, unless otherwise mentioned, we assume  $S/(N/K)$ is integer. 
It should be noted that the proposed coded shuffling algorithm for the general setting of $N \geq K$ is an extension to the one developed for the canonical setting of $N=K$.

\subsection{A Universal Delivery Scheme for General Setting: Proof of Theorem~\ref{thrm4}}
\label{sec:N_larger_K_system}
In the following, we extend the achievable scheme proposed in Section~\ref{sec:achv_scheme_optimal_N=K} for the canonical setting ($N=K$) in order to develop an achievable scheme for the general setting of the shuffling problem, that is, for any $N \geq K$.
To this end, we follow the cache placement scheme proposed in Section~\ref{sec:cache_placement}. Then, for the delivery phase, we \emph{decompose} the data shuffling problem into $N/K$ instances of sub-problems, where each sub-problem consists of $K$ worker nodes, $K$ files, and cache size of $\widehat{S}=S/(N/K)$ per worker node. Therefore, the resulting sub-problems lie in the class of the canonical setting discussed earlier in Section~\ref{sec:achv_scheme_N=K_All}. 

We first present the following lemma that is essential for the proof of Theorem~\ref{thrm4}:
\begin{lm}
	For any data shuffling problem with $K$ worker nodes and $N$ files (where $K$ divides $N$), $\mathcal{G} (V, E)$ can be decomposed into $N/K$ subgraphs $\cG_i(V, E_i)$ with $\vert E_i \vert = K$, for $i \in [N/K]$, such that
	\begin{itemize}[leftmargin=*]
		\item For each subgraph $\mathcal{G}_i (V, E_i)$, the in-degree and out-degree of each vertex in $V$ are 1.
		\item The edge sets $\{E_i: i\in[N/K]\}$ provide a partitioning for $E$, i.e., $E_i \cap E_j= \varnothing$ for any distinct pair of $i, j \in [N/K]$, and $\bigcup_{i=1}^{N/K} E_i = E$.
	\end{itemize}
	\label{lm:demposition}
\end{lm}
The proof of this lemma is based on Hall's theorem, and presented in  Appendix~\ref{app:lm:prfct_match}.

Now, we prove Theorem~\ref{thrm4} as follows. Recall from Section~\ref{sec:probForm} that the file transition graph is defined as a directed graph $\mathcal{G} (V,E)$, where an edge $e_j=(i,\ell)\in E$ with $j\in[N]$ indicates that $j\in u(i) \cap d(\ell)$, i.e., file $F^j$ is being processed by worker node $W_i$ at iteration $t$, and will be processed by worker node $W_\ell$ at iteration $t+1$.
Since each worker node processes $N/K$ files at every iteration of the distributed algorithm, the directed graph $\mathcal{G} (V,E)$ is regular since the in-degree and out-degree of each vertex in $V$ are $N/K$. The decomposition of the shuffling problem is inspired by decomposing the file transition graph. More precisely, we decompose $\mathcal{G} (V, E)$ into $N/K$ subgraphs, namely $\mathcal{G}_i (V, E_i)$ for $i \in [N/K]$, such that each subgraph induces one \emph{canonical} instance of the problem. 
The existence of such a decomposition is guaranteed by Lemma~\ref{lm:demposition}.

Each resulting subgraph $\mathcal{G}_i (V, E_i)$ induces  a data shuffling system with $K$ files (corresponding to the edges appear in the subgraph), $K$ worker nodes, and storage capacity per worker node $\widehat{S} = S/(N/K)$. 
Then, we can apply the universal coded shuffling scheme proposed in Section~\ref{sec:achv_scheme_optimal_N=K} to achieve a delivery load of $\binom{K-1}{\widehat{S}} / \binom{K-1}{\widehat{S} -1}$ for each sub-problem. 
As a result, an overall delivery load of 
\begin{IEEEeqnarray}{lCl}
	\displaystyle
	R & = & \frac{N}{K} \frac{\binom{K-1}{\widehat{S}}}{\binom{K-1}{\widehat{S}-1}}
	\nonumber
\end{IEEEeqnarray}
is achievable for any file transition graph. 
This completes the proof of Theorem~\ref{thrm4}. \hfill $\blacksquare$

The scheme is proved to be optimal for the \emph{worst-case shuffling scenario} with $K$ worker nodes and $N$ files in Section~\ref{sec:converse_proof_NlargerK}.

\subsection{A Graph-Based Delivery Scheme for General Setting}
\label{sec:N_larger_K_system_graph}
The delivery scheme proposed in Section~\ref{sec:N_larger_K_system} is based on applying the universal delivery scheme on each subgraph of the file transition graph after decomposing it. We discussed in Section~\ref{sec:achv_scheme_optimal_N=K} that the delivery load obtained by this scheme can be sub-optimum, depending on the topology of the graph, and then we proposed an opportunistic approach to slash the delivery load. 
Consequently, we can apply the graph-based delivery scheme of Section~\ref{sec:achv_scheme_optimal_N=K} on each subgraph $\cG_i(V,E_i)$, for $i \in [N/K]$, of the file transition graph, to reduce the delivery load. Let $\gamma_i$ be the number of cycles in the subgraph $\cG_i(V,E_i)$. Theorem~\ref{AchvScheme1_2} implies that the delivery load of $\frac{\binom{K-1}{\widehat{S}} - \binom{\gamma_i-1}{\widehat{S}}}{\binom{K-1}{\widehat{S}-1}}$ is achievable for the $i$-th subgraph. Therefore, we have the following corollary:
\begin{corollary}	
	\label{cor:graph-N>K}
	Consider a data shuffling system that processes $N$ files using $K$ worker nodes, each equipped with a cache of size $S$. Assume that the file transition graph $\cG(V,E)$ is decomposed into $N/K$ subgraphs $\cG_i(V,E_i)$, for $i \in [N/K]$, and denote by $\gamma_i$ the number of cycles in subgraph $\cG_i$. Then, a total delivery load of 
	\begin{IEEEeqnarray}{lCl}
		R&=&\sum_{i=1}^{N/K} \frac{\binom{K-1}{\widehat{S}} - \binom{\gamma_i-1}{\widehat{S}}}{\binom{K-1}{\widehat{S}-1}} 
		\nonumber\\
		&=& \frac{1}{\binom{K-1}{\widehat{S}-1}}\left(\frac{N}{K}\binom{K-1}{\widehat{S}} - \sum_{i=1}^{N/K} \binom{\gamma_i-1}{\widehat{S}}\right)
		\label{eq:graph-N>K}
	\end{IEEEeqnarray}
	is achievable for the file assignments given by the file transition graph. 
\end{corollary}

Next, we explain the details of the proposed shuffling schemes through an illustrative example. 

\subsection{Illustrative Example}
\label{sec:N_larger_K_system_ex}
\noindent\textbf{Example 5}:
{\it  
We consider a data shuffling system with $K=4$ worker nodes with cache size of $S=4$ files, and $N=8$ files, denoted by $\{F^1, F^2, F^3, F^4, F^5, F^6, F^7, F^8\}$. For notational simplicity, we rename the files as $\{A,B,C,D,E,F,G,H\}$, respectively. Each worker node stores $N/K = 2$ files to process at iteration $t$, and caches $S - N/K= 2$ files in the excess storage part.
Fig.~\ref{fig:ex5_fileTransGraph} depicts the underlying file transition graph $\mathcal{G} (V, E)$.
For instance, worker node $W_1$ processes two files $A$~and $E$ at iteration $t$. 
At~iteration t+1, file $A$ will be again processed by $W_1$, while file $E$ will be processed by $W_4$. Therefore, there are two directed edges outgoing from $W_1$; one from $W_1$ to $W_1$ (labeled by $A$), and the other from $W_1$ to $W_4$ (labeled by $E$).
Fig.~\ref{fig:ex5_cacheOrg} depicts the cache organization of worker nodes at iteration $t$, along with the missing subfiles that need to be processed at iteration $t+1$. Note that cache placement for the excess storage is symmetric across files and worker nodes, and does not depend on the file transition graph. 
\begin{figure}
	\centering
	\subfloat[]{\includegraphics[width=2.3in]{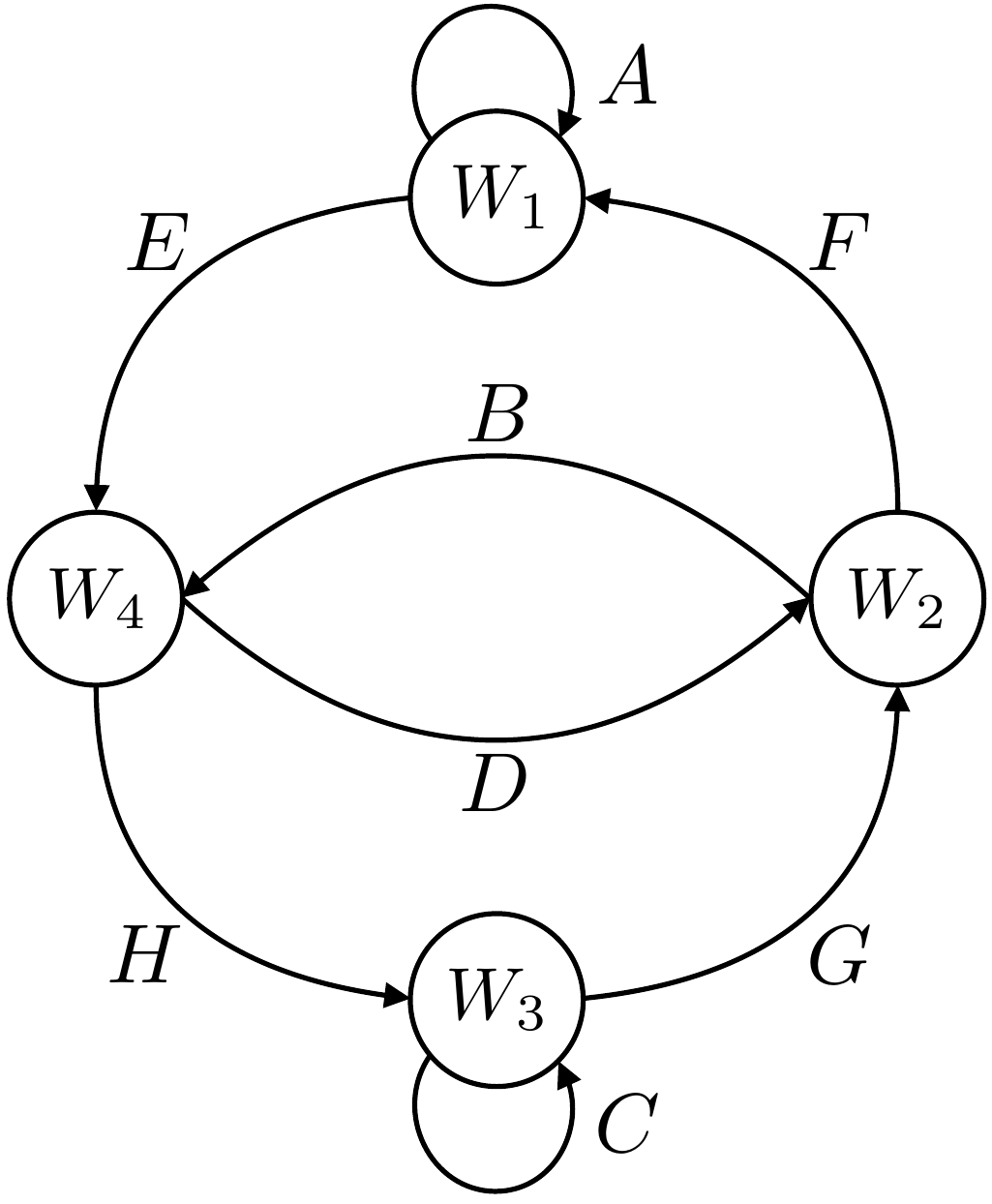}%
	\label{fig:ex5_fileTransGraph}}
	\hfil
	\subfloat[]{\includegraphics[width=2.75in]{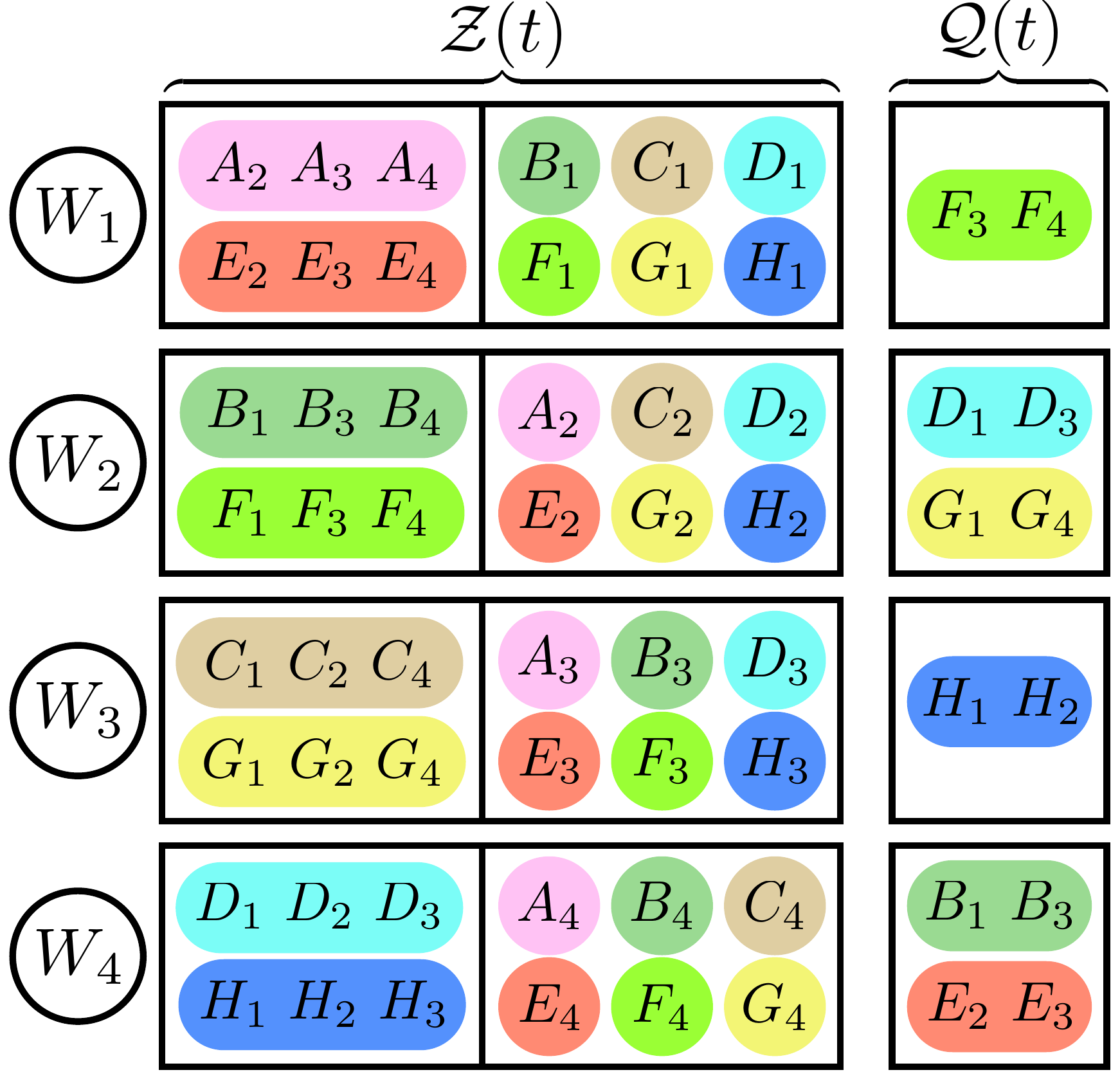}%
	\label{fig:ex5_cacheOrg}}
	\caption{
		Data shuffling system with $N = 8$, $K = 4$ and $S = 4$. 
		(a) The file transition graph
		$\mathcal{G}(V, E)$.
		(b) Cache organization of worker nodes at iteration $t$, along with the set of subfiles which are not available in the caches at iteration $t$ and need to be processed at iteration $t+1$.
	}
	\label{fig:ex5_fTG_cacheOrg}
\end{figure}

\begin{figure}
	\centering
	\subfloat[]{\includegraphics[width=3.3in]{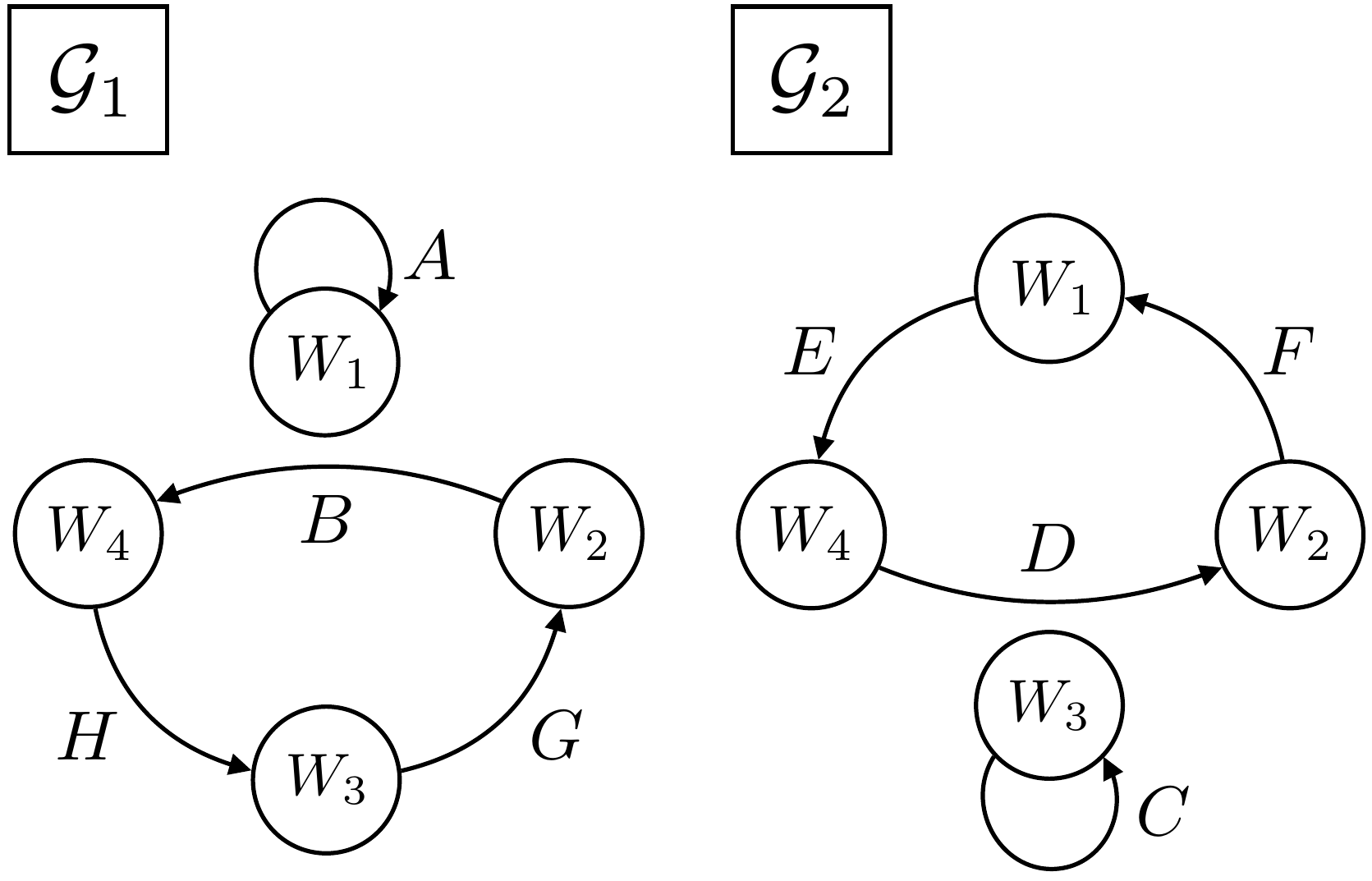}%
	\label{fig:ex5_fileTransGraph_decomp1}}
	\hspace{12mm}
	\subfloat[]{\includegraphics[width=3.3in]{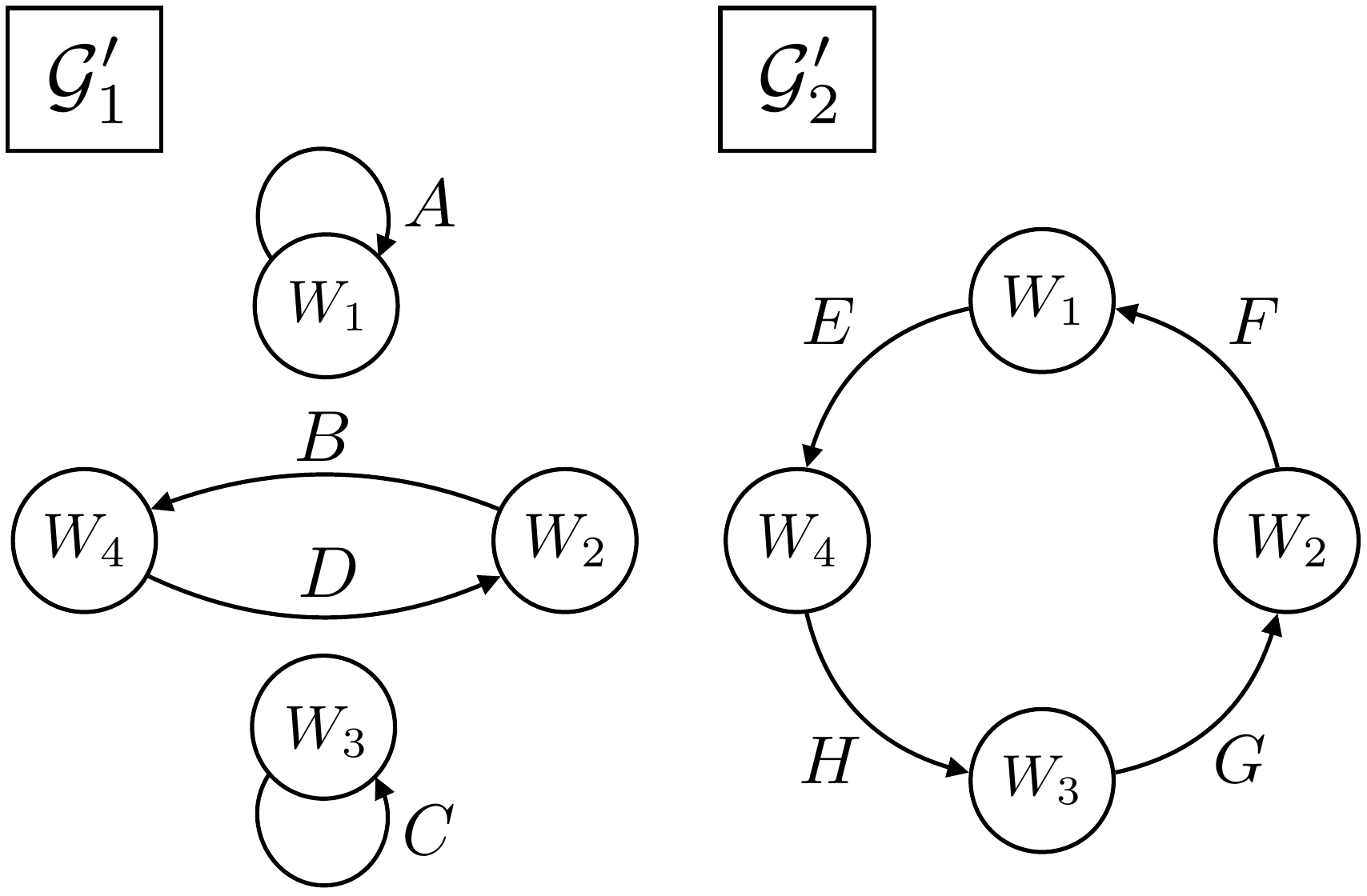}%
	\label{fig:ex5_fileTransGraph_decomp2}}
	\caption{
		(a) One possible decomposition of $\mathcal{G} (V, E)$, depicted in Fig.~\ref{fig:ex5_fTG_cacheOrg}, into $N/K = 2$ subgraphs; $\cG_1$ and $\cG_2$.
		(b) Another possible decomposition of $\mathcal{G} (V, E)$ into $N/K = 2$ subgraphs; $\cG'_1$ and $\cG'_2$.
	}
	\label{fig:ex5_fTG_decomp1_decomp2}
\end{figure}

\begin{figure}
	\centering
	\includegraphics[width=7.1in]{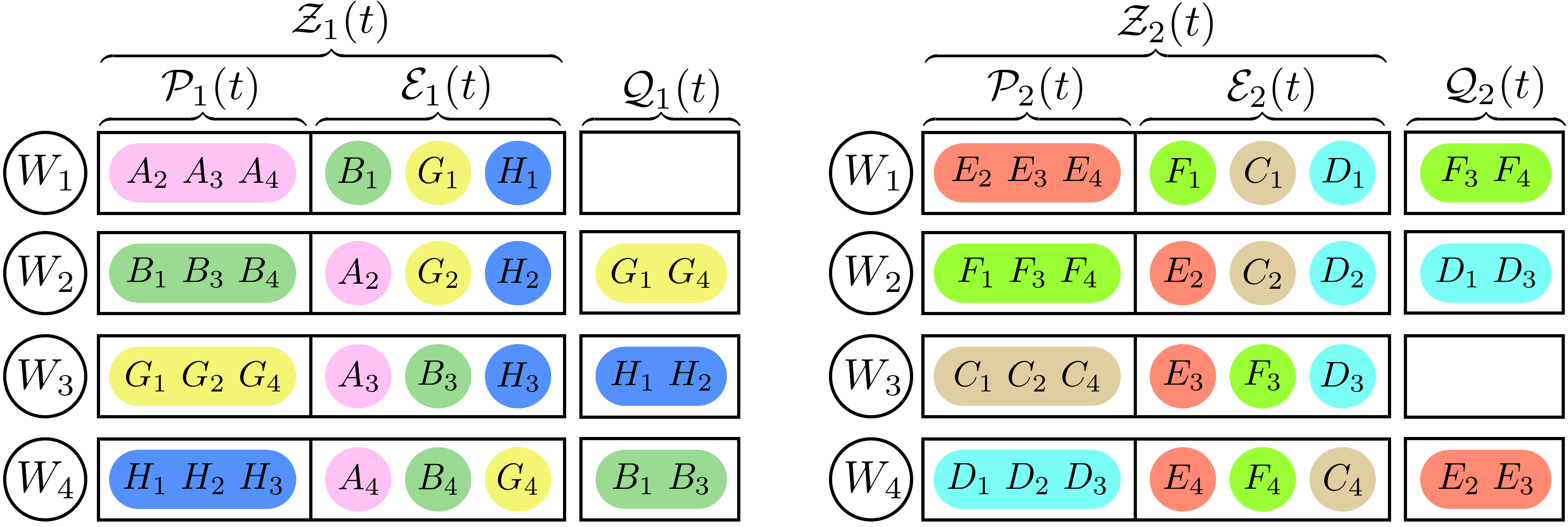}
	\caption{
		Cache organization of worker nodes at iteration $t$, along with the set of subfiles which are not available in the caches at iteration $t$ and need to be processed at iteration $t+1$, for the two subgraphs; $\cG_1$ and $\cG_2$, of the first decomposition, captured by Fig.~\ref{fig:ex5_fileTransGraph_decomp1}, of $\mathcal{G} (V, E)$.
	}
	\label{fig:ex5_decomp1_cacheOrg}
\end{figure}

After constructing of $\mathcal{G} (V, E)$, we decompose it into $N/K$ subgraphs as discussed in Section~\ref{sec:N_larger_K_system}. 
Fig.~\ref{fig:ex5_fileTransGraph_decomp1} shows a~decomposition of $\mathcal{G} (V, E)$ into $N/K = 2$ subgraphs, designated $\cG_1$ and $\cG_2$. Each of such subgraphs induces an instance of the coded shuffling problem with parameters $N_i=K=4$, for $i \in \{1,2\}$, and $\widehat{S}=2$. The corresponding cache contents and file assignments for the problems induced by $\cG_1$ and $\cG_2$ are given in Fig.~\ref{fig:ex5_decomp1_cacheOrg}. It is clear from Theorem~\ref{thrm1} that a communication load of $\binom{K-1}{\widehat{S}}/\binom{K-1}{\widehat{S}-1} = \binom{3}{2}/\binom{3}{1}=1$ is achievable for each subgraph, and hence the total delivery load is $R = R_1 + R_2=2$. 

The delivery load can be potentially reduced by exploiting the cycles in the subgraphs of file transition graph, and refraining from broadcasting redundant sub-messages. However, for subgraphs $\cG_1$ and $\cG_2$ with $\gamma_1=\gamma_2=2$ cycles, there is no communication load reduction due to a graph-based delivery scheme, since, according to Theorem~\ref{thrm2}, we have
\begin{IEEEeqnarray}{l}
	R_1 = \frac{\binom{3}{2}-\binom{1}{2}}{\binom{3}{1}} = 1,
	\quad
	R_2 = \frac{\binom{3}{2}-\binom{1}{2}}{\binom{3}{1}} = 1,
	\nonumber
\end{IEEEeqnarray}
and $R = R_1 + R_2 = 2$.

It is worth noting that the graph decomposition proposed here is not unique. In particular, another possible decomposition of $\mathcal{G} (V, E)$ is depicted in  Fig.~\ref{fig:ex5_fileTransGraph_decomp2}.
Here, $\cG'_1$ and $\cG'_2$ are subgraphs obtained by decomposing $\cG$, and they consist of  $\gamma'_1=3$ and $\gamma'_2=1$ cycles, respectively. Therefore, applying the graph-based delivery scheme of Section~\ref{sec:achv_scheme_optimal_N=K}, the delivery loads of the subgraphs are given by
\begin{IEEEeqnarray}{l}
	R'_1 = \frac{\binom{3}{2}-\binom{2}{2}}{\binom{3}{1}} = \frac{2}{3},
	\quad
	R'_2 = \frac{\binom{3}{2}-\binom{0}{2}}{\binom{3}{1}} = 1,
	\nonumber
\end{IEEEeqnarray}
and $R' = R'_1 + R'_2 = 5/3 < R$.
As a result, the second decomposition provides a lower communication load than the first decomposition. \hfill $\blacklozenge$
}

\subsection{Optimality for the Worst Case Shuffling: Proof of Theorem~\ref{thrm5}}
\label{sec:converse_proof_NlargerK}
In order to prove Theorem~\ref{thrm5}, we present one instance of the shuffling problem, for which the delivery load of Theorem~\ref{thrm4} is indeed required, and cannot be further reduced. This shows that the upper bound on the delivery load given in Theorem~\ref{thrm4} is optimum for such worst-case shuffling scenarios.

Let us consider a data shuffling problem $\mathbf{P} (N,K,S)$ with $N$ files, $K$ worker nodes, and cache memory size of $S$ files.
The size of each file is normalized to $1$ unit.
At each iteration, each worker node processes $N/K$ files out of the $N$ files, and hence $|u(i)| = |d(i)| = N/K$ for $i\in [K]$. 
The shuffling scenario we consider here is given by $d(i)=u(i+1)$ for $i \in [K-1]$ and $d(K)=u(1)$, i.e., all the files being processed by one worker node at iteration $t$ are assigned to another worker node at iteration $t+1$. 
For each worker node $W_i$, where $i \in [K]$, the expression for the cache contents $\cZ_i$, that comprises under-processing part $\proc{i}$ and excess storage part $\ex{i}{}$, at iteration $t$ is defined by \eqref{eq:Z-def}.
Moreover, the set of subfiles $\cQ_i$ to be processed by $W_i$ at iteration $t+1$ is defined by \eqref{eq:def:demand}.
Suppose, for contradiction, that there exists a shuffling scheme that achieves a delivery load $R$ where
\begin{IEEEeqnarray}{C}
	\label{eq:R_contradict}
	R(\mathbf{P}) < \frac{N}{K} \frac{  \binom{K-1}{\widehat{S}} }{\binom{K-1}{\widehat{S}-1}}.
\end{IEEEeqnarray}
Now, let us consider another data shuffling problem $\mathbf{\widetilde{P}} (\widetilde{N},\widetilde{K},\widetilde{S})$ with $\widetilde{N} = K$ files, $\widetilde{K} = K$ worker nodes, and $\widetilde{S}~=~S/(N/K) = \widehat{S}$ files. We denote the files of this instance of the problem by $\{F^1,F^2,\dots, F^{\widetilde{K}}\}$. 
Each worker node processes $1$~file at each iteration, i.e., $|u(i)| = |d(i)| = \widetilde{N}/\widetilde{K}=1$ for $i\in [\widetilde{K}]$. 
Let us divide each file in $\mathbf{\widetilde{P}} (\widetilde{N},\widetilde{K},\widetilde{S})$ into $N/K$ mini-files, each of which is of size $1/(N/K)$.
More precisely, let $F^i (r)$ be the $r^{\text{th}}$ mini-file of file $F^i$ for $i \in [\widetilde{K}]$ and $r \in [N/K]$.
Consequently, for $i \in [\widetilde{K}]$, the corresponding expressions of $\mathcal{\widetilde{Z}}_i$ and $\mathcal{\widetilde{Q}}_i$ are given by
\begin{IEEEeqnarray}{lCl}
	\label{eq:Rtilde_contradict}	
	\mathcal{\widetilde{Z}}_i & = & \mathcal{\widetilde{P}}^i \cup \mathcal{\widetilde{E}}_i = \mathcal{\widetilde{P}}^i \cup \left(\bigcup\nolimits_{\ell \in [K] \setminus i} \mathcal{\widetilde{E}}_i^\ell \right),
\end{IEEEeqnarray}
where
\begin{IEEEeqnarray}{lCl}
	\mathcal{\widetilde{P}}^i & = &
	\left\{\F{i}{\Gamma} (r): \Gamma \subseteq [K] \setminus \{i\},\: |\Gamma| = \widehat{S} - 1, \: r \in [N/K] \right\},
	\\
	\mathcal{\widetilde{E}}_i^\ell & = &
	\left\{\F{\ell}{\Gamma} (r): \ell \neq i, \: i \in \Gamma \subseteq [K], \: |\Gamma| = \widehat{S} - 1, \: r \in [N/K] \right\},
\end{IEEEeqnarray}
and
\begin{IEEEeqnarray}{lCl}
	\mathcal{\widetilde{Q}}_i & = &
	\left\{ \F{\ell}{\Gamma} (r): \ell \in d(i),\:  
	\ell \neq i,\: i \notin \Gamma,\:
	\Gamma \subseteq [K], 
	|\Gamma| = \widehat{S} - 1, \: r \in [N/K] \right\}.
\end{IEEEeqnarray}
Therefore, $\mathbf{\widetilde{P}}$ can be viewed as a data shuffling problem with $\widetilde{N} \times (N/K) = N$ mini-files each of size $1/(N/K)$, and $\widetilde{K} = K$~worker nodes where each has a cache memory to store  $\widetilde{S} \times (N/K) = S$ mini-files.
Viewing the data shuffling problem $\mathbf{\widetilde{P}} (\widetilde{N},\widetilde{K},\widetilde{S})$ as a problem similar to  $\mathbf{P}$, we can apply the delivery scheme of $\mathbf{P}$ (with mini-files of size $1/(N/K)$ instead of~$1$) and achieve a delivery load of 
\begin{IEEEeqnarray}{lCl} 
	R(\mathbf{\widetilde{P}}) &=&  \frac{1}{N/K} R(\mathbf{P})  
	\nonumber\\
	&<& 
	\frac{1}{N/K} \frac{N}{K}\frac{  \binom{K-1}{\widehat{S}} }{ \binom{K-1}{\widehat{S}-1}}
	= \frac{  \binom{K-1}{\widehat{S}} }{\binom{K-1}{\widehat{S}-1}} = \frac{  \binom{\widetilde{K}-1}{\widetilde{S}} }{\binom{\widetilde{K}-1}{\widetilde{S}-1}},
\end{IEEEeqnarray}
which contradicts Corollary~\ref{remrk_opt_NeqK}.
\hfill $\blacksquare$

\subsection{On the Sub-Optimality of Decomposition-Based Delivery}
The delivery load proposed in Corollary~\ref{cor:graph-N>K} depends on the decomposition of the file transition graph $\cG(V,E)$. As discussed in Example~5, such decomposition is not unique. Therefore, one can minimize the delivery load in \eqref{eq:graph-N>K} by exhaustively searching over all possible decompositions of the file transition graph $\mathcal{G} (V, E)$. A natural question is whether the delivery load obtained by such a best decomposition is optimum. 

\begin{figure}
	\centering
	\subfloat[]{\includegraphics[height=2in]{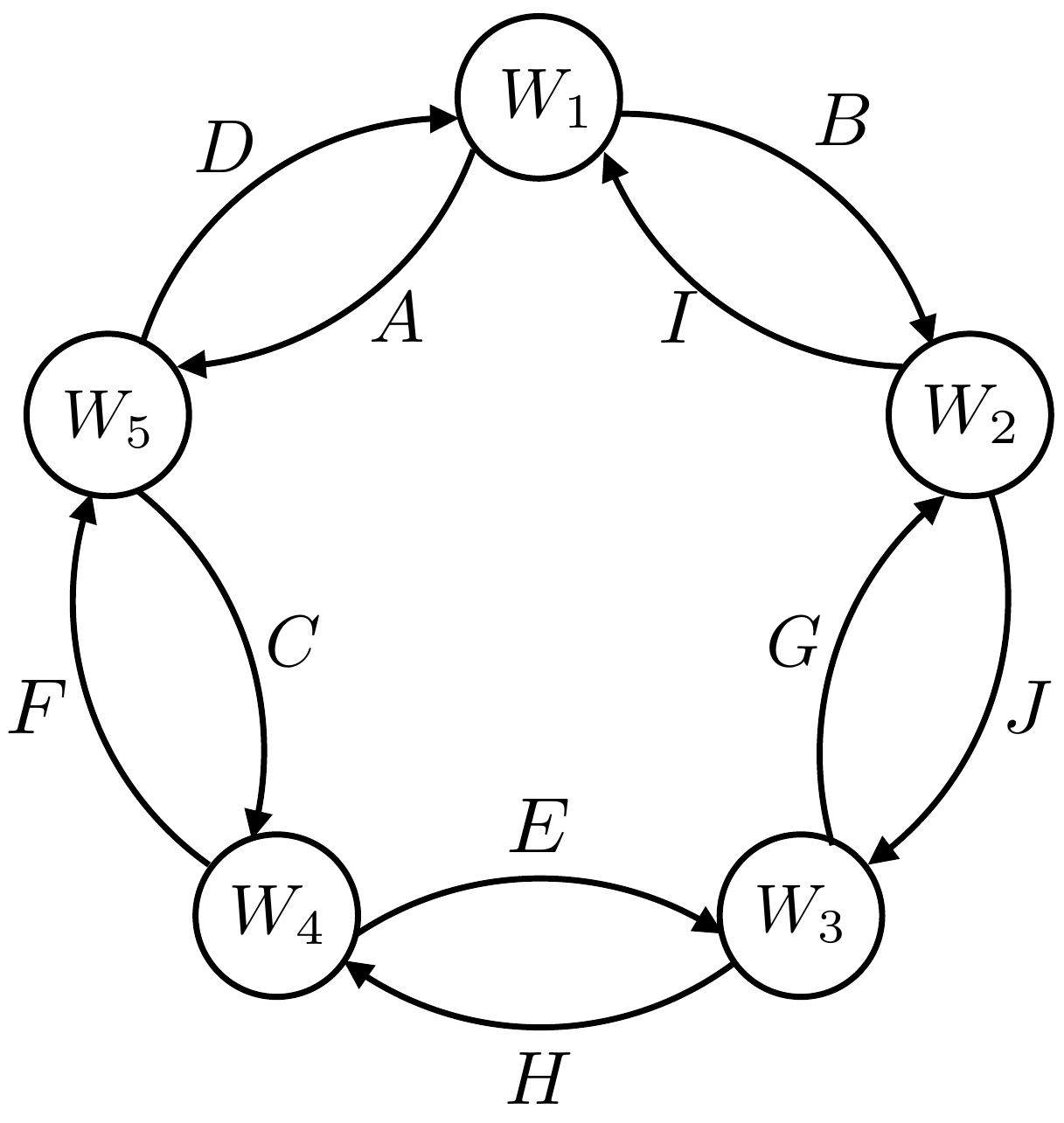}%
	\label{fig:counterEx_fileTransGraph}}
	\hspace{20mm}
	\subfloat[]{\includegraphics[height=2in]{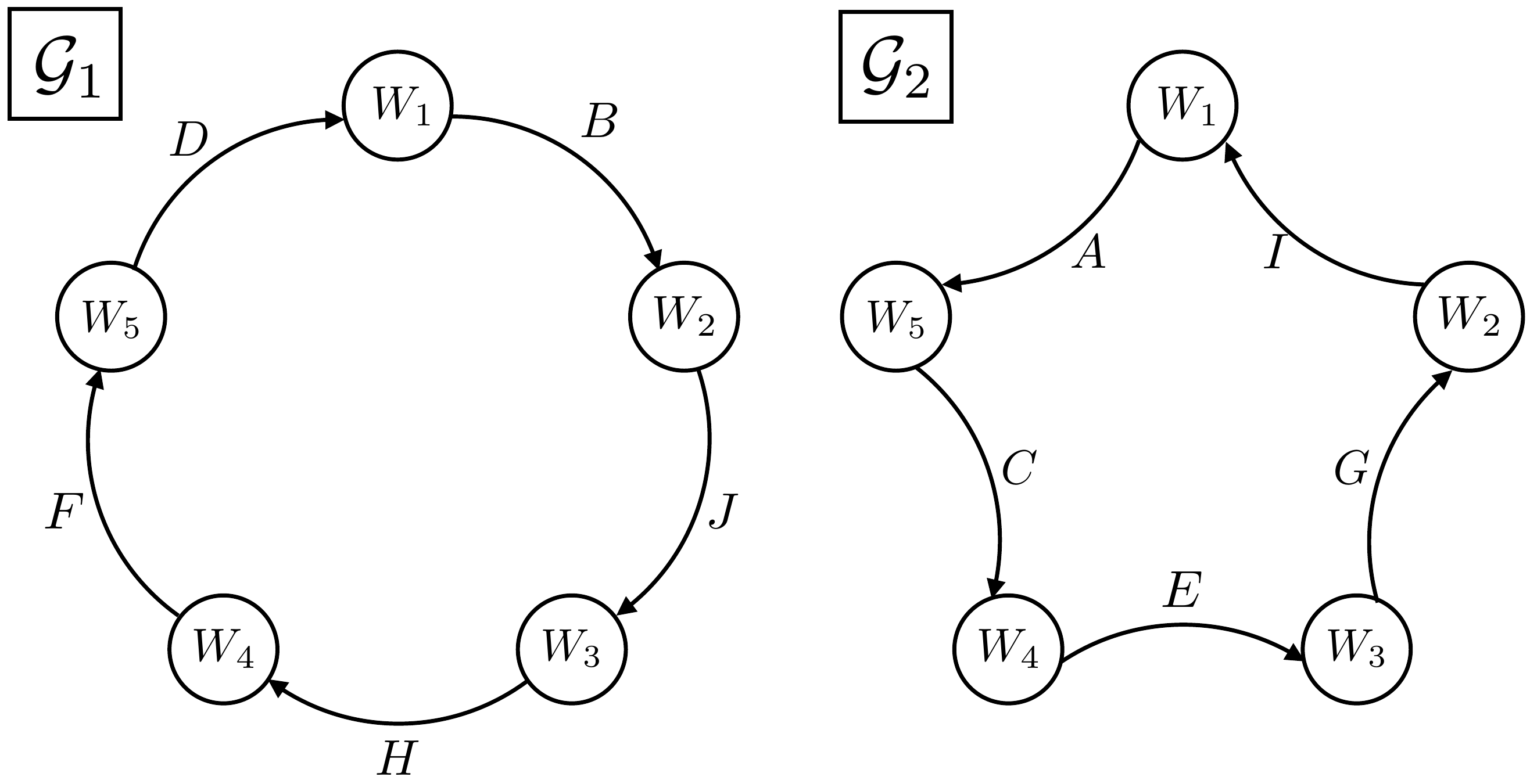}%
	\label{fig:counterEx_fileTransGraph_decomp1}}
	\caption{
		Data shuffling system with $N = 10$, $K = 5$ and $S = 2$. 
		(a) The file transition graph $\mathcal{G}(V, E)$.
		(b) Decomposition of $\mathcal{G} (V, E)$ into $N/K = 2$ subgraphs; $\cG_1$ and $\cG_2$.
	}
	\label{fig:counterEx_fTG_decomp1}
\end{figure}

\begin{figure}
	\centerline{
		\includegraphics[height=2in]{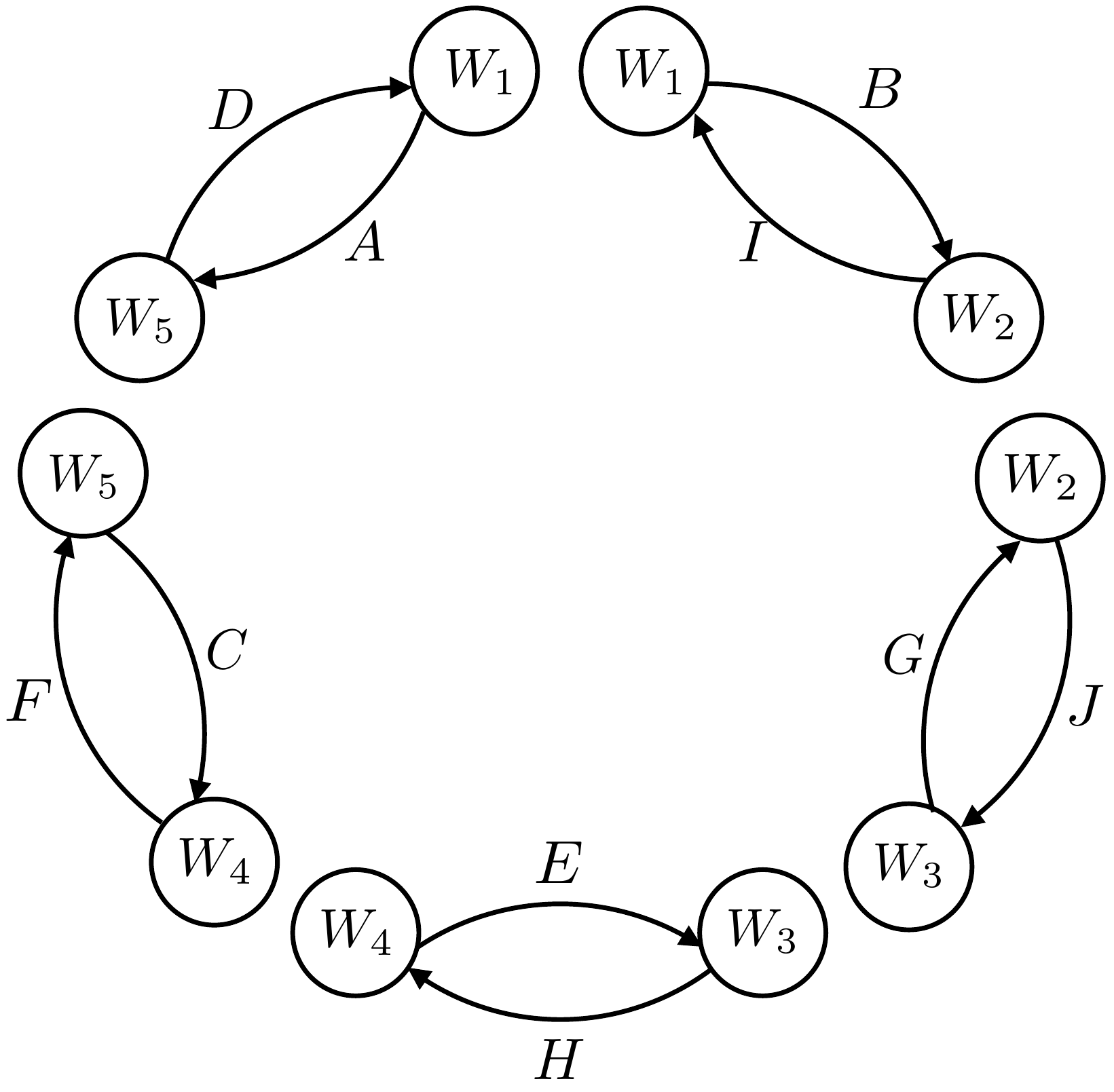}
	}
	\caption{
		A \emph{different} decomposition of the file transition graph $\mathcal{G}(V, E)$, shown in Fig.~\ref{fig:counterEx_fileTransGraph}.
	}
	\label{fig:counterEx_fileTransGraph_decomp2}
\end{figure}	

In the following example, we show that the answer to this question is ``No'', and a decomposition-based delivery scheme can be sub-optimum for a general file transition graph.
Consider a data shuffling system with $K=5$ worker nodes, and $N=10$~files, denoted by $\{A,B,C,D,E,F,G,H,H,I,J\}$.
The cache available at each worker node is $S=2$, i.e., $\widehat{S}=S/(N/K)=1$, which implies that there is no excess storage. The file transition graph of this problem is depicted in Fig.~\ref{fig:counterEx_fileTransGraph}. It turns out that there is only one possible decomposition of this graph, shown in Fig.~\ref{fig:counterEx_fileTransGraph_decomp1}, that satisfies the conditions given in Section~\ref{sec:N_larger_K_system}. It is clear that each subgraph has only one cycle, i.e., $\gamma_1=\gamma_2$. Hence, from Corollary~\ref{cor:graph-N>K}, the delivery load is given by
\begin{IEEEeqnarray}{lCl}
	R &=& 
	\frac{\binom{K-1}{\widehat{S}} - \binom{\gamma_1-1}{\widehat{S}}}{\binom{K-1}{\widehat{S}-1}} 
	+ \frac{\binom{K-1}{\widehat{S}} - \binom{\gamma_2-1}{\widehat{S}}}{\binom{K-1}{\widehat{S}-1}} 
	= \frac{\binom{4}{1}- \binom{0}{1}}{\binom{4}{0}} + \frac{\binom{4}{1}- \binom{0}{1}}{\binom{4}{0}}
	= 8.
	\nonumber
\end{IEEEeqnarray}

Now, let us consider an alternative transmission strategy as follows:
\begin{IEEEeqnarray}{l}
	\mathcal{X} = \left\{A \oplus D,\: B \oplus I,\: C \oplus F,\: E \oplus H,\: G \oplus J\right\}.
	\nonumber
\end{IEEEeqnarray}
It is easy to check that all the worker nodes can recover their assigned files from their cache contents and $\cX$. Since we transmit $5$ sub-messages, each of which is of size $1$, then the corresponding delivery load is $R = 5$, which is strictly less than $R=8$, can be achieved. Indeed, this delivery scheme is inspired by a different decomposition of the graph $\cG(V,E)$, as depicted in Fig.~\ref{fig:counterEx_fileTransGraph_decomp2}. This decomposition consists of $5$ subgraphs, and the corresponding data shuffling problems do not lie in the class of canonical problems discussed in Section~\ref{sec:N_larger_K_system}. This example shows that the upper bound on the delivery load, given in Corollary~\ref{cor:graph-N>K}, is loose in general.

\section{Simulation Results}
\begin{figure}
	\centering
	\includegraphics[width=0.48\textwidth]{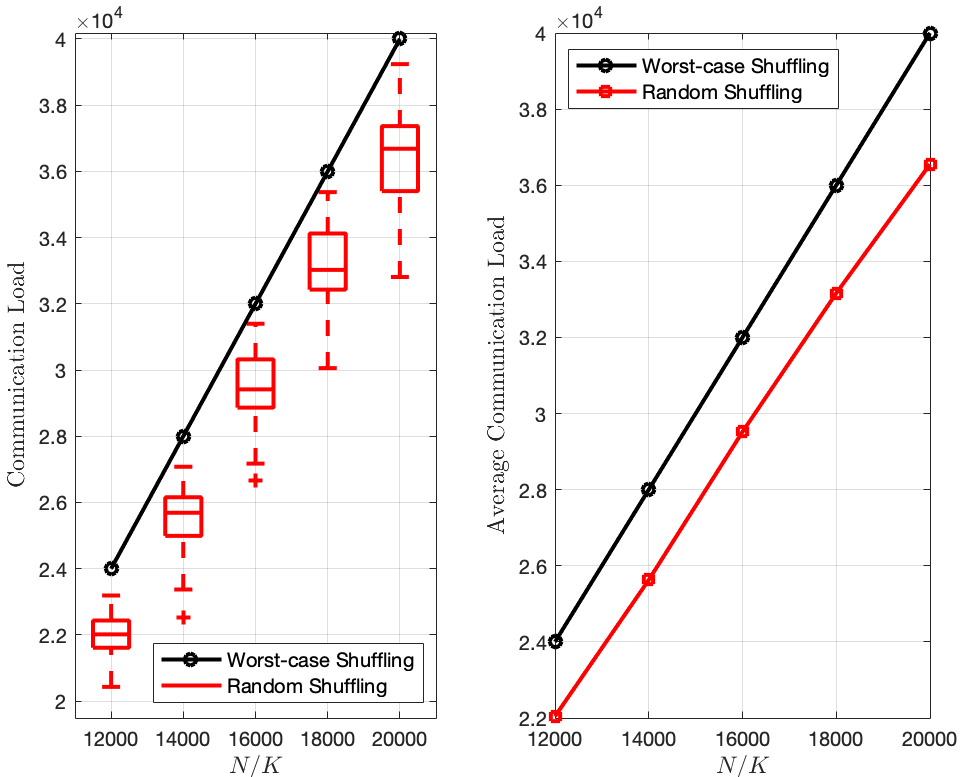}
	\caption{
		The trade-off between the communication load versus $N/K$ for a data shuffling problem with $K = 6$ and $\widehat{S} = S/(N/K) = 2$. The left subplot depicts range of achieved communication loads over $10^3$ shuffling iterations, while the right subplot shows the corresponding average communication loads. The black curve is for the worst-case shuffling, while the red curve is for the random shuffling.
	}
	\label{fig:simRes_K6_S2}
\end{figure}

\begin{figure}
	\centering
	\includegraphics[width=0.48\textwidth]{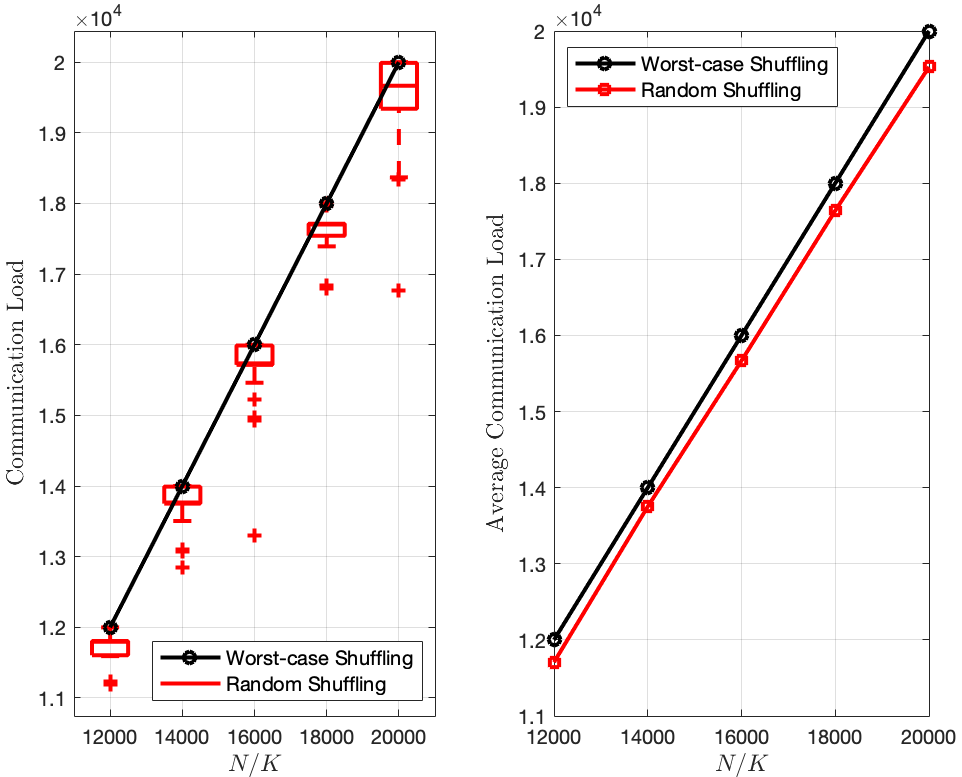}
	\caption{
		The trade-off between the communication load versus $N/K$ for a data shuffling problem with $K = 6$ and $\widehat{S} = S/(N/K) = 3$. The left subplot depicts range of achieved communication loads over $10^3$ shuffling iterations, while the right subplot shows the corresponding average communication loads. The black curve is for the worst-case shuffling, while the red curve is for the random shuffling.
	}
	\label{fig:simRes_K6_S3}
\end{figure}

Next, we present simulation results for the achieved communication load when $N>K$ for random shuffling.
In the below figures, we plot the communication load for different values of $N/K$. 
We set the number of worker nodes to $K=6$, and evaluate the communication load as a function of $N/K$. We also set the normalized storage size to $\widehat{S}=2$ in Fig.~\ref{fig:simRes_K6_S2}, and  $\widehat{S}=3$ in Fig.~\ref{fig:simRes_K6_S3}.
The red box-plot in the left subplots of the figures depict the range of the achieved communication loads over $10^3$ random shuffling scenarios. This shows that the communication load can significantly vary depending on the underlying random shuffling. 
The black curves, however, are the achieved communication loads for the worst-case shuffling scenario.
On the other hand, the right subplots  depict the corresponding average communication loads over $10^3$ random shuffling scenarios (red curves), in comparison with the achieved communication loads for the worst-case shuffling scenario (black curves).

It is clearly evident that the proposed algorithm achieves a lower communication load compared to the one achieved for worst-case scenario in all figures. Moreover, the performance gap is more significant for smaller values of  $\widehat{S}$. This is consistent with our theoretical result in Corollary 2: The saving in the communication load (compared to the worst-case shuffling) is proportional to 
\[
\sum_{i=1}^{N/K} \binom{\gamma_i-1}{\widehat{S}}.
\]
Therefore, each term in the summation gets smaller as $\widehat{S}$ increases. However, as $N/K$ increases, we have a larger number of terms contributing to the saving.

It should be noted that the algorithm used in our simulations for $N/K$ graph decompositions (or more specifically $N/K$~perfect matchings) hinges on the Hungarian algorithm \cite{kuhn1955hungarian,munkres1957algorithms}. 
The pseudocode of the graph decomposition algorithm is given in Algorithm~\ref{algo_decomposeGraph}, presented in Appendix~\ref{app:psuedocodes_p1}. In~Remark~\ref{rmrk_bipartite}, given in Appendix~\ref{app:lm:prfct_match}, we present a brief discussion about the different algorithms for finding a perfect matching, along with their time complexities.

\newpage
\section{Conclusion}
\label{sec:conclusion}
In this paper, we proposed a novel deterministic coded shuffling scheme which improves the state of the art by achieving a~lower communication load for any shuffling when the number of files is equal to the number of worker nodes. Furthermore, the optimality of the proposed coded shuffling scheme was demonstrated through a matching converse proof. We showed that the placement phase of the proposed scheme, assuming uncoded prefetching, is optimal. Then, we exploited this canonical setting as a building block, and proposed a shuffling strategy for the general problem setting when the number of files is greater than or equal to the number of worker nodes. Moreover, we proved that the delivery load is optimum for worst-case shuffling. The characterization of the optimum trade-off for a given file transition graph is, however, still an open problem.

Promising directions for future research include the following. To complete our understanding of the problem, a topic of future work is to characterize the exact load-memory trade-off for any shuffling for the general setting of the shuffling problem. Moreover, in this work, we optimized the communication load of the data shuffling procedure for any two consecutive iterations, i.e., one-round shuffling. A more general framework consists of multiple consecutive shuffling iterations, and the design of a~shuffling mechanism in order to achieve an enhanced overall delivery load would be of practical interest.

\newpage
\appendices
\section{Proof of Lemma~\ref{lm:decode-all-OFC}}
\label{app:lm:dec-all}
We will prove that all the subfiles intended for $W_{\ell}$, $\ell\in [K-1]$, can be recovered from the family of broadcast sub-messages $\cX$ and cache contents $\cZ_\ell$.  Recall that all such subfiles are indexed by $\F{d(\ell)}{\Gamma}$ for some $\Gamma\subseteq [K]\setminus \{\ell, d(\ell)\}$ with $|\Gamma|=S-1$, otherwise either the subfile is already cached at the worker node (when $\ell\in \Gamma$), or the subfile is zero (if $d(\ell)\in \Gamma$).  We~distinguish the following two cases:

\subsection{$K\notin \Gamma$}
\label{sec:dec-all-notK}
The condition $K\notin \Gamma$ implies that $\{\ell\} \cup \Gamma \subseteq [K-1]$ and $|\{\ell\} \cup \gamma| = S$. Hence, $X_{\{\ell\} \cup \Gamma} \in \cX$, as defined in \eqref{eq:X_all}, i.e., $X_{\{\ell\} \cup \Gamma}$ is one of the sub-messages broadcast by the master node. We can recover $\F{d(\ell)}{\Gamma}$ from $X_{\{\ell\} \cup \Gamma}$ as follows: 
\begin{IEEEeqnarray}{lCl}
	X_{\{\ell\} \cup \Gamma} & = &
	\bigoplus_{i \in \{\ell\} \cup \Gamma} \left(\F{i}{(\{\ell\} \cup \Gamma) \setminus \{i\}} 
	\oplus
	\F{d(i)}{(\{\ell\} \cup \Gamma) \setminus \{d(i)\}}
	\oplus\left(\bigoplus_{ j \in [K] \setminus (\{\ell\} \cup \Gamma) } 
	\F{d(i)}{(\{j,\ell\} \cup \Gamma)\setminus \{i, d(i)\}} \right)
    \right) \nonumber\\
	&  = & \bigoplus_{i \in \{\ell\} \cup \Gamma}  Y_i,
	\label{eq:ach_proof:1}
\end{IEEEeqnarray}
where we denote the summand for $i$ by $Y_i$. 

\begin{itemize}[leftmargin=*]
	\item First note that if $i\neq \ell$ and $d(i)\neq \ell$  then we have $\ell \in (\{\ell\} \cup \Gamma) \setminus \{i\}$, $\ell \in (\{\ell\} \cup \Gamma) \setminus \{d(i)\}$, and $\ell \in (\{j,\ell\} \cup \Gamma)\setminus \{i, d(i)\}$. Hence, each term in 
	\begin{IEEEeqnarray}{lCl}
		Y_i &= & \F{i}{(\{\ell\} \cup \Gamma) \setminus \{i\}} 
		\oplus
		\F{d(i)}{(\{\ell\} \cup \Gamma) \setminus \{d(i)\}} 
		\oplus	
		\left(\bigoplus_{ j \in [K] \setminus (\{\ell\}\cup \Gamma) } 
		\F{d(i)}{(\{j,\ell\} \cup \Gamma)\setminus \{i, d(i)\}}\right)
	\end{IEEEeqnarray}
	exists in the excess storage of worker node $\ell$ (see definition of $\ex{\ell}{}$ in \eqref{eq:E-def}) and hence $Y_i$ can be removed from $X_{\{\ell\} \cup \Gamma}$ using the cache $\cZ_\ell$. 
	\item Next, for $i$ with $i\neq \ell$ but $d(i)=\ell$ we have $\ell \in (\{\ell\} \cup \Gamma) \setminus \{i\}$ which implies $\F{i}{(\{\ell\} \cup \Gamma) \setminus \{i\}} \in \ex{\ell}{}\subset \cZ_\ell$. Moreover, from $d(i)=\ell$ we have 
	\begin{IEEEeqnarray}{lCl}
		Y_i &=& \F{i}{(\{\ell\} \cup \Gamma) \setminus \{i\}}
		\oplus
		\F{d(i)}{(\{\ell\} \cup \Gamma) \setminus \{d(i)\}}
		\oplus \left(\bigoplus_{ j \in [K] \setminus (\{\ell\}\cup \Gamma) } 
		\F{d(i)}{(\{j,\ell\} \cup \Gamma)\setminus \{i, d(i)\}}\right)\nonumber\\
		&=&
		\underbrace{\F{i}{(\{\ell\} \cup \Gamma) \setminus \{i\}}}_{\in \ex{\ell}{}}
		\oplus
		\underbrace{\F{\ell}{\Gamma}}_{\in \proc{\ell}}
		\oplus	
		\left(\bigoplus_{ j \in [K] \setminus (\{\ell\}\cup \Gamma) } 
		\underbrace{\F{\ell}{(\{j \} \cup \Gamma)\setminus \{i \}}}_{\in \proc{\ell}} \right),
		\nonumber
	\end{IEEEeqnarray}
	which shows $Y_i$ can be fully removed using the cache contents $\cZ_\ell$. 
	\item Finally, for $i=\ell$ with $d(i) \neq \ell$, we have
	\begin{IEEEeqnarray}{lCl}
		Y_i &=& \F{i}{(\{\ell\} \cup \Gamma) \setminus \{i\}} 
		\oplus
		\F{d(i)}{(\{\ell\} \cup \Gamma) \setminus \{d(i)\}}
		\left( \bigoplus_{ j \in [K] \setminus (\{\ell\}\cup \Gamma) } 
		\F{d(i)}{(\{j,\ell\} \cup \Gamma)\setminus \{i, d(i)\}} \right) \nonumber\\
		&=&
		\F{\ell}{\Gamma} 
		\oplus
		\F{d(\ell)}{(\{\ell\} \cup \Gamma) \setminus \{d(\ell)\}}
		\oplus	
		\bigoplus_{ j \in [K] \setminus (\{\ell\}\cup \Gamma) } 
		\F{d(\ell)}{(\{j\} \cup \Gamma)\setminus \{d(\ell)\}}\nonumber\\
		&=&
		\F{\ell}{\Gamma} 
		\oplus	
		\bigoplus_{ j \in [K] \setminus  \Gamma } 
		\underbrace{\ \F{d(\ell)}{(\{j\} \cup \Gamma)\setminus \{d(\ell)\}\ }
		}_{=0 \textrm{ if $d(\ell)\notin \{j\}\cup \Gamma$}}\nonumber\\
		&=&
		\F{\ell}{\Gamma} 
		\oplus	
		\bigoplus_{ j =d(\ell) } 
		\F{d(\ell)}{(\{j\} \cup \Gamma)\setminus \{d(\ell)\}\ }\label{eq:ach-1}\\
		&=&
		\F{\ell}{\Gamma} 
		\oplus	
		\F{d(\ell)}{\Gamma},
	\end{IEEEeqnarray}
	where in \eqref{eq:ach-1} we have used the fact that  $\F{d(\ell)}{(\{j\} \cup \Gamma)\setminus \{d(\ell)\}}$ is non-zero only if $d(\ell)\in \{j\} \cup \Gamma$, because otherwise $|(\{j\} \cup \Gamma)\setminus \{d(\ell)\}|=S>S-1$. On the other hand, we know $d(\ell)\notin \Gamma$. Therefore, the only non-zero term is the one corresponding to $j=d(\ell)$. Also note that $F^\ell_{\Gamma} \in \proc{\ell}$ exists in the cache of worker node $\ell$. 
\end{itemize}
Therefore, $X_{\{\ell\} \cup \Gamma}$ can be written as
\begin{IEEEeqnarray*}{l}
X_{\{\ell\} \cup \Gamma} = \zeta_0 + \F{d(\ell)}{\Gamma},
\end{IEEEeqnarray*}
where the interference term $\zeta_0$ can be completely removed using the cache contents $\cZ_\ell$. This implies  $\F{d(\ell)}{\Gamma}$ can be recovered from the received sub-message $X_{\{\ell\} \cup \Gamma}$ and cache contents of $W_{\ell}$. 

\subsection{$K\in \Gamma$}
\label{sec:dec-all-K}
When $\Gamma$ is a set of indices that includes $K$, the desired subfile cannot be directly recovered from one  single transmit sub-message, since the proposed broadcast strategy does not send any sub-message $X_\set$ with $K\in \set$. However, we will show that $\F{d(\ell)}{\Gamma}$ can be still recovered from the summation of 
subfiles from the cache contents and previously sub-messages decoded by~$W_{\ell}$. Define $\sett = (\Gamma \setminus \{K\})\cup \{d(\ell)\}$. Recall that we assume $d(\ell)\neq K$, otherwise $\F{d(\ell)}{\Gamma}=0$ is a dummy subfile. This implies $| \{\ell\} \cup \sett|=S$.  Then we have 
\begin{IEEEeqnarray}{lCl}
	X_{\{\ell\} \cup \sett} &=&
	\bigoplus_{i \in \{\ell\} \cup \sett} 
	\left( 
	\F{i}{( \{\ell\} \cup \sett) \setminus \{i\}} 
	\oplus
	\F{d(i)}{(\{\ell\} \cup \sett) \setminus \{d(i)\}}
	\oplus	
	\left(\bigoplus_{ j \in [K] \setminus (\{\ell\} \cup \sett)} 
	\F{d(i)}{(\{j,\ell\} \cup \sett) \setminus\{i, d(i)\}} \right)  
	\right)	
	\nonumber\\
	&=& \bigoplus_{i \in \{\ell\} \cup \sett} Y_i,
	\label{eq:ach_proof:2}
\end{IEEEeqnarray} 
where $Y_i$ is the summand for the corresponding $i$ in \eqref{eq:ach_proof:2}. 
Similar to Appendix~\ref{sec:dec-all-notK}, we can identify the following three cases:
\begin{itemize}[leftmargin=*]
	\item If $i$ is such that $i\neq \ell$ and $d(i)\neq \ell$, then $\ell \in ( \{\ell\} \cup \sett) \setminus \{i\}$, $\ell \in ( \{\ell\} \cup \sett) \setminus \{d(i)\}$ and $\ell \in (\{j,\ell\} \cup \sett) \setminus\{i, d(i)\}$, i.e., all the subfiles added in $Y_i$ are indexed by a set that includes $\ell$. All such subfiles are cached in  $\cZ_\ell$ (see \eqref{eq:Z-def}), and hence $Y_i$~can be reconstructed and removed from the summation in \eqref{eq:ach_proof:1}. 
	\item For $i$ with $d(i)=\ell$ but $i \neq \ell$, we have
	\begin{IEEEeqnarray*}{l}
	Y_i = 
	\underbrace{\F{i}{(\{\ell\} \cup \sett) \setminus \{i\}}}_{\in \ex{\ell}{}}
	\oplus
	\underbrace{\F{\ell}{\sett}}_{\in \proc{\ell}}
	\oplus
	\left(\bigoplus_{ j \in [K] \setminus (\{\ell\} \cup \sett)}  
	\underbrace{\F{\ell}{(\{j,\ell\} \cup \sett)\setminus \{i, d(i)\}}}_{\in \proc{\ell}} \right),
	\end{IEEEeqnarray*} 
	which implies that the entire $Y_i$ can be reconstructed from~$\cZ_\ell = \ex{\ell}{} \cup \proc{\ell}$.
	
	\item Finally, for $i=\ell$ with $d(i) \neq i$, we can simplify $Y_i$ as
	\begin{IEEEeqnarray}{lCl}
		Y_i &=& \F{\ell}{\sett} 
		\oplus
		\F{d(\ell)}{(\{\ell\} \cup \sett) \setminus \{d(\ell)\}}
		\oplus
		\left(\bigoplus_{ j \in [K] \setminus (\{\ell\} \cup \sett)} 
		\F{d(\ell)}{(\{j\} \cup \sett)\setminus \{ d(\ell)\}} \right)
		\nonumber\\
		&=&
		\F{\ell}{\sett} 
		\oplus	
		\left(\bigoplus_{ j \in [K] \setminus \sett}  
		\F{d(\ell)}{(\{j\} \cup \sett)\setminus \{ d(\ell)\}} \right)\label{eq:dec-all-notK-2}\\
		&=&\F{\ell}{\Gamma} 
		\oplus	
		\left(\bigoplus_{ j \in [K-1] \setminus \sett}  
		\F{d(\ell)}{(\{j\} \cup \sett)\setminus \{ d(\ell)\}} \right)
		\oplus 
		\left(\bigoplus_{ j =K}  
		\F{d(\ell)}{(\{j\} \cup \sett)\setminus \{ d(\ell)\}} \right)	\label{eq:dec-all-notK-3}\\
		&=& \F{\ell}{\Gamma}
		\oplus	
		\left(\bigoplus_{ j \in [K-1] \setminus \sett}  
		\F{d(\ell)}{(\{j\} \cup \sett)\setminus \{ d(\ell)\}} \right)
		\oplus 
		\F{d(\ell)}{\Gamma}\label{eq:dec-all-notK-4},
	\end{IEEEeqnarray}
	where in \eqref{eq:dec-all-notK-2} we have merged two summations over $j=\ell$ and $j \in [K] \setminus (\{\ell\} \cup \sett)$, and again split it to $j\in [K-1] \setminus \sett$ and $j=K$ in \eqref{eq:dec-all-notK-3}. In \eqref{eq:dec-all-notK-4}, we use the fact that
	\[
	(\{K\} \cup \sett)\setminus \{ d(\ell)\} = (\{K,d(\ell)\} \cup \Gamma\setminus\{K\})\setminus \{ d(\ell)\} = \Gamma.
	\]
	Also note that $\F{\ell}{\Gamma} \in \proc{\ell} \subseteq \cZ_\ell$. 
\end{itemize}

Therefore, we get 
\begin{IEEEeqnarray*}{lCl}
	X_{\{\ell\} \cup \sett} = \zeta_1 + \underbrace{\left(\bigoplus_{ j \in [K-1] \setminus \sett}  
	\F{d(\ell)}{(\{j\} \cup \sett)\setminus \{ d(\ell)\}} \right)
	}_{\textrm{already decoded}} + \F{d(\ell)}{\Gamma},
\end{IEEEeqnarray*}
where $\zeta_1$ is the sum of some subfiles that are cached in $\cZ_\ell$. Note that all subfiles in the second term are indexed by sets $(\{j\} \cup \sett)\setminus \{ d(\ell)\}$ which do not include $K$, and hence already decoded by $W_{\ell}$ as explained in Appendix~\ref{sec:dec-all-notK}. Thus, the desired subfile $\F{d(\ell)}{\Gamma}$ can be recovered from $X_{\{\ell\} \cup \sett}$ by removing the interference using cached data (interference suppression), as well as the previously decoded subfiles (successive interference cancellation).  This completes the proof of Lemma~\ref{lm:decode-all-OFC}. \hfill $\blacksquare$

\section{Proof of Lemma~\ref{lm:decode-K-OFC}}
\label{app:lm:decode-K-OFC}
We will show the decodability of subfiles assigned to the ignored worker node $W_K$ by the master node at iteration $t+1$. More precisely, we need to show $\F{d(K)}{}$ can be recovered from $(\cX, \cZ_K)$. First, note that if $d(K)=K$, then $\F{d(K)}{}=\F{K}{}$ is already cached at $W_K$ and the claim clearly holds.   Next, note that the subfiles of $\F{d(K)}{}$ indexed by $\Gamma$ with $K\in \Gamma$ are cached in $\ex{K}{d(K)} = \{\F{d(K)}{\Gamma}: K\in \Gamma, \Gamma \subseteq [K] \setminus \{d(K)\}, |\Gamma| = S-1\}$. Therefore, we can assume $d(K)\neq K$ and restrict our attention to $\F{d(K)}{\Gamma}$ with $K\notin \Gamma$. 
Worker node $W_K$ can add up the received sub-messages $X_{\{\ell\} \cup \Gamma}$ over all $\ell\in [K-1]\setminus \Gamma$ to obtain
\begin{IEEEeqnarray}{l}
	\bigoplus_{\ell\in [K-1]\setminus \Gamma} X_{\{\ell\} \cup \Gamma} 
	\nonumber\\
	=
	\bigoplus_{\ell\in [K-1]\setminus \Gamma} \:
	\bigoplus_{i \in \{\ell\} \cup \Gamma} 
	\left( 
	\F{i}{(\{\ell\} \cup \Gamma) \setminus \{i\}} 
	\oplus
	\F{d(i)}{(\{\ell\} \cup \Gamma) \setminus \{d(i)\}}
	\oplus	
	\bigoplus_{ j \in [K] \setminus (\{\ell\} \cup \Gamma) } 
	\F{d(i)}{(\{j\} \cup \{\ell\} \cup \Gamma)\setminus \{i, d(i)\}} 
	\right)	\nonumber\\
	=
	\bigoplus_{\ell\in [K-1]\setminus \Gamma} \:
	\bigoplus_{i \in \Gamma} 
	\left( 
	\F{i}{(\{\ell\} \cup \Gamma) \setminus \{i\}} 
	\oplus
	\F{d(i)}{(\{\ell\} \cup \Gamma) \setminus \{d(i)\}}
	\oplus	
	\bigoplus_{ j \in [K] \setminus (\{\ell\} \cup \Gamma) } 
	\F{d(i)}{(\{j, \ell\} \cup \Gamma)\setminus \{i, d(i)\}} 
	\right)\nonumber\\
	\phantom{=}
	\oplus 
	\bigoplus_{\ell\in [K-1]\setminus \Gamma} \:
	\bigoplus_{i \in \{\ell\}} 
	\left( 
	\F{i}{(\{\ell\} \cup \Gamma) \setminus \{i\}} 
	\oplus
	\F{d(i)}{(\{\ell\} \cup \Gamma) \setminus \{d(i)\}}
	\oplus	
	\bigoplus_{ j \in [K] \setminus (\{\ell\} \cup \Gamma) } 
	\F{d(i)}{(\{j, \ell\} \cup \Gamma)\setminus \{i, d(i)\}} 
	\right)
	\label{eq:ignored-1}\\
	=
	\bigoplus_{i \in \Gamma} \:
	\bigoplus_{\ell\in [K-1]\setminus \Gamma}
	\left( 
	\F{i}{(\{\ell\} \cup \Gamma) \setminus \{i\}} 
	\oplus
	\F{d(i)}{(\{\ell\} \cup \Gamma) \setminus \{d(i)\}}
	\oplus	
	\bigoplus_{ j \in [K] \setminus (\{\ell\} \cup \Gamma) } 
	\F{d(i)}{(\{j, \ell\} \cup \Gamma)\setminus \{i, d(i)\}} 
	\right)\nonumber\\
	\phantom{=}
	\oplus 
	\bigoplus_{\ell\in [K-1]\setminus \Gamma}
	\left( 
	\F{\ell}{(\{\ell\} \cup \Gamma) \setminus \{\ell\}} 
	\oplus
	\F{d(\ell)}{(\{\ell\} \cup \Gamma) \setminus \{d(\ell)\}}
	\oplus	
	\bigoplus_{ j \in [K] \setminus (\{\ell\} \cup \Gamma) } 
	\F{d(\ell)}{(\{j, \ell\} \cup \Gamma)\setminus \{\ell, d(\ell)\}} 
	\right)
	\label{eq:ignored-1_2}\\
	=
	\underbrace{\left(
		\bigoplus_{i \in \Gamma}
		\bigoplus_{\ell\in [K-1]\setminus \Gamma}
		\F{i}{(\{\ell\} \cup \Gamma) \setminus \{i\}}
		\right)}_{\mathsf{Term_1}} 
	\oplus
	\underbrace{
		\left(
		\bigoplus_{i \in \Gamma}
		\bigoplus_{\ell\in [K-1]\setminus \Gamma}
		\F{d(i)}{(\{\ell\} \cup \Gamma) \setminus \{d(i)\}}
		\right)}_{\mathsf{Term_2}} 
	\oplus
	\underbrace{
		\left(
		\bigoplus_{i \in \Gamma}
		\bigoplus_{\ell\in [K-1]\setminus \Gamma}
		\bigoplus_{ j \in [K] \setminus (\{\ell\} \cup \Gamma) } 
		\F{d(i)}{(\{j, \ell\} \cup \Gamma)\setminus \{i, d(i)\}} 
		\right)}_{\mathsf{Term_3}} 
	\nonumber\\
	\phantom{=}
	\oplus 
	\underbrace{
		\left( 
		\bigoplus_{\ell\in [K-1]\setminus \Gamma}
		\F{\ell}{\Gamma } 
		\right)}_{\mathsf{Term_4}} 
	\oplus
	\underbrace{
		\left(
		\bigoplus_{\ell\in [K-1]\setminus \Gamma}
		\left(
		\F{d(\ell)}{(\{\ell\} \cup \Gamma) \setminus \{d(\ell)\}}
		\oplus
		\left(	
		\bigoplus_{ j \in [K] \setminus (\{\ell\} \cup \Gamma) } 
		\left(	
		\F{d(\ell)}{(\{j\} \cup \Gamma)\setminus \{ d(\ell)\}} 
		\right)
		\right)\right)
		\right)}_{\mathsf{Term_5}},
	\label{eq:ignored-0}	
\end{IEEEeqnarray}
where
\begin{itemize}[leftmargin=*]
	\item in \eqref{eq:ignored-1}, the inner summation over $i\in \{\ell\} \cup \Gamma$ is broken into $i\in \Gamma$ and $i=\ell$;
	\item in \eqref{eq:ignored-1_2}, the order of summation over $\ell$ and $i$ is reversed;
	\item and finally, the double sum in \eqref{eq:ignored-1_2} is decomposed into $\mathsf{Term_1} \oplus \mathsf{Term_2} \oplus \mathsf{Term_3}$, and the last summation in \eqref{eq:ignored-1_2} is decomposed into $\mathsf{Term_4} \oplus \mathsf{Term_5}$. 
\end{itemize}

Next, we can rewrite $\mathsf{Term_2}$ in~\eqref{eq:ignored-0}	as
\begin{IEEEeqnarray}{lCl}
	\mathsf{Term_2}
	&=&
	\left(		
	\bigoplus_{i \in \Gamma} 
	\bigoplus_{\ell\in [K-1]\setminus \Gamma}
	\F{d(i)}{(\{\ell\} \cup \Gamma) \setminus \{d(i)\}}
	\right)
	\nonumber\\
	&=&
	\left(
	\bigoplus_{\ell \in \Gamma} 
	\bigoplus_{j\in [K-1]\setminus \Gamma}
	\F{d(\ell)}{(\{j\} \cup \Gamma) \setminus \{d(\ell)\}}
	\right)
	\label{eq:term2_1}\\
	& = &
	\left(	
	\bigoplus_{j\in [K-1]\setminus \Gamma}
	\bigoplus_{\ell \in \Gamma} 
	\F{d(\ell)}{(\{j\} \cup \Gamma) \setminus \{d(\ell)\}}
	\right),
	\label{eq:term2_2}
\end{IEEEeqnarray}
where
\begin{itemize}[leftmargin=*]
	\item in \eqref{eq:term2_1}, we change the name of variables $i$ and $\ell$ to $\ell$ and~$j$, respectively;
	\item and in \eqref{eq:term2_2}, the order of summation over $\ell$ and $j$ is reversed.
\end{itemize}
Moreover, $\mathsf{Term_5}$ in \eqref{eq:ignored-0} can be expanded as
\begin{IEEEeqnarray}{lCl}
	\mathsf{Term_5} 
	&=&
	\left( 
	\bigoplus_{\ell\in [K-1]\setminus \Gamma}
	\left( 
	\F{d(\ell)}{(\{\ell\} \cup \Gamma) \setminus \{d(\ell)\}}
	\oplus
	\left(	
	\bigoplus_{ j \in [K] \setminus (\{\ell\} \cup \Gamma) } 
	\F{d(\ell)}{(\{j\} \cup \Gamma)\setminus \{ d(\ell)\}} 
	\right)
	\right) 
	\right)
	\nonumber\\
	&=&
	\left(	
	\bigoplus_{\ell\in [K-1]\setminus \Gamma}
	\left(
	\bigoplus_{ j \in [K] \setminus \Gamma } 
	\F{d(\ell)}{(\{j\} \cup \Gamma)\setminus \{ d(\ell)\}} 
	\right)	
	\right)
	\label{eq:term3_1}\\
	&=&
	\left(
	\bigoplus_{ j \in [K] \setminus \Gamma }
	\bigoplus_{\ell\in [K-1]\setminus \Gamma}
	\F{d(\ell)}{(\{j\} \cup \Gamma)\setminus \{ d(\ell)\}}
	\right)\label{eq:term3_2}\\
	&=&
	\left(
	\bigoplus_{ j \in [K-1] \setminus \Gamma }
	\bigoplus_{\ell\in [K-1]\setminus \Gamma}
	\F{d(\ell)}{(\{j\} \cup \Gamma)\setminus \{ d(\ell)\}}
	\right)
	\oplus 
	\left(
	\bigoplus_{\ell\in [K-1]\setminus \Gamma}
	\F{d(\ell)}{(\{K\} \cup \Gamma)\setminus \{ d(\ell)\}}
	\right),
	\label{eq:term3_3}
\end{IEEEeqnarray}
where
\begin{itemize}[leftmargin=*]
	\item in \eqref{eq:term3_1}, the inner summations over $j=\ell$ and $j\in [K]\setminus(\{\ell\}\cup\Gamma)$ are merged into a single summation over $j\in [K]\setminus \Gamma$;
	\item in \eqref{eq:term3_2}, the order of summation over $\ell$ and $j$ is reversed;
	\item and finally in \eqref{eq:term3_3}, the summation over $j$ is broken into $j\in [K-1]\setminus \Gamma$ and $j=K$.
\end{itemize}
Combining $\mathsf{Term_2}$ and $\mathsf{Term_5}$, we obtain
\begin{IEEEeqnarray}{lCl}
	\mathsf{Term_2} \oplus \mathsf{Term_5} 
	&=&
	\left(
	\bigoplus_{j\in [K-1]\setminus \Gamma}
	\bigoplus_{\ell \in [K-1]} 
	\F{d(\ell)}{(\{j\} \cup \Gamma) \setminus \{d(\ell)\}}
	\right)
	\oplus 
	\underbrace{\left(
		\bigoplus_{\ell\in [K-1]\setminus \Gamma}
		\F{d(\ell)}{(\{K\} \cup \Gamma)\setminus \{ d(\ell)\}}
		\right)}_{\zeta_2}
	\label{eq:ignored-merge-ell}\\
	&=&
	\left(
	\bigoplus_{j\in [K-1]\setminus \Gamma}
	\bigoplus_{\ell \in [K]} 
	\F{d(\ell)}{(\{j\} \cup \Gamma) \setminus \{d(\ell)\}} 
	\right)
	\oplus
	\left(
	\bigoplus_{j\in [K-1]\setminus \Gamma}
	\F{d(K)}{(\{j\} \cup \Gamma) \setminus \{d(K)\}}
	\right)
	\oplus 
	\zeta_2
	\label{eq:ignored-add-subtract}\\
	&=&
	\bigoplus_{j\in [K-1]\setminus \Gamma}
	\left(
	\bigoplus_{u  \in [K]} \ 
	\underbrace{\ \ \F{u}{(\{j\} \cup \Gamma) \setminus \{u\}}\ \ }_{=0 \  \textrm{if $u\notin\{j\}\cup\Gamma$ }}
	\right)
	\oplus
	\left(
	\bigoplus_{j\in [K-1]\setminus \Gamma}\ 
	\underbrace{\ \F{d(K)}{(\{j\} \cup \Gamma) \setminus \{d(K)\}}\ }_{=0 \ \textrm{if $d(K)\notin\{j\}\cup\Gamma$}}
	\right)
	\oplus 
	\zeta_2
	\label{eq:ignored-d-bijective}\\
	&=&
	\bigoplus_{j\in [K-1]\setminus \Gamma}
	\left(
	\bigoplus_{u  \in \{j\}\cup \Gamma} 
	\F{u}{(\{j\} \cup \Gamma) \setminus \{u\}}
	\right)
	\oplus
	\left(
	\F{d(K)}{\Gamma}
	\right)
	\oplus 
	\zeta_2
	\label{eq:ignored-cardinality}\\
	&=& 
	\bigoplus_{j\in [K-1]\setminus \Gamma}
	\left(
	\bigoplus_{u  \in \Gamma} 
	\F{u}{(\{j\} \cup \Gamma) \setminus \{u\}}
	\right)
	\oplus
	\left(
	\bigoplus_{j\in [K-1]\setminus \Gamma}
	\F{j}{\Gamma}
	\right)
	\oplus
	\left(
	\F{d(K)}{\Gamma}
	\right)
	\oplus 
	\zeta_2
	\label{eq:ignored-break-u}\\
	&=& 
	\left(
	\bigoplus_{u  \in \Gamma} 
	\bigoplus_{j\in [K-1]\setminus \Gamma}
	\F{u}{(\{j\} \cup \Gamma) \setminus \{u\}}
	\right)
	\oplus
	\left(
	\bigoplus_{j\in [K-1]\setminus \Gamma}
	\F{j}{\Gamma}
	\right)
	\oplus
	\left(
	\F{d(K)}{\Gamma}
	\right)
	\oplus 
	\zeta_2
	\label{eq:ignored-swap-2}\\
	&=& \mathsf{Term_1} \oplus \mathsf{Term_4} \oplus
	\F{d(K)}{\Gamma}
	\oplus 
	\zeta_2,
	\label{eq:ignored-term-2-5}
\end{IEEEeqnarray}
where
\begin{itemize}[leftmargin=*]
	\item in \eqref{eq:ignored-merge-ell}, $\mathsf{Term_2}$ with summation over $\ell\in \Gamma$ and the first parentheses of $\mathsf{Term_5}$ with summation over $\ell\in [K-1]\setminus \Gamma$ are merged to a single summation over $\ell\in [K-1]$;
	\item in \eqref{eq:ignored-add-subtract}, the term 
	$\displaystyle \bigoplus_{j\in [K-1]\setminus \Gamma} \bigoplus_{\ell=K} 
	\F{d(\ell)}{(\{j\} \cup \Gamma) \setminus \{d(\ell)\}}$ is added and subtracted, i.e., the term is XORed twice;
	\item in \eqref{eq:ignored-d-bijective}, we use the fact that $d(\cdot)$ is a bijective map over $[K]$, and hence replaced summation over $\ell\in[K]$ with another summation over $u\in [K]$ where $u=d(\ell)$;
	\item in \eqref{eq:ignored-cardinality}, we know that $|\Gamma|=S-1$ and $j\notin \Gamma$. Hence, $|(\{j\} \cup \Gamma)\setminus \{ u\}|=S-1$ if and only if $u\in \{j\} \cup \Gamma$. Similarly, $|(\{j\}\cup \Gamma)\setminus\{d(K)\}|=S-1$ if and only if $d(K)\in \{j\} \cup \Gamma$. Moreover, we know that $d(K)\notin \Gamma$, since we assume $\F{d(K)}{\Gamma}$~is a non-trivial subfile requested by $W_K$ at iteration $t+1$. Hence, the only valid choice for $d(K)$ is $d(K)=j$. After all simplifications, we get the expression in \eqref{eq:ignored-cardinality};
	\item in \eqref{eq:ignored-break-u}, we break the summation over $u\in \{j\}\cup \Gamma$ into two summations, one over $u\in \Gamma$ and the other over~$u=j$;
	\item and finally in \eqref{eq:ignored-swap-2}, the order of the two summations of the first term is reversed in order to identify expressions similar to $\mathsf{Term_1}$ and $\mathsf{Term_4}$ in \eqref{eq:ignored-0}.
\end{itemize}
It should be noted that every non-zero subfile that appears in  $\zeta_2$ in \eqref{eq:ignored-term-2-5} is either a subfile of $\F{K}{}$ (for $\ell$ satisfying $d(\ell)=K$) or a subfile indexed by set $\left(\{K\} \cup \Gamma \right) \setminus \{d(\ell)\}$ that includes $K$. Both groups of such subfiles are cached at worker node $W_K$~by definition of $\cZ_K$ in \eqref{eq:Z-def}. Therefore, $\zeta_2$ can be completely recovered from $\cZ_K$. 

Furthermore, for each $i\in \Gamma$, the inner term in $\mathsf{Term_3}$ can be rewritten as  
\begin{IEEEeqnarray}{l}
	\bigoplus_{\ell\in [K-1]\setminus \Gamma} \: 
	\bigoplus_{ j \in [K] \setminus (\{\ell\} \cup \Gamma) } 
	\F{d(i)}{(\{j, \ell\} \cup \Gamma)\setminus \{ i,d(i)\}} 
	\nonumber\\
	\hspace{0.25in}
	=
	\underbrace{\left(
		\bigoplus_{\ell\in [K-1]\setminus \Gamma} \:
		\bigoplus_{ j \in [K-1] \setminus (\{\ell\} \cup \Gamma) } 
		\F{d(i)}{(\{j,\ell\} \cup \Gamma)\setminus \{ i, d(i)\}}
		\right)}_{=0}
	\oplus 
	\underbrace{\left( \bigoplus_{\ell\in [K-1]\setminus \Gamma} 
		\F{d(i)}{(\{K, \ell\} \cup \Gamma)\setminus \{ i, d(i)\}}
		\right)}_{\zeta_3(i)},
	\label{eq:ignored-term3}
\end{IEEEeqnarray}
where the first term is zero since every subfile $\F{d(i)}{(\{a,b\} \cup \Gamma)\setminus \{ i, d(i)\}}$ appears exactly twice in the summation: once for $(\ell=a, j=b)$ and another time for $(\ell=b,j=a)$, where $a \neq b$. Hence, their contributions will be canceled when they are XORed. Moreover, we can show that all the subfiles appearing in $\zeta_3(i)$ are already cached in $\cZ_K$, and can be recovered by $W_K$. To see this, recall that $i\in \Gamma$ and $K\notin \Gamma$, which imply $i\neq K$. Therefore, either $d(i)=K$, or subscript $(\{K, \ell\} \cup \Gamma)\setminus \{ i, d(i)\}$ includes $K$. In the former case we have  $\F{d(i)}{(\{K, \ell\} \cup \Gamma)\setminus \{ i, d(i)\}} = \F{K}{(\{K, \ell\} \cup \Gamma)\setminus \{ i, d(i)\}}\in \proc{K}\subseteq \cZ_K$, while in the latter case we have $\F{d(i)}{(\{K, \ell\} \cup \Gamma)\setminus \{ i, d(i)\}} \in \ex{K}{} \subset \cZ_K$.  

Plugging \eqref{eq:ignored-term-2-5} and \eqref{eq:ignored-term3} into \eqref{eq:ignored-0}, we get
\begin{IEEEeqnarray*}{lCl}
	\bigoplus_{\ell\in [K-1]\setminus \Gamma} X_{\{\ell\} \cup \Gamma} 
	&=& \left(\mathsf{Term_2} \oplus \mathsf{Term_5}\right) \oplus  \left(\mathsf{Term_1} \oplus  \mathsf{Term_4}\right) \oplus  \mathsf{Term_3} 
	\nonumber\\
	&=& \left(\mathsf{Term_1} \oplus \mathsf{Term_4} \oplus
	\F{d(K)}{\Gamma} \oplus
	\zeta_2 \right)
	\oplus \left(\mathsf{Term_1} \oplus  \mathsf{Term_4}\right) 
	\oplus \bigoplus_{i\in \Gamma} \zeta_3(i) 
	\nonumber\\ 
	&=& \left(\zeta_2 \oplus  \bigoplus_{i\in \Gamma} \zeta_3(i)\right) \oplus  \F{d(K)}{\Gamma}, 
\end{IEEEeqnarray*}
where $\left(\zeta_2 \oplus  \bigoplus_{i\in \Gamma} \zeta_3(i)\right)$ can be reconstructed from $\cZ_K$. Consequently, the subfile $\F{d(K)}{\Gamma}$ can be recovered from the cache content $\cZ_K$ and $\bigoplus_{\ell\in [K-1]\setminus \Gamma} X_{\{\ell\} \cup \Gamma}$. This completes the proof of Lemma~\ref{lm:decode-K-OFC}. \hfill $\blacksquare$

\section{Proof of Lemma~\ref{lm:OptAchvScheme}}
\label{app:opt_achvScheme}
We first define some notation.
Without loss of generality, we assume $u(i)=i$ for $i\in [K]$. Let us partition $\mathcal{W}$, the set of $K$~worker nodes, into $\gamma$ disjoint subsets $\mathcal{W}_i$ according to the cycles of the file transition graph, where $\mathcal{W}_i$ is the set of worker nodes that belong to cycle $i$, for $i \in [\gamma]$. Hence, we have
\begin{IEEEeqnarray}{lCl}
	\cW &=& \bigcup_{i=1}^\gamma \cW_i,
\end{IEEEeqnarray}
where
\begin{IEEEeqnarray}{l}
	\label{eqn:W_intersect}
	\mathcal{W}_i \cap \mathcal{W}_j = \varnothing, \quad
	\forall i \neq j, \: i, j\in [\gamma].
\end{IEEEeqnarray}
In what follows, without loss of generality, let us consider the first $\gamma-1$ cycles, and designate the  $\gamma$th cycle as the \emph{ignored cycle}. 
Moreover, assume that the ignored worker node $W_K$ belongs to the ignored cycle $\gamma$. 	
Let $\gamma-1 \geq S$, and consider an arbitrary subset $\Psi$ of $S$ cycles, that is $\Psi \subseteq [\gamma-1]$ and $|\Psi| = S$. We define $\mathcal{W}^{\otimes\Psi}$ to be the Cartesian product of the corresponding $S$ sets of worker nodes, that is defined as
\begin{IEEEeqnarray*}{l}
	\mathcal{W}^{\otimes\Psi} = 
	\bigotimes_{i \in \Psi} \mathcal{W}_i = \{\set \subseteq [K]: \set \cap \cW_i =1, \forall i\in \Psi \}, 
\end{IEEEeqnarray*}	
 for every $\Psi \subseteq [\gamma-1]$ with $\vert \Psi \vert = S$. 
Note that each element set in $\mathcal{W}^{\otimes\Psi}$ consists of a tuple of $S$ worker nodes, each belongs to a different cycle. 

Recall that each sub-message $X_{\set}$, defined in \eqref{eq:X_Delta}, is designed for a subset of $S$ worker nodes, i.e., $\set \subseteq [K-1]$ and $|\set|=S$. The sub-message $X_\set$ is given by
\begin{IEEEeqnarray}{l}	
	\label{eqn:Xtheta_again}
	X_{\set}  = 
	\bigoplus_{i \in \Delta} \left( \F{i}{\Delta \setminus \{i\}} 
	\oplus
	\F{d(i)}{\Delta \setminus \{d(i)\}}
	\oplus	
	\bigoplus_{ j \in [K] \setminus \set } 
	\F{d(i)}{(\{j\} \cup \set)\setminus \{i, d(i)\}}
	\right).
\end{IEEEeqnarray}
In the following, we consider some $\set \in \mathcal{W}^{\otimes\Psi}$, and expand the corresponding sub-message $X_\set$. Note that since $\set \in \mathcal{W}^{\otimes\Psi}$, it has only one worker node from each cycle. hence, for any $i\in \set$, we have either $d(i)\notin \set$ or $d(i)=i$ (because otherwise $W_i$~and $W_{d(i)}$ will be two distinct and consecutive worker nodes that belong to the same cycle). This allows us to expand $X_{\set}$~as
\begin{IEEEeqnarray}{lCl}	
	X_{\set} & = &
	\left(\bigoplus_{\substack{i \in \Delta \\ d(i) \notin \set}} 
	\F{i}{\Delta \setminus \{i\}}\right) 
	\oplus
	\left(\bigoplus_{\substack{i \in \Delta \\ d(i) =i}} 
	\F{i}{\Delta \setminus \{i\}}\right) 
	\oplus
	\left(\bigoplus_{\substack{i \in \Delta \\ d(i)\notin \set}} 
	\F{d(i)}{\Delta \setminus \{d(i)\}}\right) 
	\oplus
	\left(\bigoplus_{\substack{i \in \Delta \\ d(i)=i}} \F{d(i)}{\Delta \setminus \{d(i)\}}\right)
	\nonumber\\
	& &
	\oplus	
	\left(\bigoplus_{\substack{i \in \Delta \\ d(i) \notin \set}} 
	\: \bigoplus_{ j \in [K] \setminus \set } 
	\F{d(i)}{(\{j\} \cup \set)\setminus \{i, d(i)\}}\right)
	\oplus
	\left(\bigoplus_{\substack{i \in \Delta \\ d(i) =i}} 
	\: \bigoplus_{ j \in [K] \setminus \set } 
	\F{d(i)}{(\{j\} \cup \set)\setminus \{i, d(i)\}}\right)
	\label{eqn:Xset_MPC_1}
	\\
	& = &
	\left(\bigoplus_{\substack{i \in \Delta \\ d(i) \notin \set}} 
	\F{i}{\Delta \setminus \{i\}}\right) 
	\oplus
	\left(\bigoplus_{\substack{i \in \Delta \\ d(i) =i}} 
	\F{i}{\Delta \setminus \{i\}}\right) 
	\oplus
	\left(\bigoplus_{\substack{i \in \Delta \\ d(i)\notin \set}} 
	\F{d(i)}{\Delta \setminus \{d(i)\}}\right) 
	\oplus
	\left(\bigoplus_{\substack{i \in \Delta \\ d(i) =i}} \F{i}{\Delta \setminus \{i\}}\right)
	\nonumber\\
	& &
	\oplus	
	\left(\bigoplus_{\substack{i \in \Delta \\ d(i) \notin \set}} 
	\: \bigoplus_{ j \in [K] \setminus \set } 
	\F{d(i)}{(\{j\} \cup \set)\setminus \{i, d(i)\}}\right)
	\oplus
	\left(\bigoplus_{\substack{i \in \Delta \\ d(i) =i}} 
	\: \bigoplus_{ j \in [K] \setminus \set } 
	\F{i}{(\{j\} \cup \set)\setminus \{i\}}\right)
	\label{eqn:Xset_MPC_1_2}
	\\
	&=&
	\left(\bigoplus_{\substack{i \in \Delta \\ d(i) \notin \set}} 
	\F{i}{\Delta \setminus \{i\}}\right) 
	\oplus
	\left(\bigoplus_{\substack{i \in \Delta \\ d(i) \notin \set}} 
	\: \bigoplus_{ j \in [K] \setminus \set } 
	\F{d(i)}{(\{j\} \cup \set)\setminus \{i, d(i)\}}\right)
	\label{eqn:Xset_MPC_2}
	\\
	& =&
	\left(\bigoplus_{\substack{i \in \Delta \\  d(i) \notin \set}} 
	\F{i}{\Delta \setminus \{i\}}\right) 
	\oplus
	\left(\bigoplus_{\substack{i \in \Delta \\ d(i) \notin \set}} 
	\: \bigoplus_{\substack{j \in [K] \setminus \set \\ j \neq d(i)}}
	\F{d(i)}{(\{j\} \cup \set)\setminus \{i, d(i)\}}\right)
	\oplus
	\left(\bigoplus_{\substack{i \in \Delta \\ d(i) \notin \set}} 
	\bigoplus_{j=d(i)}
	\F{d(i)}{(\{j\} \cup \set)\setminus \{i, d(i)\}}\right)
	\label{eqn:Xset_MPC_3}
	\\
	& = &
	\left( \bigoplus_{\substack{
			i \in \set \\
			d(i) \notin \set
	}} 
	\F{i}{\set \setminus \{i\}}
	\right)
	\oplus 
	\left(
	\bigoplus_{\substack{
			i \in \set \\
			d(i) \notin \set}} \:
	F^{d(i)}_{\set{\setminus \{i\}}}
	\right),
	\label{eqn:Xset_MPC_4}
\end{IEEEeqnarray}
where
\begin{itemize}[leftmargin=*]
	\item in \eqref{eqn:Xset_MPC_1}, the first summation over $i\in \set$ in \eqref{eqn:Xtheta_again} is split into $(i \in \Delta, d(i) \notin \set)$ and $(i \in \Delta, d(i)=i )$;
	\item in \eqref{eqn:Xset_MPC_1_2}, we have used the fact that if $i,d(i)\in \set$ then $i=d(i)$ in the forth and sixth summations;
	\item in \eqref{eqn:Xset_MPC_2}, we have canceled out the identical quantities in the second and the forth summations of \eqref{eqn:Xset_MPC_1_2}. Moreover, in the third term of \eqref{eqn:Xset_MPC_1_2},  each subfile is indexed by $\set\setminus \{d(i)\}$, while $d(i)\notin \set$. Note that $|\set \setminus\{d(i)\}|=S\neq S-1$, and thus such subfiles are defined to be zero.
	This implies that the third summation is zero. Similarly, the sixth summation in \eqref{eqn:Xset_MPC_1_2} is zero, due the fact that $ |(\{j\} \cup \set)\setminus \{i\}| =  1+ S -1 = S >S-1$,
	and hence all the subfiles in the sixth summation are zero. Therefore, only the first and the fifth terms of \eqref{eqn:Xset_MPC_1_2} survive;
	\item in \eqref{eqn:Xset_MPC_3}, the summation over $j$ is split into $j \in [K] \setminus \set, j \neq d(i)$ and $j = d(i)$;
	\item and finally, the second term in \eqref{eqn:Xset_MPC_3} is set to zero in \eqref{eqn:Xset_MPC_4}. This is due to the fact that  for $d(i) \notin \set$ and $d(i) \neq j$, we have  $\vert (\{j\} \cup \set)\setminus \{i, d(i)\} \vert = \vert (\{j\} \cup \set)\setminus \{i\} \vert  = S > S-1$, and hence, $\F{d(i)}{(\{j\} \cup \set)\setminus \{i, d(i)\}} = 0$. 
\end{itemize}

Next, we prove that there exists one linearly dependent sub-message in the set of sub-messages $\{X_\set: \set \in \mathcal{W}^{\otimes\Psi}\}$ for each group of $S$ cycles, determined by $\Psi\subseteq [\gamma-1]$. More~precisely, we claim that
\begin{IEEEeqnarray}{lCl}
	\bigoplus_{\set \in \mathcal{W}^{\otimes \Psi}} X_{\set}
	& = &
	\bigoplus_{\set \in \mathcal{W}^{\otimes \Psi}} 
	\left(
	\left( \bigoplus_{\substack{
			i \in \set \\
			d(i) \neq i
	}} 
	\F{i}{\set \setminus \{i\}}
	\right)
	\oplus
	\left(
	\bigoplus_{\substack{
			i \in \set \\
			d(i) \notin \set}} \:
	F^{d(i)}_{\set{\setminus \{i\}}}
	\right)
	\right)\nonumber\\
	& = & 
	\bigoplus_{\Gamma \in \mathcal{W}^{\otimes\Psi}}
	\left( \bigoplus_{\substack{
			j \in \Gamma \\
			d(j) \neq j
	}} 
	\F{j}{\Gamma \setminus \{j\}}
	\right)
	\oplus 
	\bigoplus_{\set \in \mathcal{W}_{\Psi}^{\otimes}}
	\left(
	\bigoplus_{\substack{
			i \in \set \\
			d(i) \notin \set}} 
	F^{d(i)}_{\set{\setminus \{i\}}}
	\right)\nonumber\\
	&=&
	0.
	\label{eqn:sum_Xset_0}
\end{IEEEeqnarray}
We prove \eqref{eqn:sum_Xset_0} by showing that each subfile appears exactly twice in the XOR equation which immediately yields that the summation in \eqref{eqn:sum_Xset_0} equals to zero.
To this end,  consider a pair of $(\Delta,i)$ with $\set \in \mathcal{W}^{\otimes\Psi}$, $i\in \Delta$ and $d(i)\neq i$. Therefore, we have  a subfile $F^{d(i)}_{\Delta\setminus\{i\}}$ that appears in the second summation in \eqref{eqn:sum_Xset_0}. Now, define $j=d(i)$ and $\Gamma = (\Delta \cup \{j\} ) \setminus \{i\}$. It is clear that $i$ and $d(i)$ belong to the same cycle, and hence $\Gamma \in \mathcal{W}^{\otimes\Psi}$. On the other hand, since $d(\cdot)$ is a bijective map, the fact that $d(i)\neq i$ implies that $d(d(i)) \neq d(i)$, or $d(j) \neq j$. Therefore, the pair $(\Gamma,j)$ satisfies the three conditions in the first summation, namely $\Gamma \in \mathcal{W}^{\otimes\Psi}$, $j\in \Gamma$, and $d(j)\neq j$. Hence, the corresponding term $F^j_{\Gamma\setminus\{j\}}$ appears in the second summation. However, by plugging in $j=d(i)$ and $\Gamma = (\Delta \cup \{j\} ) \setminus \{i\}$, we get $F^j_{\Gamma\setminus\{j\}} = F^{d(i)}_{\set \setminus \{i\}}$. This shows that the terms $F^j_{\Gamma\setminus\{j\}}$ in the first summation and $F^{d(i)}_{\Delta\setminus\{i\}}$ in the second summation cancel out each other in the entire summation. This applies to each term, and shows that \eqref{eqn:sum_Xset_0} holds. 

As a result, there is (at least) one linearly dependent sub-message in the set of sub-messages $\{X_{\set} : \set \in \mathcal{W}_{\Psi}^{\otimes}\}$, and any of these sub-messages can be reconstructed by summing the others. Therefore, we can refrain from sending one of them in the delivery phase. Moreover, the groups of sub-messages $\{X_{\set} : \set \in \mathcal{W}_{\Psi}^{\otimes}\}$ are disjoint for different groups of cycles $\Psi$. 
Since there are $\binom{\gamma-1}{S}$ such groups of cycles, therefore one can refrain from sending $\binom{\gamma-1}{S}$ redundant sub-messages out of the $\binom{K-1}{S}$ sub-messages that the master node should broadcast to the worker nodes (according to the proposed delivery scheme in Section~\ref{sec:achv_scheme_N=K}). 
Consequently, the delivery load given by \eqref{eq:R_achv2} is achievable.
This completes the proof of Lemma~\ref{lm:OptAchvScheme}. \hfill $\blacksquare$

\section{Proof of Lemma~\ref{lemma_virtualNode}}
\label{app:lemma_virtualNode}
We want to prove that the broadcast message $\mathcal{X}$ and the cache contents of the virtual worker node $\mathcal{Z}_\star$ enable us to perfectly recover all the files of the shuffling system. First, note that according to the one-to-one positional labeling, we have 
\begin{IEEEeqnarray}{lCl}
	\label{eq:alpha_order}
	\{F^{i}: i \in \{1,2,\dots, K\}\} 
	&\equiv& \{F^{(c,p)}: c \in \{1,2,\dots, \gamma\}, \: p \in \{1,2, \dots, \ell_c\}\}.\nonumber
\end{IEEEeqnarray}
Hence, in order to prove the lemma, it suffices to show that 
\begin{IEEEeqnarray*}{lCl}
	H\left(\{\{F^{(c,p)}\}_{p=1}^{\ell_c}\}_{c=1}^{\gamma} \Big| \mathcal{X}, \cZ_\star \right) = 0.
\end{IEEEeqnarray*}
To this end, we consider two cases: $p=1$ and $p>1$. For $p=1$, from \eqref{eq:Z_star} we simply have 
$F^{(c,1)} \subseteq \cZ_\star$ for each $c\in [\gamma]$, which implies $H\left( \{ F^{(c,1)}\}_{c=1}^{\gamma}| \cX,  \cZ_\star \right) \leq H\left( \{ F^{(c,1)}\}_{c=1}^{\gamma}|  \cZ_\star \right)=0$. Therefore, we have
\begin{IEEEeqnarray}{lCl}
	H\left( \{\{F^{(c,p)}\}_{p=1}^{\ell_c}\}_{c=1}^{\gamma} \Big| \mathcal{X}, \cZ_\star \right) 
	&=& H\left( \{\{F^{(c,p)}\}_{p=2}^{\ell_c}\}_{c=1}^{\gamma} \Big|  \{F^{(x,1)}\}_{x=1}^{\gamma}, \mathcal{X}, \cZ_\star \right) 
	\nonumber\\
	&=& \sum_{c=1}^\gamma \sum_{p=1}^{\ell_c-1} 
	H\left( F^{(c,p+1)} \Big| \{ F^{(c',p')} \}_{(c',p')\preceq (c,p)} ,
	\{F^{(x,1)}\}_{x=1}^{\gamma}, \mathcal{X}, \cZ_\star \right),
	\label{eq:pr:lemma_virtualNode:1}
\end{IEEEeqnarray}
where we have used the chain rule in the last equation. Recall from \eqref{eq:Z_positional} that for any pair $(c,p)$ with $p<\ell_c$, we have
\begin{IEEEeqnarray}{lCl}
	\cZ_{(c,p)} 
	&=& F^{(c,p)} \cup \left(\bigcup_{(c',p') \neq (c,p)} \fu{(c',p')}{(c,p)}\right) 
	\nonumber\\
	&=& \underbrace{F^{(c,p)} \cup \left(\bigcup_{(c',p') \prec (c,p)} \fu{(c',p')}{(c,p)}\right)}_{\subseteq \bigcup_{(c',p')\preceq (c,p)} F^{(c',p')}} 
	\cup \underbrace{\left( \bigcup_{c'\in [\gamma]} \fu{(c',1)}{(c,p)} \right) \cup \left(\bigcup_{\substack{(c',p') \succ (c,p)\\ p'>1}} \fu{(c',p')}{(c,p)}\right)}_{\subseteq \cZ_\star}.
	\label{eq:recursive_cache_const}
\end{IEEEeqnarray}
Therefore, each term in \eqref{eq:pr:lemma_virtualNode:1} can be bounded by
\begin{IEEEeqnarray}{lCl}
	H\Big( F^{(c,p+1)} \Big| \{ F^{(c',p')} \}_{(c',p')\preceq (c,p)},  \mathcal{X}, \cZ_\star \Big) 
	&\leq& H\left( F^{(c,p+1)} , \cZ_{(c,p)} \Big| \{ F^{(c',p')} \}_{(c',p')\preceq (c,p)},  \mathcal{X}, \cZ_\star \right)
	\nonumber\\
	&\leq& H\left( \cZ_{(c,p)} \Big| \{ F^{(c',p')} \}_{(c',p')\preceq (c,p)}, \cZ_\star \right)
	+ H\left( F^{(c,p+1)} \Big| \cZ_{(c,p)}, \mathcal{X} \right)=0,
	\label{eq:pr:lemma_virtualNode:2}
\end{IEEEeqnarray}
where the first term in \eqref{eq:pr:lemma_virtualNode:2} is zero due to \eqref{eq:recursive_cache_const}, and the second term is zero due to the fact that the worker node at position $(c,p)$ should be able to recover its assigned file $F^{(c,p+1)}$ from its cache $\cZ_{(c,p)}$ and the broadcast message $\cX$. Plugging \eqref{eq:pr:lemma_virtualNode:2} into \eqref{eq:pr:lemma_virtualNode:1}, we conclude the claim of Lemma~\ref{lemma_virtualNode}. \hfill $\blacksquare$

\section{Proof of Lemma~\ref{lemma_R}}
\label{app:lemma_R}
The next step in the converse proof is to provide a lower bound on the communication load for a given instance of the shuffling problem as follows.

Consider a directed file transition graph $\cG(V,E)$, that characterizes one instance of the problem. We can start~with
\begin{IEEEeqnarray}{lCl}
	K &=& H \left( \{F^{i}\}_{i=1}^{K} \right)
	\nonumber\\
	& \leq & H\left(\{F^{i}\}_{i=1}^{K} ,\cX,\cZ_\star \right) 
	\nonumber\\
	& = & H(\cX,\cZ_\star) 
	+ H\left( \{F^{i}\}_{i=1}^{K} | \cX, \cZ_\star\right)
	\nonumber\\
	& = & H(\cX,\cZ_\star) 
	\label{eq:lowBound_P1_1}
	\\
	& = &
	H \left(\cX, 
	\bigcup_{c\in [\gamma]} F^{(c,1)}   , 
	\bigcup_{c \in \left[\gamma\right], p >1} 
	\left(\bigcup_{\substack{(c',p') \prec (c,p)\\ p'<\ell_{c'}}} \fu{(c,p)}{(c',p')} \right)\right)
	\label{eq:lowBound_P1_2}
	\\
	& \leq &
	H(X) + \sum_{c=1}^{\gamma}  H(F^{(c,1)}) + 
	\sum_{c=1}^{\gamma} \sum_{p=2}^{\ell_{c}}
	H \left(\bigcup_{\substack{(c',p') \prec(c,p)\\ p'<\ell_{c'}}} \fu{(c,p)}{(c',p')} 
	\right)
	\nonumber\\
	& = &
	R(d) + \gamma + 
	\sum_{c=1}^{\gamma} \sum_{p=2}^{\ell_c}  
	\mu^{(c,p)}_{\Phi(c,p)},
	\label{eq:lowBound_P1}
\end{IEEEeqnarray}
where $\Phi(c,p)=\{(c',p'): (c',p')\prec (c,p), p'<\ell_{c'}\}$.  Note that \eqref{eq:lowBound_P1_1} holds due to Lemma~\ref{lemma_virtualNode}, and in \eqref{eq:lowBound_P1_2} we use the definition of $\cZ_\star$ in \eqref{eq:Z_star}.  
This implies a lower bound on the communication load $R$ that is given by 
\begin{IEEEeqnarray}{lCl}
	R(d) & \geq &
	K - \gamma -  
	\sum_{c=1}^{\gamma} \sum_{p=2}^{\ell_c}
	\mu^{(c,p)}_{\Phi(c,p)}.
	\label{eq:R-LB-fixed-order}
\end{IEEEeqnarray}

A similar argument holds for any instance of the problem, characterized by an assignment function $d(\cdot)$ whose file transition graph is isomorphic to $\cG(V,E)$. Let $V=\{(c,p): c \in [\gamma], \ p \in [\ell_c]\}$ be the set of vertices of the graph, and $\Pi: V \rightarrow V$ be the set of all possible permutations on the vertices of the graph. For each $\pi\in \Pi$,  we have an instance of the shuffling problem characterized by the same file transition graph $\cG(V,E)$, in which node $W_i$ which was at position $(c,p)$ in the original problem, is now positioned at $\pi(c,p)$. For the new instance of the problem, we have 
\begin{IEEEeqnarray*}{l}
	\cZ_{\pi(c,p)} =  F^{\pi(c,p)} \cup \left(\bigcup_{(c',p') \neq (c,p)} \fu{\pi(c',p')}{\pi(c,p)}\right). 
\end{IEEEeqnarray*}
Hence, following an argument similar to that of \eqref{eq:R-LB-fixed-order}, we obtain 
\begin{IEEEeqnarray}{lCl}
	R(\cG) & =& \frac{1}{|\cD_\cG| } \sum_{d\in \cD_\cG} R(d)
	\nonumber\\
	& \geq & 
	K - \gamma -  
	\sum_{c=1}^{\gamma} \sum_{p=2}^{\ell_c}
	\mu^{\pi(c,p)}_{\pi(\Phi(c,p))},
	\label{eq:R-LB-permuted-order}
\end{IEEEeqnarray}
where $\pi(\Phi(c,p)) = \{\pi(c',p'): (c,'p')\in \Phi(c,p)\}$.  
Next, by averaging \eqref{eq:R-LB-permuted-order} over all $\pi\in \Pi$, we get 
\begin{IEEEeqnarray}{lCl}
	R(\cG) & \geq &
	K - \gamma -  
	\frac{1}{K! }  \sum_{\pi \in \Pi} \sum_{c=1}^{\gamma} \sum_{p=2}^{\ell_c}
	\mu^{\pi(c,p)}_{\pi(\Phi(c,p))},
	\label{eq:R-averaged}
\end{IEEEeqnarray}
where 
\begin{align}
	\frac{1}{K!} \sum_{\pi \in \Pi}  \sum_{c=1}^{\gamma} \sum_{p=2}^{\ell_c}
	\mu^{\pi(c,p)}_{\pi(\Phi(c,p))} 
	& = \frac{1}{K!}  \sum_{c=1}^{\gamma} \sum_{p=2}^{\ell_c} \sum_{k \in V}  \:\sum_{\substack{ \cA: \cA\subseteq V \setminus\{k\} \\ |\cA| = |\Phi(c,p)|}} \:\sum_{\substack{\pi\in \Pi: \pi(c,p) = k \\ \pi(\Phi(c,p))= \cA}}  
	\mu^{k}_{\cA}\nonumber\\
	& = \frac{1}{K!}  \sum_{c=1}^{\gamma} \sum_{p=2}^{\ell_c} \sum_{k\in [K]}  \: \sum_{\substack{ \cA: \cA\subseteq V \setminus\{k\} \\ |\cA| = |\Phi(c,p)|}} 1! |\cA|! (K-|\cA|-1)! 
	\mu^{k}_{\cA}
	\label{eq:mu-averaged_0}
	\\
	& = \frac{1}{K\binom{K-1}{|\cA|}}  \sum_{c=1}^{\gamma} \sum_{p=2}^{\ell_c} \sum_{k\in [K]}  \sum_{\substack{ \cA: \cA\subseteq V \setminus\{k\} \\ |\cA| = |\Phi(c,p)|}} 
	\mu^{k}_{\cA}\nonumber\\
	& = \sum_{c=1}^{\gamma} \sum_{p=2}^{\ell_c} 
	\mu_{|\Phi(c,p)|},
	\label{eq:mu-averaged}
\end{align}
where in \eqref{eq:mu-averaged_0} the innermost summation is evaluated by counting the number of permutations $\pi$ that satisfy $\pi(c,p) = k$ and $\pi(\Phi(c,p))= \cA$. More precisely, there are $|\cA|!$ ways to map entries of $\Phi(c,p)$ to $\cA$, and there is only one way to map $(c,p)$ to $k$. There are a total of $K-|\cA|-1$ remaining entries, which can be mapped in $(K-|\cA|-1)!$ ways. Moreover, in~\eqref{eq:mu-averaged}~we have used the definition of $\mu_\eta$ in \eqref{eqn:mu_eta}. Recall from the definition of $\Phi(c,p)$ that $|\Phi(c,p)| = \left(\sum_{t=1}^{c-1} \ell_t-1 \right) + (p - 1)$. It is easy to see that $|\Phi(c,p)| \neq |\Phi(c',p')|$  for any distinct pair of $(c,p)$ and $(c',p')$. Moreover, $|\Phi(c,p)| \geq 1$ (for $p\geq 1$) and $|\Phi(c,p)| \leq K-\gamma$. These together imply that $\{|\Phi(c,p)|: c \in \{1,2,\dots,\gamma\},\ p \in \{2,3,\dots, \ell_c\}\} = \{1,2,\dots, K-\gamma\}$. In~other words, for each integer $i$, there exists exactly one pair of $(c,p)$ such that $|\Phi(c,p)|=i$. Hence, the RHS of \eqref{eq:mu-averaged} can be simplified to $\sum_{i=1}^{K-\gamma} \mu_i$. Plugging this into \eqref{eq:R-averaged}, we~obtain
\begin{IEEEeqnarray}{lCl}
	R(\cG) & \geq &
	K - \gamma -  \sum_{i=1}^{K - \gamma}\mu_{i}.
\end{IEEEeqnarray}
This completes the proof of Lemma~\ref{lemma_R}. \hfill $\blacksquare$

\section{Proof of Lemma~\ref{lemma_opt_mu}}
\label{app:lemma_lemma_opt_mu}
This appendix is dedicated to prove an upper bound on $\mu_{\alpha}$'s in order to obtain a lower bound on $R(\cG)$. Note that variables $\{\mu_\alpha\}_{\alpha=0}^{K-1}$ defined in \eqref{eqn:mu_eta} can be generalized to any arbitrary family of sets. 

Consider an arbitrary family of sets $\{A_1,A_2,\ldots, A_{K-1}\}$, and define 
\begin{IEEEeqnarray*}{l}
	\theta_\alpha = \frac{1}{\binom{K-1}{\alpha}} \sum_{\substack{\cJ\subseteq [K-1] \\ |\cJ|=\alpha}} \left| A_{\cJ}\right|,
\end{IEEEeqnarray*}
where $A_{\cJ} = \bigcup_{j\in \cJ} A_j$. 

We need to derive some preliminary results in order to prove the lemma. The proofs of these claims are presented at the end of this appendix. The first lemma below establishes a core inequality on linear combinations of $\theta_\alpha$'s.

\begin{lm}
	For any family of sets $\{A_1,A_2,\dots, A_{K-1}\}$, and an integer  $T\in \{0, 1,2,\dots, K-2\}$, we have 
	\begin{IEEEeqnarray}{lCl}
		\sum_{j = T}^{K-1} 
		(-1)^{j-T} \binom{K-1-T}{j-T} \theta_j 
		&\leq & 0.
		\label{eq:submod_constraint}
	\end{IEEEeqnarray}
	\label{lm:inc_exc} 
\end{lm}

The following propositions will be used later in the proof of Lemma~\ref{lemma_opt_mu}:
\begin{prop}
	\label{indentity1}
	For given integers $0\leq q \leq p$  and any sequence of real numbers $(\theta_0, \theta_2, \ldots, \theta_p)$, we have
	\begin{IEEEeqnarray}{lCl}
		\sum_{x=0}^{p}\sum_{y=0}^{p} 
		(-1)^{x-y} \binom{p-q}{x-y,\: p-x,\: y-q}  \theta_x 
		&=& \theta_q.
		\label{eq:indentity1}
	\end{IEEEeqnarray}
\end{prop}

\begin{prop}
	\label{prop_induction}
	For given $T, \alpha, \beta \in \mathbb{Z}^+$, we have
	\begin{IEEEeqnarray}{lCl}
		\label{eqn:claim1_V}
		(\alpha-1) \binom{T}{\alpha} - \beta \binom{T-1}{\alpha-1}
		& \geq & - \binom{\beta}{\alpha}.
	\end{IEEEeqnarray}
\end{prop}

For $\alpha \in \{1,2,\ldots,K-1\}$ and $\beta \in \{0,1,\ldots,K-1\}$, we define  $V_{\alpha,\beta}$ as 
\begin{IEEEeqnarray}{lCl}
	V_{\alpha,\beta} &\triangleq & 
	\binom{K-\alpha-1}{K-\beta-1} + 
	\frac{\binom{K-\alpha-1}{S-1}\binom{K-1}{K-\beta-1}}
	{\binom{K-1}{S-1}}
	\left((\alpha-1)-\frac{\alpha \beta}{K-S}\right).
	\label{eqn:V_ab}
\end{IEEEeqnarray}
The following proposition shows that  $V_{\alpha,\beta}$ is non-negative for any choice of $\alpha$ and $\beta$:
\begin{prop}
	\label{claim1_nonNeg}
	For $\alpha \in \{1,2,\ldots,K-1\}$ and $\beta \in \{0,1,\ldots,K-1\}$, we have $V_{\alpha,\beta} \geq 0$.
\end{prop}

Now, we are ready to prove Lemma~\ref{lemma_opt_mu}. 
\begin{IEEEproof}[Proof of Lemma~\ref{lemma_opt_mu}]
	We start from Lemma~\ref{lm:inc_exc}, and write 
	\begin{IEEEeqnarray}{lCl}
		\sum_{i = \beta}^{K-1} 
		(-1)^{i-\beta} \binom{K-\beta-1}{i-\beta} \theta_i 
		&\leq & 0,\quad \beta \in \{0,1,\dots, K-2\}. \nonumber
	\end{IEEEeqnarray}
	Now, fix some $\alpha\in \{1,2,\dots, K-1\}$ and recall from Proposition~\ref{claim1_nonNeg} that $V_{\alpha,\beta} \geq 0$. Multiplying both sides of this inequality by non-negative coefficients $V_{\alpha,\beta}$  and summing over all values of $\beta$ to obtain
	\begin{IEEEeqnarray}{lCl}
		\label{eqn:claim2_0}
		\sum_{\beta = 0}^{K-2} V_{\alpha, \beta}
		\sum_{i = \beta}^{K-1} 
		(-1)^{i-\beta} \binom{K-\beta-1}{i-\beta} \theta_i 
		& \:\leq\: & 
		\sum_{\beta=0}^{K-2} V_{\alpha, \beta} \times 0 = 0.
	\end{IEEEeqnarray}
	Moreover, when $\beta=K-1$, we have 
	\begin{IEEEeqnarray}{lCl}
		\label{eqn:claim2_0:1}
		V_{\alpha, K-1} 
		\sum_{i = K-1}^{K-1} 
		(-1)^{i-(K-1)} \binom{0}{i-(K-1)} \theta_i = V_{\alpha, K-1}  \theta_{K-1}. 
	\end{IEEEeqnarray}
	Summing \eqref{eqn:claim2_0} and \eqref{eqn:claim2_0:1}, we get
	\begin{IEEEeqnarray}{lCl}
		V_{\alpha, K-1}  \theta_{K-1} 
		\nonumber\\
		\hspace{0.25in}
		\geq
		\sum_{\beta = 0}^{K-1} V_{\alpha, \beta}
		\sum_{i = 0}^{K-1} 
		(-1)^{i-\beta} \binom{K-\beta-1}{i-\beta} \theta_i
		\label{eqn:LHS_claim2_3_2}\\
		\hspace{0.25in}
		=
		\sum_{i = 0}^{K-1} 
		\sum_{\beta = 0}^{K-1} 
		(-1)^{i-\beta} \binom{K-\beta-1}{i-\beta} V_{\alpha, \beta} \theta_i
		\nonumber\\
		\hspace{0.25in}
		=
		\sum_{i = 0}^{K-1} 
		\sum_{\beta = 0}^{K-1} 
		(-1)^{i-\beta} \binom{K-\beta-1}{i-\beta}
		\binom{K-\alpha-1}{K-\beta-1} \theta_i
		\:+\: (\alpha-1) \frac{\binom{K-\alpha-1}{S-1}}{\binom{K-1}{S-1}}
		\sum_{i = 0}^{K-1} 
		\sum_{\beta = 0}^{K-1} 
		(-1)^{i-\beta} \binom{K-\beta-1}{i-\beta}
		\binom{K-1}{K-\beta-1} \theta_i	
		\nonumber\\
		\hspace{0.25in}
		\phantom{=}
		- \frac{ \alpha}{K-S} \frac{\binom{K-\alpha-1}{S-1}}{\binom{K-1}{S-1}}
		\sum_{i = 0}^{K-1} 
		\sum_{\beta = 0}^{K-1} 
		(-1)^{i-\beta} \binom{K-\beta-1}{i-\beta}
		\binom{K-1}{K-\beta-1} \beta \theta_i
		\label{eqn:LHS_claim2_3}\\
		\hspace{0.25in}
		=
		\underbrace{\sum_{i = 0}^{K-1} 
			\sum_{\beta = 0}^{K-1} 
			(-1)^{i-\beta} 
			\binom{K-1-\alpha}{i-\beta,  K-1-i, \beta-\alpha} \theta_i}_{\mathsf{Term_1}}
		\:+\: (\alpha-1) \frac{\binom{K-\alpha-1}{S-1}}{\binom{K-1}{S-1}}
		\underbrace{\sum_{i = 0}^{K-1} 
			\sum_{\beta = 0}^{K-1} 
			(-1)^{i-\beta} 
			\binom{K-1}{i-\beta, K-1-i, \beta}\theta_i}_{\mathsf{Term_2}}	\nonumber\\
		\hspace{0.25in}
		\phantom{=}
		- \frac{ \alpha}{K-S} \frac{\binom{K-\alpha-1}{S-1}}{\binom{K-1}{S-1}}
		\underbrace{\sum_{i = 0}^{K-1} 
			\sum_{\beta = 0}^{K-1} 
			(-1)^{i-\beta} \binom{K-\beta-1}{i-\beta}
			\binom{K-1}{\beta} \beta \theta_i}_{\mathsf{Term_3}}.
		\label{eqn:LHS_claim2_3_1}
	\end{IEEEeqnarray}	
 	Note that \eqref{eqn:LHS_claim2_3_2} holds because the binomial term is zero for $i<\beta$; and in \eqref{eqn:LHS_claim2_3} we used the definition of $V_{\alpha, \beta}$ given by~$\eqref{eqn:V_ab}$. 
	Each term in \eqref{eqn:LHS_claim2_3_1} can be simplified  as follows. 
	First, we can use Proposition~\ref{indentity1} for $(p,q,x,y)=(K-1,\alpha, i, \beta)$ to get 
	\begin{IEEEeqnarray}{lCl}
		\mathsf{Term_1} &=&
		\sum_{i = 0}^{K-1} 
		\sum_{\beta = 0}^{K-1} 
		(-1)^{i-\beta} 
		\binom{K-1-\alpha}{i-\beta,\:  K-1-i,\: \beta-\alpha} \theta_i
		=\theta_\alpha.
		\label{eqn:term1_Claim2}
	\end{IEEEeqnarray}
	Similarly, setting $(p,q,x,y)=(K-1,0, i, \beta)$ in Proposition~\ref{indentity1}, we obtain
	\begin{IEEEeqnarray}{lCl}
		\mathsf{Term_2} &=&
		\sum_{i = 0}^{K-1} 
		\sum_{\beta = 0}^{K-1} 
		(-1)^{i-\beta} 
		\binom{K-1}{i-\beta,\: K-1-i,\: \beta}\theta_i
		=
		\theta_0 = 0.
		\label{eqn:term2_Claim2}
	\end{IEEEeqnarray}
	Lastly, for $\mathsf{Term_3}$, we have 
	\begin{IEEEeqnarray}{lCl}
		\mathsf{Term_3}
		&=&
		\sum_{i = 0}^{K-1} 
		\sum_{\beta = 0}^{K-1} 
		(-1)^{i-\beta} \binom{K-\beta-1}{i-\beta} 
		\binom{K-1}{\beta} \beta \theta_i
		\nonumber\\
		&=&
		\sum_{i = 0}^{K-1} 
		\sum_{\beta = 0}^{K-1} 
		(-1)^{i-\beta} \binom{K-\beta-1}{i-\beta} 
		\binom{K-2}{\beta-1} \frac{K-1}{\beta} \beta \theta_i
		\nonumber\\
		&=&
		(K-1) \sum_{i = 0}^{K-1} 
		\sum_{\beta = 0}^{K-1} 
		(-1)^{i-\beta} \binom{K-2}{i-\beta,\: K-1-i,\: \beta-1}  \theta_i
		\nonumber\\
		&=&
		(K-1)\theta_1,
		\label{eqn:term3_Claim2_4}
	\end{IEEEeqnarray}
	where \eqref{eqn:term3_Claim2_4} follows from Proposition~\ref{indentity1} by setting $(p,q,x,y)=(K-1,1, i, \beta)$. 
	Plugging \eqref{eqn:term1_Claim2}, \eqref{eqn:term2_Claim2} and \eqref{eqn:term3_Claim2_4} into \eqref{eqn:LHS_claim2_3_1}, we obtain 
	\begin{IEEEeqnarray}{l}		
		\theta_\alpha  			
		- \frac{ \alpha}{K-S} \frac{\binom{K-\alpha-1}{S-1}}{\binom{K-1}{S-1}}
		(K-1)\theta_1 \:\leq\: V_{\alpha, K-1}  \theta_{K-1}, \nonumber
	\end{IEEEeqnarray}
	or equivalently, 
	\begin{IEEEeqnarray}{lCl}		
		\frac{1}{\binom{K-1}{\alpha}} \sum_{\substack{\cJ\subseteq [K-1] \\ |\cJ|=\alpha}} \left|\bigcup_{j\in \cJ} A_j\right|
		&\:\leq\:&  
		\frac{ \alpha (K-1)}{K-S} \frac{\binom{K-\alpha-1}{S-1}}{\binom{K-1}{S-1}} \frac{1}{\binom{K-1}{1}} \sum_{j\in [K-1]} |A_j| 
		+ V_{\alpha, K-1}  \frac{1}{\binom{K-1}{K-1}} \left|\bigcup_{j\in [K-1]} A_j\right|.
		\label{eqn:theta_bound}
	\end{IEEEeqnarray}	
	
	It should be noted that \eqref{eqn:theta_bound} holds for any choice of $\{A_1,A_2,\ldots, A_{K}\}$. In particular, applying \eqref{eqn:theta_bound} to the family of sets $\left\{\fu{i}{j}: j\in[K] \setminus\{i\}\right\}$ for a fixed $i$, we get  
	\begin{IEEEeqnarray}{l}	
		\frac{1}{\binom{K-1}{\alpha}} \sum_{\substack{\cJ\subseteq [K]\setminus \{i\} \\ |\cJ|=\alpha}} \left|\bigcup_{j\in \cJ} \fu{i}{j}\right| 
		\:\leq\:  
		\frac{ \alpha(K-1)}{K-S} \frac{\binom{K-\alpha-1}{S-1}}{\binom{K-1}{S-1}}
		\frac{1}{\binom{K-1}{1}} \sum_{j\in [K]\setminus\{i\}} \left| \fu{i}{j} \right| 
		+ V_{\alpha, K-1}  \frac{1}{\binom{K-1}{K-1}} \left|\bigcup\nolimits_{j\in [K]\setminus\{i\}} \fu{i}{j}\right|,
		\label{eqn:theta_bound_of_F}
	\end{IEEEeqnarray}
	for $i \in \{1,2,\ldots, K\}$. Averaging \eqref{eqn:theta_bound_of_F} over $i \in \{1,2,\ldots, K\}$, we obtain 
	\begin{IEEEeqnarray}{l}	
		\frac{1}{K\binom{K-1}{\alpha}} \sum_{i\in [K]}\sum_{\substack{\cJ\subseteq [K]\setminus \{i\} \\ |\cJ|=\alpha}} \left|\bigcup_{j\in \cJ} \fu{i}{j}\right| 
		\:\leq\:
		\frac{ \alpha(K-1)}{K-S} \frac{\binom{K-\alpha-1}{S-1}}{\binom{K-1}{S-1}}
		\frac{1}{K\binom{K-1}{1}} \sum_{i\in [K]}\sum_{j\in [K]\setminus\{i\}} \!\!\left| \fu{i}{j} \right| 
		+ V_{\alpha, K-1} \frac{1}{K\binom{K-1}{K-1}} \sum_{i\in [K]}\left|\bigcup_{j\in [K]\setminus\{i\}} \!\!\!\!\fu{i}{j}\right|,\nonumber
	\end{IEEEeqnarray}
	or equivalently, 
	\begin{IEEEeqnarray}{lCl}	
		\mu_\alpha \leq  \frac{ \alpha(K-1)}{K-S} \frac{\binom{K-\alpha-1}{S-1}}{\binom{K-1}{S-1}}
		\mu_1 + V_{\alpha, K-1} \mu_{K-1}.
		\label{eq:UB_mu_1,K-1}
	\end{IEEEeqnarray}
	Hence, it remains to upper bound $\mu_1$ and $\mu_{K-1}$. Recall from definition of $\mu_i$'s in \eqref{eqn:mu_eta} that
	\begin{IEEEeqnarray}{lCl}
		\mu_{1} &=&
		\frac{1}{K (K-1)} \sum_{i \in [K]} \sum_{j\in [K]\setminus \{i\}}
		\left|  \tF^i_j \right| 
		\nonumber\\
		&=& 
		\frac{1}{K (K-1)} \sum_{j \in [K]} \sum_{i\in [K]\setminus\{j\}}	\left|  \tF^i_j \right| \nonumber\\
		&=& \frac{1}{K (K-1)} \sum_{j \in [K]} \left|\cZ_j\setminus \proc{j}\right| 
		\nonumber\\
		&\leq&  \frac{1}{K (K-1)} \sum_{j \in [K]} (S-1) = \frac{S-1}{K-1}, 
	\end{IEEEeqnarray}
	and
	\begin{IEEEeqnarray}{lCl}
		\label{eq:mu_1}
		\mu_{K-1} &=& 
		\frac{1}{K} \sum_{i \in [K]} \left| \bigcup\nolimits_{j\in [K]\setminus \{i\}} \tF^i_j \right| 
		\nonumber\\
		&\leq& 
		\frac{1}{K} \sum_{i \in [K]} \min\left\{\sum_{j\in [K]\setminus \{i\}}|\tF^i_j |,\: |F^i|\right\}  \nonumber\\
		&=&
		\frac{1}{K}\min \left\{\frac{1}{K} \sum_{j\in [K]}\sum_{i\in [K]\setminus \{j\} } |\tF^i_j |,\: \frac{1}{K}\sum_{i\in [K]} 1\right\}
		\nonumber\\
		&=& \min\{S-1, 1\} \label{eq:mu_K-1}. 
	\end{IEEEeqnarray}
	Plugging \eqref{eq:mu_1} and \eqref{eq:mu_K-1} into \eqref{eq:UB_mu_1,K-1}, we get
	\begin{IEEEeqnarray}{lCl}
		\mu_\alpha 
		& \leq & 
		\frac{ \alpha(K-1)}{K-S} \frac{\binom{K-\alpha-1}{S-1}}{\binom{K-1}{S-1}}
		\frac{S-1}{K-1} + V_{\alpha, K-1} \cdot  1 \nonumber\\
		& = & 
		\frac{\binom{K-\alpha-1}{S-1}}{\binom{K-1}{S-1}}
		\frac{S-1}{K-S}\alpha 
		+ \left(1 + \frac{\binom{K-\alpha-1}{S-1}}{\binom{K-1}{S-1}} \left( (\alpha-1 ) - \frac{\alpha(K-1)}{K-S}\right)\right) 	\nonumber\\
		&=&
		1 + 	\frac{\binom{K-\alpha-1}{S-1}}{\binom{K-1}{S-1}} \left( \frac{S-1}{K-S} \alpha - \frac{K-1}{K-S}\alpha + \alpha-1\right)\nonumber\\
		&=& 
		1 - 	\frac{\binom{K-\alpha-1}{S-1}}{\binom{K-1}{S-1}}. 
	\end{IEEEeqnarray}
	This completes the proof of Lemma~\ref{lemma_opt_mu}.
\end{IEEEproof}

It remains to prove Lemma~\ref{lm:inc_exc} and Propositions \ref{indentity1}, \ref{prop_induction}, and \ref{claim1_nonNeg}, whose proofs are presented as follows.

\begin{IEEEproof}[Proof of Lemma~\ref{lm:inc_exc}]
	Fix a $\cT\subseteq [K-1]$ be a set of indices of size $|\cT|=T$. Define  $\cX \triangleq A_\cT = \bigcup_{t\in T} A_t$. Clearly, we have $\left(\cX \cap \bigcap_{\ell\in \cT\c} A_\ell \right) \subseteq \left(\bigcap_{\ell\in \cT\c} A_\ell\right)$, which implies 
	\begin{IEEEeqnarray}{lCl}
		\left|  \cX \cap  \bigcap_{\ell\in \cT\c} A_\ell  \right| - \left|\bigcap_{\ell\in \cT\c} A_\ell\right| \leq 0.
		\label{eq:set-diff}
	\end{IEEEeqnarray}
	Each term in the LHS of \eqref{eq:set-diff} can be expanded using the inclusion-exclusion principle as follows:
	\begin{IEEEeqnarray}{lCl}
		\left|\bigcap_{\ell\in \cT\c} A_\ell\right| &=& \sum_{\substack{\cU\subset \cT\c \\ |\cU|=1}} |A_{\cU}| - \sum_{\substack{\cU\subset \cT\c \\ |\cU|=2}} |A_{\cU}| + \sum_{\substack{\cU\subset \cT\c \\ |\cU|=3}} |A_{\cU}| - \cdots 
		+ (-1)^{|\cT\c|-1} \sum_{\substack{\cU\subset \cT\c \\ |\cU|=|\cT\c|}} |A_{\cU}|.
		\label{eq:set-diff-1}
	\end{IEEEeqnarray}
	In a similar way, we have 
	\begin{IEEEeqnarray}{lCl}
		\left| \cX \cap  \bigcap_{\ell\in \cT\c} A_\ell \right|  
		&=&
		\left[|\cX| + \sum_{\substack{\cU\subset \cT\c \\ |\cU|=1}} |A_{\cU}| \right] - \left[ \sum_{\substack{\cU\subset \cT\c \\ |\cU|=1}} |\cX \cup A_\cU| + \sum_{\substack{\cU\subset \cT\c \\ |\cU|=2}} |A_{\cU}| \right] + \left[\sum_{\substack{\cU\subset \cT\c \\ |\cU|=2}} |\cX \cup A_\cU| + \sum_{\substack{\cU\subset \cT\c \\ |\cU|=3}} |A_{\cU}|\right] 
		\nonumber\\
		& \phantom{=}  &
		- \cdots 
		+ (-1)^{|\cT\c|-1} \left[\sum_{\substack{\cU\subset \cT\c \\ |\cU|=|\cT\c|-1}} |\cX \cup A_\cU| + \sum_{\substack{\cU\subset \cT\c \\ |\cU|=|\cT\c|}}  |A_{\cU}|\right] 	
		+ (-1)^{|\cT\c|} \sum_{\substack{\cU\subset \cT\c \\ |\cU|=|\cT\c|}} \left|\cX \cup A_\cU\right|.
		\label{eq:set-diff-2}
	\end{IEEEeqnarray}	

	Substituting \eqref{eq:set-diff-1} and \eqref{eq:set-diff-2} in \eqref{eq:set-diff}, we get 
	\begin{IEEEeqnarray}{lCl}
		0 &\geq&  \left|  \cX \cap  \bigcap_{\ell\in \cT\c} A_\ell  \right| -\left|\bigcap_{\ell\in \cT\c} A_\ell\right|
		\nonumber\\
		& = &
		|\cX| -  \sum_{\substack{\cU\subset \cT\c \\ |\cU|=1}} |\cX \cup A_\cU|  
		+  \sum_{\substack{\cU\subset \cT\c \\ |\cU|=2}} |\cX \cup A_\cU|  - \cdots 	
		+  (-1)^{|\cT\c|} \sum_{\substack{\cU\subset \cT\c \\ |\cU|=|\cT^c|}} |\cX \cup A_\cU|  
		\nonumber\\
		& = &
		|A_\cT| -  \sum_{\substack{\cU\subset \cT\c \\ |\cU|=1}} |A_\cT \cup A_\cU|  
		+  \sum_{\substack{\cU\subset \cT\c \\ |\cU|=2}} |A_\cT \cup A_\cU|  - \cdots 
		+  (-1)^{|\cT\c|} \sum_{\substack{\cU\subset \cT\c \\ |\cU|=|\cT^c|}} |A_\cT \cup A_\cU|.
		\label{eq:set-diff-3}
	\end{IEEEeqnarray}
	Note that \eqref{eq:set-diff-3} holds for  any subset of indices $\cT$ of size $|\cT|=T$. Averaging this inequality over all choices of $\cT \subseteq [K-1]$ with $|\cT|=T$, we obtain
	\begin{align}
		0 &\geq 
		\frac{1}{\binom{K-1}{T}} \sum_{\substack{ \cT \subseteq [K-1] \\ |\cT| =T }}  \left[
		|A_\cT| -  \sum_{\substack{\cU\subset \cT\c \\ |\cU|=1}} |A_\cT \cup A_\cU|  
		+  \sum_{\substack{\cU\subset \cT\c \\ |\cU|=2}} |A_\cT \cup A_\cU|  - \cdots 
		+  (-1)^{|\cT\c|} \sum_{\substack{\cU\subset \cT\c \\ |\cU|=|\cT^c|}} |A_\cT \cup A_\cU|
		\right]
		\nonumber\\
		& = 
		\frac{1}{\binom{K-1}{T}} \sum_{\substack{ \cT \subseteq [K-1] \\ |\cT| =T }}  \left|A_{\cT } \right|
		- \frac{1}{\binom{K-1}{T}} \sum_{\substack{ \cT \subseteq [K-1] \\ |\cT| =T }} \: \sum_{\substack{\cU\subset \cT\c \\ |\cU|=1}} |A_\cT \cup A_\cU|    
		+  \frac{1}{\binom{K-1}{T}} \sum_{\substack{ \cT \subseteq [K-1] \\ |\cT| =T }} \: \sum_{\substack{\cU\subset \cT\c \\ |\cU|=2}} |A_\cT \cup A_\cU| - \cdots 	
		\nonumber\\
		& \phantom{=}
		+ (-1)^{|\cT\c|} \frac{1}{\binom{K-1}{T}} \sum_{\substack{ \cT \subseteq [K-1] \\ |\cT| =T }}  \sum_{\substack{\cU\subset \cT\c \\ |\cU|=|\cT\c|}} |A_\cT \cup A_\cU|   
		\nonumber\\
		& =
		\frac{1}{\binom{K-1}{T}} \sum_{\substack{ \cV \subseteq [K-1] \\ |\cV| =T }}  \left|A_{\cV } \right|
		- \frac{\binom{T+1}{T}}{\binom{K-1}{T}} \sum_{\substack{ \cV \subseteq 	[K-1] \\ |\cV| =T+1 }}  |A_\cV |    
		+  \frac{\binom{T+2}{T}}{\binom{K-1}{T}} \sum_{\substack{ \cV \subseteq [K-1] \\ |\cV| =T+2 }} |A_\cV |  
		- \cdots
		+ (-1)^{|\cT\c|} \frac{\binom{T+|\cT\c|}{T}}{\binom{K-1}{T}} \sum_{\substack{ \cV \subseteq [K-1] \\ |\cV| =T + |\cT\c| }}  |A_\cV |   
		\label{eq:long-1_1}
		\\
		& =
		\frac{1}{\binom{K-1}{T}} \sum_{\substack{ \cV \subseteq [K-1] \\ |\cV| =T }}  \left|A_{\cV } \right|
		- \frac{\binom{K-1-T}{1}}{\binom{K-1}{T+1}} \sum_{\substack{ \cV \subseteq 	[K-1] \\ |\cV| =T+1 }}  |A_\cV |    
		+  \frac{\binom{K-1-T}{2}}{\binom{K-1}{T+2}} \sum_{\substack{ \cV \subseteq [K-1] \\ |\cV| =T+2 }} |A_\cV |  
		- \cdots
		+ (-1)^{|\cT\c|} \frac{\binom{K-1-T}{|\cT\c|}}{\binom{K-1}{T+|\cT\c|}} \sum_{\substack{ \cV \subseteq [K-1] \\ |\cV| =T + |\cT\c| }}  |A_\cV |   
		\label{eq:long-1_2}
		\\
		& = 
		\theta_T
		- \binom{K-1-T}{1} \theta_{T+1}
		+ \binom{K-1-T}{2} \theta_{T+2}
		- \cdots  
		+ (-1)^{|\cT\c|} \binom{K-1-T}{|\cT\c|} \theta_{T + |\cT\c| }  
		\nonumber\\
		& =
		\sum_{j=T}^{K-1}  (-1)^{j-T} \binom{K-1-T}{j-T} \theta_{j}.
		\label{eq:long-1}
	\end{align}
	Note that in \eqref{eq:long-1_1} we have replaced $A_{\cT} \cup A_{\cU}$ by $A_{\cV} = A_{\cT \cup \cU}$; in \eqref{eq:long-1_2} equality holds since $\binom{x}{y}  \binom{y}{z}  =  \binom{x}{z} \binom{x-z}{y-z}$ for any triple of integers $(x,y,z)$; and in \eqref{eq:long-1}  we have $|\cT\c| = K-1-T$. 
	Hence, \eqref{eq:submod_constraint} readily follows for all values of $T \in \{0,1,\dots, K-2\}$.
	This completes the proof of Lemma~\ref{lm:inc_exc}.
\end{IEEEproof}

\begin{IEEEproof}[Proof of Proposition~\ref{indentity1}]
	\begin{IEEEeqnarray}{lCl}
		\sum_{x=0}^{p}  \sum_{y=0}^{p} 
		(-1)^{x-y} \binom{p-q}{x-y,\: p-x,\: y-q} \theta_x
		&=&
		\sum_{x=0}^{p}  \sum_{y=q}^{x} 
		(-1)^{x-y} \binom{p-q}{p-x} \binom{x-q}{x-y} \theta_x \label{eq:a_eq_y_0}\\
		&=&
		\sum_{x=0}^{p} \left[\sum_{z=0}^{x-q} (-1)^{z}  \binom{x-q}{z}\right] \binom{p-q}{p-x} \theta_x 
		\label{eq:a_eq_y_1}\\ 
		&=&
		\sum_{x=0}^{p} (1-1)^{x-q} \binom{p-q}{p-x} \theta_x 
		\\ 
		&=&
		\binom{p-q}{p-q} \theta_q = \theta_q, \label{eq:a_eq_y}
	\end{IEEEeqnarray}
	where in \eqref{eq:a_eq_y_0} we have used the fact that the multinomial term is zero whenever $y<q$ or $y>x$; in \eqref{eq:a_eq_y_1} we substitute $x-y$ by $z$ that takes values in $\{0,1,\dots, x-q\}$; and 
	\eqref{eq:a_eq_y} holds since $(1-1)^{x-q}$ is zero except for $x=q$. 
	This complete the proof of Proposition~\ref{indentity1}.
\end{IEEEproof}

\begin{IEEEproof}[Proof of Proposition~\ref{prop_induction}]
	Let us define 
	\begin{align*}
	    f(T) = (\alpha-1) \binom{T}{\alpha} - \beta \binom{T-1}{\alpha-1},
	\end{align*}
	 for fixed $\alpha$ and $\beta$. We prove that $f(T)\geq -\binom{\beta}{\alpha}$ by a two-sided induction on $T$, i.e., for $T \geq \beta$ and $T \leq \beta$. First, note that for $T=\beta$, we have
	\begin{IEEEeqnarray}{lCl}
		f(\beta) &=& (\alpha-1) \binom{\beta}{\alpha} - \beta \binom{\beta-1}{\alpha-1} 
		\nonumber\\
		&=& (\alpha-1) \binom{\beta}{\alpha} - \alpha \binom{\beta}{\alpha} =  - \binom{\beta}{\alpha},
		\label{eqn:app:prop2_inductionBasis}
	\end{IEEEeqnarray}
	which shows the inequality in \eqref{eqn:claim1_V} holds with equality. Let \eqref{eqn:app:prop2_inductionBasis} be the base case for induction. Next, we consider two individual cases for $T \geq \beta$ and $T \leq \beta$.
	\paragraph{$T \geq \beta$}
	Assume that \eqref{eqn:claim1_V}  holds for $T = t\geq \beta$, i.e, $f(t)\geq -\binom{\beta}{\alpha}$. In what follows, we prove that \eqref{eqn:claim1_V} holds for~$T = t+1$:
	\begin{IEEEeqnarray}{lCl}
		f(t+1) 
		&=& 
		(\alpha-1) \binom{t+1}{\alpha} - \beta \binom{t}{\alpha-1}
		\nonumber\\
		&=& 
		(\alpha-1) \left[ \binom{t}{\alpha} +  \binom{t}{\alpha-1} \right]
		- \beta \left[\binom{t-1}{\alpha-1} +  \binom{t-1}{\alpha-2}\right]
		\nonumber\\
		&=& 
		\left[(\alpha-1) \binom{t}{\alpha} - \beta \binom{t-1}{\alpha-1}\right]
		+ \left[(\alpha-1) \binom{t}{\alpha-1} - \beta \binom{t-1}{\alpha-2}\right]\nonumber\\
		&=& 
		\left[(\alpha-1) \binom{t}{\alpha} - \beta \binom{t-1}{\alpha-1}\right]
		+ \left[t \binom{t-1}{\alpha-2} - \beta \binom{t-1}{\alpha-2}\right]
		\nonumber \\
		&=& f(t) + (t - \beta) \binom{t-2}{\alpha-2}
		\geq 
		- \binom{\beta}{\alpha},
		\label{eqn:app:induction_part1}
	\end{IEEEeqnarray}
	where the inequality in \eqref{eqn:app:induction_part1} holds due to $f(t) \geq -\binom{\beta}{\alpha}$ by the induction hypothesis, and the fact that $t\geq \beta$. 
	
	\paragraph{$T \leq \beta$}
	Similar to the previous case, assume that \eqref{eqn:claim1_V}  holds for $T = t \leq \beta$, i.e., $f(t) \geq -\binom{\beta}{\alpha}$. For $T = t-1$, we can write
	\begin{IEEEeqnarray}{lCl}
		f(t-1) 
		&=& (\alpha-1) \binom{t-1}{\alpha} - \beta \binom{t-2}{\alpha-1}
		\nonumber\\
		&=& 
		(\alpha-1) \left[\binom{t}{\alpha} -\binom{t-1}{\alpha-1}\right] - \beta \left[ \binom{t-1}{\alpha-1} 
		- \binom{t-2}{\alpha-2}\right]
		\nonumber\\
		&=&
		\left[(\alpha-1) \binom{t}{\alpha} - \beta \binom{t-1}{\alpha-1}\right]
		+ \left[ \beta \binom{t-2}{\alpha-2} - (\alpha-1) \binom{t-1}{\alpha-1}  \right] 
		\nonumber\\
		&=&
		\left[(\alpha-1) \binom{t}{\alpha} - \beta \binom{t-1}{\alpha-1}\right]
		+ \left[\beta \binom{t-2}{\alpha-2} - (t-1) \binom{t-2}{\alpha-2} \right]
		\nonumber\\
		&=&
		f(t) + (\beta - (t-1)) \binom{t-2}{\alpha-2}	
		\geq 
		- \binom{\beta}{\alpha},
		\label{eqn:app:induction_part2}
	\end{IEEEeqnarray}
	where the inequality in \eqref{eqn:app:induction_part2} holds since $f(t) \geq -\binom{\beta}{\alpha}$ by the induction hypothesis, and $\beta \geq t-1$. This completes the proof of Proposition~\ref{prop_induction} for any arbitrary $T$.
\end{IEEEproof}

\begin{IEEEproof}[Proof of Proposition~\ref{claim1_nonNeg}]
	In order to prove the inequality, we can rewrite  $V_{\alpha,\beta}$ defined in \eqref{eqn:V_ab}  as
	\begin{IEEEeqnarray}{lCl}
		V_{\alpha,\beta}
		&=&
		\binom{K-\alpha-1}{K-\beta-1}+ 
		\frac{\binom{K-\alpha-1}{S-1}\binom{K-1}{K-\beta-1}}
		{\binom{K-1}{S-1}}
		\left((\alpha-1) - \frac{\alpha \beta}{K-S}\right) 
		\nonumber\\
		&=&
		\binom{K-\alpha-1}{K-\beta-1} 
		+ 
		\frac{(K-\alpha-1)! \, (K-S)!}{(K-\alpha-S)! \, (K-\beta-1)! \, \beta!}
		\cdot 
		\frac{\alpha! \, (\beta - \alpha)!}{\alpha! \, (\beta - \alpha)!} 
		\left((\alpha-1) - \frac{\alpha \beta}{K-S}\right)
		\nonumber\\
		&=&
		\binom{K-\alpha-1}{K-\beta-1}+ 
		\frac{\binom{K-\alpha-1}{K-\beta-1}\binom{K-S}{\alpha}}
		{\binom{\beta}{\alpha}}
		\left((\alpha-1)-\frac{\alpha \beta}{K-S}\right)
		\nonumber\\
		&=&
		\binom{K-\alpha-1}{K-\beta-1}
		\left(
		1+ \frac{\binom{K-S}{\alpha}}{\binom{\beta}{\alpha}}
		\left((\alpha-1)-\frac{\alpha \beta}{K-S}\right)
		\right)
		\nonumber\\
		&=&
		\displaystyle
		\frac{\binom{K-\alpha-1}{K-\beta-1}}{\binom{\beta}{\alpha}}
		\left(\binom{\beta}{\alpha} + (\alpha-1) \binom{K-S}{\alpha} - \frac{\alpha \beta}{K-S} \binom{K-S}{\alpha}\right)
		\nonumber\\
		&=&
		\frac{\binom{K-\alpha-1}{K-\beta-1}}{\binom{\beta}{\alpha}}
		\left( \binom{\beta}{\alpha} + 
		(\alpha-1) \binom{K-S}{\alpha} - \beta \binom{K-S-1}{\alpha-1}
		\right) 
		\nonumber\\
		&\geq& 0,
		\label{eqn:app:prop3_ineq}
	\end{IEEEeqnarray}
	where the inequality in \eqref{eqn:app:prop3_ineq} holds due to the claim of Proposition~\ref{prop_induction} for $T=K-S$, that is
	\begin{IEEEeqnarray*}{l}
	f(T-S) = (\alpha-1) \binom{K-S}{\alpha} - \beta \binom{K-S-1}{\alpha-1} \geq -\binom{\beta}{\alpha}.
	\end{IEEEeqnarray*}
	This completes the proof of Proposition~\ref{claim1_nonNeg}.
\end{IEEEproof}

\section{Decomposition of File Transition Graph: Proof of Lemma~\ref{lm:demposition}}
\label{app:lm:prfct_match} 
Given a file transition graph $\mathcal{G} (V,E)$, let us construct an undirected bipartite graph $\mathcal{H} (V',E')$ (or more accurately bipartite multigraph, since there might be multiple edges between two nodes) by distinguishing worker nodes at iterations $t$ and $t+1$. More precisely, 
the vertex set $V'$ is partitioned into two disjoint subsets $\cW^{(t)}= \bigl\{W_i^{(t)}: i\in [K]\bigr\}$ and $\cW^{(t+1)} = \bigl\{W_i^{(t+1)}: i\in [K]\bigr\}$. For every file $F^j$ with $j\in u(i) \cap d(\ell)$, we include one edge $e_j$ between $W_i^{(t)}$ and $W_\ell^{(t+1)}$. Note that if there is more than one file processed by $W_i$ at iteration $t$ and assigned to $W_\ell$ at iteration $t+1$, then there are multiple edges between the two vertices $W_i^{(t)}$ and $W_\ell^{(t+1)}$ in the graph, one for each file.
From the aforementioned description of $\mathcal{H} (V',E')$, it is evident that $\mathcal{H} (V', E')$ is an $N/K$-regular graph bipartite multigraph. It is easy to see that the original file transition graph $\mathcal{G} (V,E)$~can be recovered from $\mathcal{H} (V',E')$ by collapsing nodes $W_i^{(t)}$ and $W_i^{(t+1)}$ for each $i\in [K]$. More precisely, each edge between $W_i^{(t)}$~and $W_\ell^{(t+1)}$ in $\mathcal{H} (V', E')$ is corresponding to a directed edge from $W_i$ to $W_\ell$ in $\mathcal{G} (V, E)$. This shows a one-to-one mapping between the directed graph $\mathcal{G} (V, E)$ and the undirected bipartite graph $\mathcal{H} (V', E')$. 

Our proposed decomposition for the graph $\cG(V,E)$ is based on the decomposition of the bipartite graph $\cH(V',E')$ into perfect matchings. Assume $\cH(V',E')$ can be decomposed into $N/K$ perfect matchings between $\cW^{(t)}$ and $\cW^{(t+1)}$, namely $\cH_j(V',E'_j)$, where $j \in [N/K]$ and $|E'_j|=K$, and the degree of each node in $\cH_j(V',E'_j)$ is exactly one. Then, by collapsing nodes $W_i^{(t)}$ and $W_i^{(t+1)}$, for $i \in [K]$, in $\cH_j(V',E'_j)$, we get a directed graph $\cG_j(V,E_j)$ over $V=\{W_1,W_2,\ldots, W_K\}$ where the in-degree and out-degree of each node $W_i$ in $\cG_j(V,E_j)$ are equal to those of $W_i^{(t)}$ and $W_i^{(t+1)}$ in $\cH_j(V',E'_j)$, respectively, which are both equal to $1$. 

The fact that the bipartite graph $\cH(V',E')$  can be decomposed into $N/K$ perfect matchings follows 
from a recursive application of Hall's theorem.
We use Hall's theorem to find perfect matchings in $\cH(V',E')$ as follows. 

\begin{theorem}[Theorem~1 in \cite{hall1935representatives}]
	\label{thm:prfctMatch}
	A bipartite graph $\cH(V',E')$, with vertex set $V'= \cW^{(t)} \cup \cW^{(t+1)}$, contains a complete matching
	from $\cW^{(t)}$ to $\cW^{(t+1)}$  if and only if 
	\begin{IEEEeqnarray}{lCl}
		\vert \mathcal{N} (T) \vert &\geq& \vert T \vert,
		\label{eqn:HallCond}
	\end{IEEEeqnarray}
	for every non-empty subset $T\subseteq \cW^{(t)}$. 
\end{theorem}

In order to apply Hall's theorem to $\cH(V',E')$, we need to test the necessary and sufficient condition. Let 
$T$ be an arbitrary set of vertices in $\cW^{(t)}$, i.e., $T \subset \cW^{(t)}$, and $\mathcal{N} (T)$ denote the set of neighbors of $T$ in $\cW^{(t+1)}$. 
Recall that the degree of each vertex in $T$ is $N/K$. Hence, there are a total of $\vert T \vert N/K$ edges exit $T$ to reach nodes in $\cN(T)$. However, note that the total number of edges incoming to $\cN(T)$ cannot exceed $\vert \cN(T) \vert N/K$, because each node in $\cN(T)$ has degree $N/K$. The set of edges outgoing $T$ is a subset of edges incoming to $\cN(T)$. Therefore, we have $\vert T \vert N/K \leq \vert \cN(T) \vert N/K$, which implies $\vert T \vert \leq \vert \cN(T) \vert$. Thus, the condition of Hall's theorem holds and there exists one perfect matching between the nodes in  $\cW^{(t)}$ and $\cW^{(t+1)}$. Including the set of edges that constitute such a perfect matching  in the set $E'_{N/K}$, we obtain the subgraph $\cH_{N/K}(V',E'_{N/K})$. 

Next, one can remove the edges in $E'_{N/K}$ from $E'$. This reduces the degree of each node by $1$, resulting in a new regular bipartite graph, with degree $N/K-1$. Repeating the same argument on each residual graph, we can find a perfect matching, and then we remove its edges from the bipartite graph $\cH(V',E')$ until no edge is left. This provides us with $N/K$ perfect matchings in $\cH (V',E')$, and each perfect matching corresponds to a subgraph of the file transition graph $\cG(V,E)$. This completes the proof of Lemma~\ref{lm:demposition}. \hfill $\blacksquare$

\begin{remark}
	\label{rmrk_bipartite}
	The problem of finding a perfect matching in an $N/K$-regular bipartite graph $\mathcal{H}(V',E')$, with $|V'|=2K$ vertices and $|E'| = K \times N/K=N$ edges, is well-studied in the graph theory literature.
	For regular bipartite graphs, a perfect matching is known to be computable in $O(N)$ time \cite{cole2001edge}.
	Recently, the authors in \cite{goel2013perfect} have proposed a randomized algorithm that finds a perfect matching in an $N/K$-regular  bipartite graph. The resulting runtime is $O(K \log K)$ (both in expectation and with high probability).
	We refer the interested reader to the aforementioned references for further details.
\end{remark}

\begin{figure}
	\centering
	\subfloat[]{\includegraphics[height=2.12in]{ex5_fileTransGraph.pdf}%
	\label{fig:ex6_fileTransGraph}}
	\hspace{4cm}
	\subfloat[]{\includegraphics[height=2.12in]{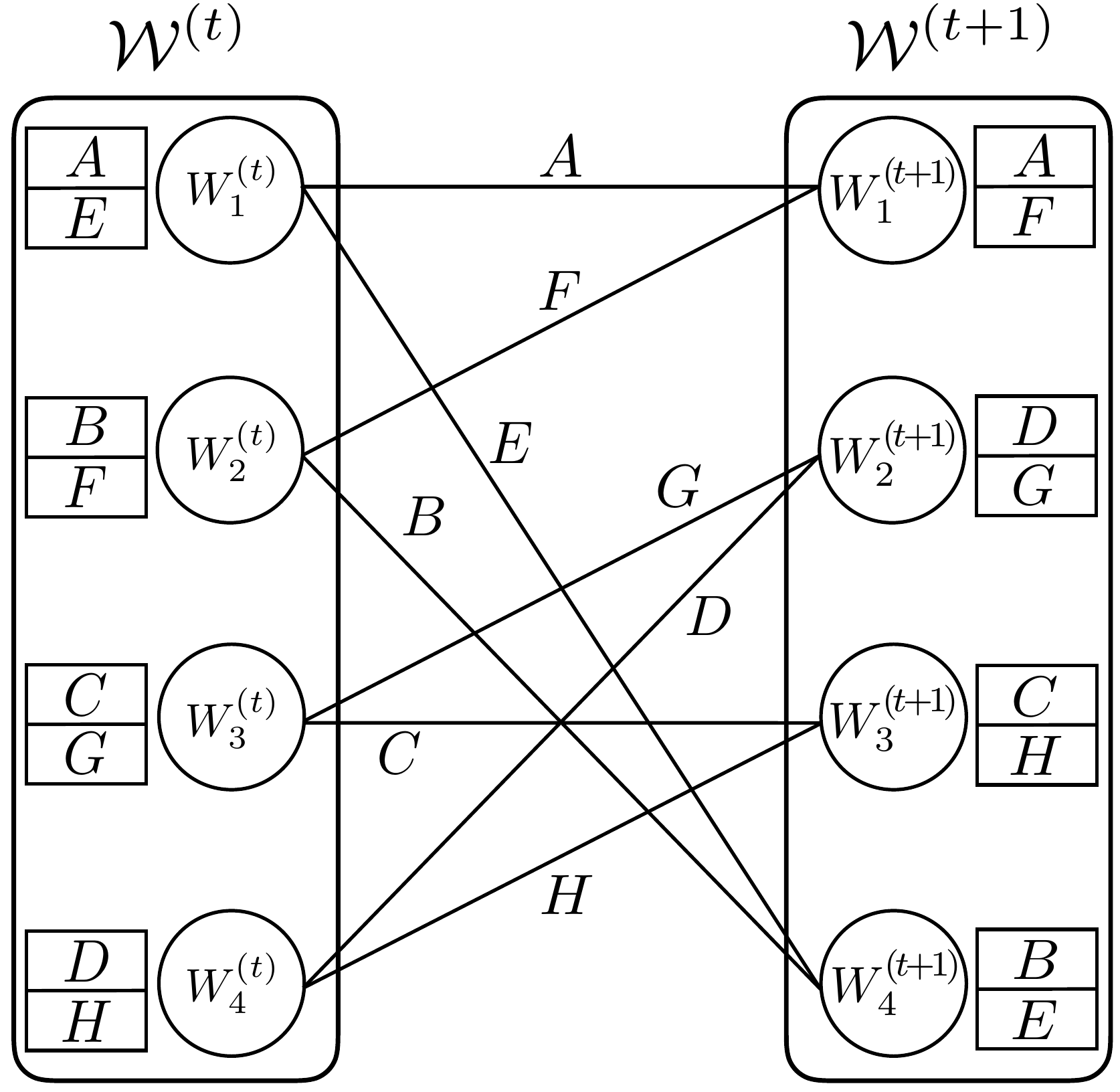}%
	\label{fig:ex6_bipartiteGraph}}
	\caption{
		(a) The file transition graph $\cG(V,E)$ for the data shuffling system of Example~5.  
		(b) The corresponding bipartite graph $\cH(V',E')$. 
	}
	\label{fig:ex6_fTG_bipartiteG}
\end{figure}

\begin{figure}
	\centering
	\subfloat[]{\includegraphics[height=2.35in]{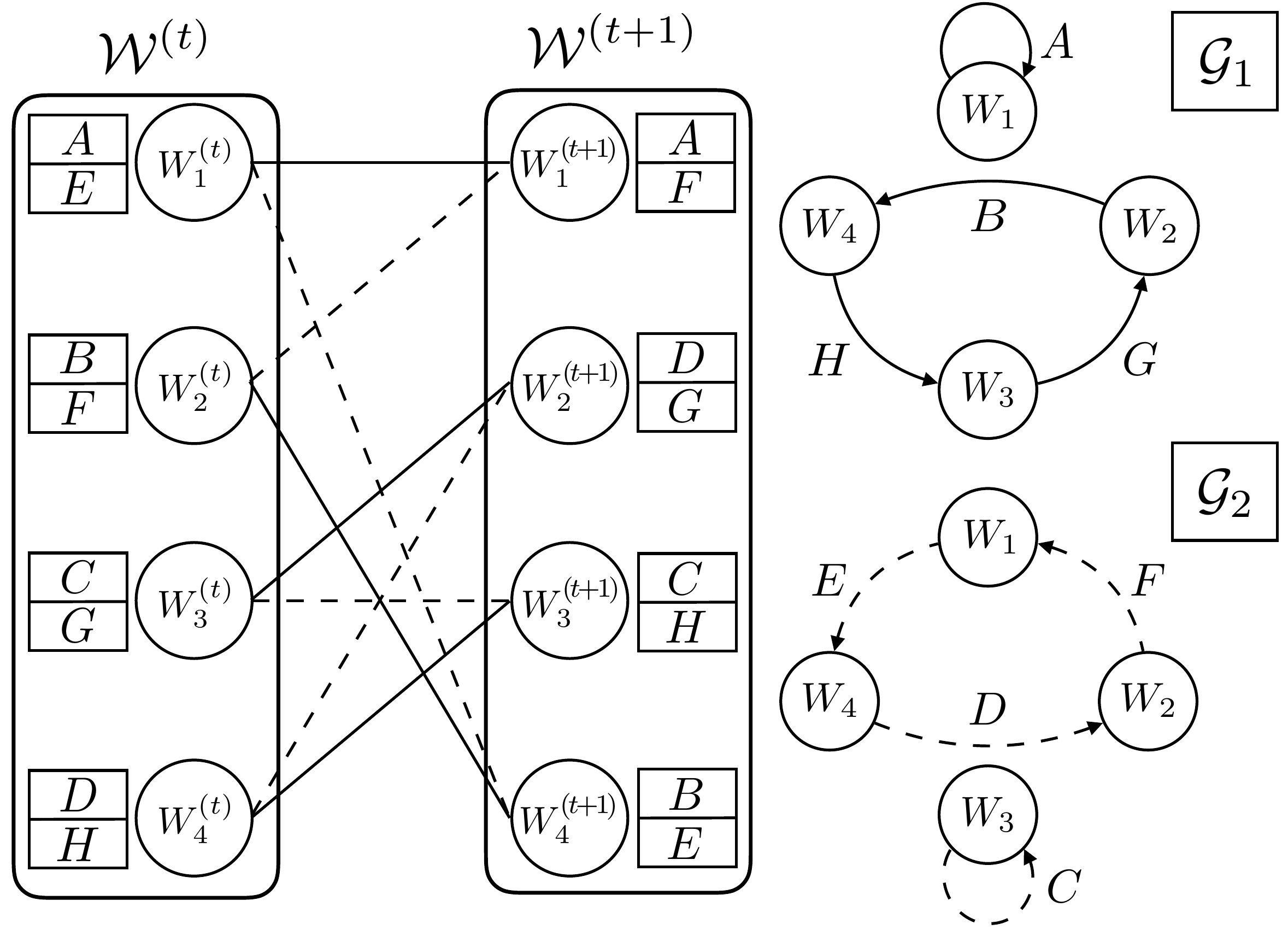}%
	\label{fig:ex6_decomp1_bipartiteGraph}}
	\hspace{12mm}
	\subfloat[]{\includegraphics[height=2.35in]{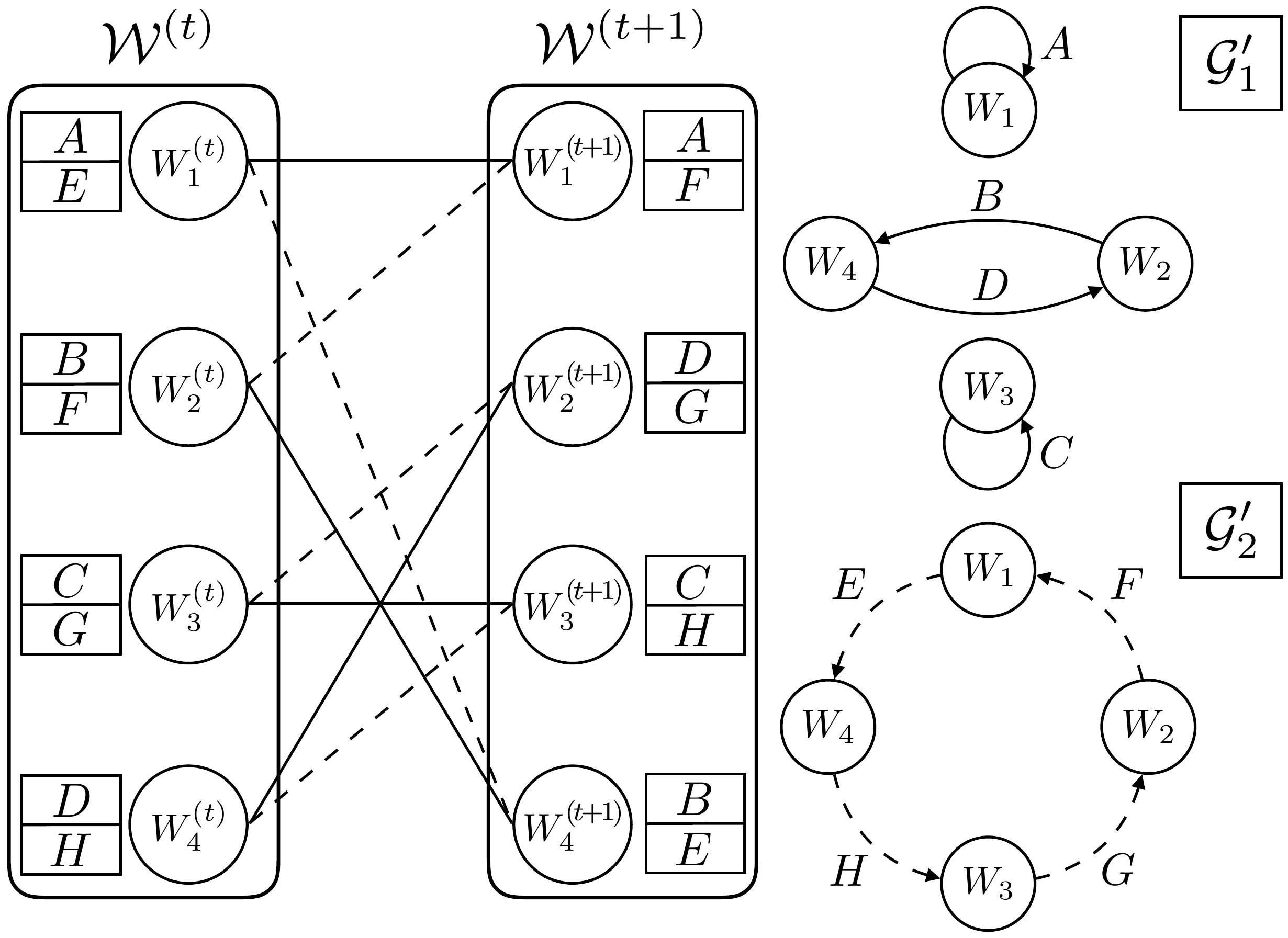}%
	\label{fig:ex6_decomp2_bipartiteGraph}}
	\caption{
		(a) One decomposition of $\mathcal{H} (V', E')$, shown in Fig.~\ref{fig:ex6_bipartiteGraph}, into $N/K = 2$ perfect matchings, designated by solid and dashed lines, and the corresponding decomposition of $\mathcal{G} (V, E)$, shown in Fig.~\ref{fig:ex6_fileTransGraph}, into two canonical subgraphs. 
		(b) Another decomposition of $\mathcal{H} (V', E')$ into $N/K = 2$ perfect matchings, and the corresponding decomposition of $\mathcal{G} (V, E)$ into canonical subgraphs. 
	}
	\label{fig:ex6_decomp1_decomp2_bipartiteG}
	\vspace{-1mm}
\end{figure}

\noindent\textbf{Example 6 (Example 5 Continued)}: 
{\it 
Let us revisit the data shuffling system studied in Example~5, with  $K=4$ worker nodes, and $N=8$ files, denoted by $\{A,B,C,D,E,F,G,H\}$. 
Fig.~\ref{fig:ex6_fileTransGraph} captures the file transition graph $\cG(V,E)$, and Fig.~\ref{fig:ex6_bipartiteGraph} depicts the corresponding  bipartite  graph $\mathcal{H}(V', E')$.
For example, the directed edge, labeled by $E$, from $W_1$ to $W_4$ in $\cG(V,E)$ indicates that file $E$ is processed by worker nodes $W_1$ and $W_4$ at iterations $t$ and $t+1$, respectively.  Accordingly, there is an edge, labeled by $E$, between $W_1^{(t)}$ and $W_4^{(t+1)}$ that shows an assignment of file $E$ to worker node $W_1$ at iteration~$t$, and to worker node $W_4$ at iteration $t+1$. 

After constructing $\mathcal{H} (V', E')$, we decompose it into $N/K=2$ 
perfect matchings between  $\cW^{(t)}$ and $\cW^{(t+1)}$, designated by solid and dashed lines in
Fig.~\ref{fig:ex6_decomp1_bipartiteGraph}. The corresponding canonical subgraphs of $\cG(V,E)$ are also shown in the figure.  
Another possible decomposition of $\mathcal{H} (V', E')$ is depicted by Fig.~\ref{fig:ex6_decomp2_bipartiteGraph}, along with the corresponding decomposition of $\cG(V,E)$. 
The existence of the decomposition of $\cH(V',E')$ is guaranteed 
by Hall's theorem, which results in the feasibility of the decomposition of $\cG(V,E)$. However,  the graph decomposition is not unique, and there may be more than one decomposition for one instance of the data shuffling problem. 
\hfill $\blacklozenge$
}

\section{Pseudocodes}
\label{app:psuedocodes_p1}
Algorithm~\ref{AchvScheme1_1} describes the file partitioning and labeling, while Algorithm~\ref{AchvScheme1_2} presents the cache placement. 
Next, Algorithms~\ref{AchvScheme1_3},~\ref{AchvScheme1_4}, and \ref{AchvScheme1_5} describe the encoding, decoding, and cache updating and subfile relabeling, respectively.
Finally, Algorithm~\ref{algo_decomposeGraph} presents the graph decomposition that is based on the Hungarian algorithm \cite{kuhn1955hungarian,munkres1957algorithms}.

\begin{algorithm}[H] 
	\caption{\partitionFiles}
	\label{AchvScheme1_1}
	\begin{algorithmic}[1]
		\State {\bfseries Input:} $\displaystyle N, K, S, \left\{F^j : j \in [N] \right\}$
		\State {\bfseries Output:} $\Big\{ \F{j}{\Gamma}: j \in [N], i \in [K], j\in u(i), \Gamma\subseteq [K] \setminus \{i\},$ $|\Gamma|=S/(N/K)-1\Big\}$
		\State $\widehat{S} \gets S/(N/K)$
		\For{$i \gets 1$ {\bfseries to} $K$}
		\ForAll{$j\in u(i)$}
		\State Worker node $W_i$ partitions $\F{j}{}$ into $\binom{K-1}{\widehat{S}-1}$ subfiles of equal sizes: $\left\{\F{j}{\Gamma}: \Gamma\subseteq [K] \setminus \{i\}, |\Gamma|=\widehat{S}-1\right\}$
		\EndFor
		\EndFor
	\end{algorithmic}
\end{algorithm}

\begin{algorithm}[H] 
	\caption{\placeCache}
	\label{AchvScheme1_2}
	\begin{algorithmic}[1]
		\State {\bfseries Input:} $\displaystyle N, K, S, t, \Big\{ \F{j}{\Gamma}: j \in [N], i \in [K], j\in u(i), \Gamma\subseteq [K] \setminus \{i\}, |\Gamma|=S/(N/K)-1\Big\}$
		\State {\bfseries Output:} $\left\{\cZ_i(t) : i \in [K]\right\}$
		\State $\widehat{S} \gets S/(N/K)$
		\For{$i \gets 1$ {\bfseries to} $K$}
		\State $\proc{i}(t) \gets \left\{\F{j}{\Gamma}: j \in u(i),\: \Gamma \subseteq [K] \setminus \{i\},\: |\Gamma| = \widehat{S}-1\right\}$
		\Comment{i.e., $\proc{i}(t) \gets \left\{\F{j}{} : j \in u(i)\right\}$}
		\ForAll{$\ell \in [N] \setminus u(i)$}
		\State $\ex{i}{\ell} \gets \left\{\F{\ell}{\Gamma}:  i \in \Gamma \subseteq [K], \: |\Gamma| = \widehat{S} - 1\right\}$
		\EndFor
		\State $\ex{i}{}(t) \gets \bigcup\nolimits_{\ell \in [N] \setminus u(i)} \ex{i}{\ell}$
		\State $\cZ_i(t) \gets \proc{i}(t) \cup \ex{i}{}(t)$
		\EndFor
	\end{algorithmic}
\end{algorithm}

\begin{algorithm}[H] 
	\caption{\encodeSubmessages}
	\label{AchvScheme1_3}
	\begin{algorithmic}[1]
		\State {\bfseries Input:} $\displaystyle K, S, \left\{\F{i}{\Gamma} : i \in [K], \: \Gamma\subseteq [K]\setminus\{i\},\: |\Gamma|=S-1\right\}$, $\left\{d(i): i \in [K]\right\}$
		\State {\bfseries Output:} $\left\{X_\set : \set  \subseteq [K-1],\: |\set|=S\right\}$
		\ForAll{$\set  \subseteq [K-1]$  \AND $|\set|=S$}
		\State $\displaystyle X_{\Delta} \gets 
		\bigoplus_{i \in \Delta}  \left( \F{i}{\Delta \setminus \{i\}} 
		\oplus
		\F{d(i)}{\Delta \setminus \{d(i)\}}
		\oplus	
		\bigoplus_{ j \in [K] \setminus \set } 
		\F{d(i)}{(\{j\} \cup \set)\setminus \{i, d(i)\}} 
		\right)$
		\EndFor
		\State $\cX \gets \{X_\set : \set  \subseteq [K-1],\: |\set|=S \}$ 
		\State The master node broadcasts the message $\cX$ to the worker nodes
	\end{algorithmic}
\end{algorithm}

\vspace{-2.3mm}
\begin{algorithm}[H] 
	\caption{$\decodeSubfiles$}
	\label{AchvScheme1_4}
	\begin{algorithmic}[1]
		\State {\bfseries Input:} $\displaystyle K, S, t, \left\{d(i) : i \in [K]\right\}, \left\{\cZ_i(t) : i \in [K]\right\}$, $\left\{X_\set : \set  \subseteq [K-1],\: |\set|=S\right\}$
		\State {\bfseries Output:} $\left\{\cQ_i(t) : i \in [K]\right\} $
		\For{$i \gets 1$ {\bfseries to} $K$}
		\If{$i < K$}
		\ForAll{$\Gamma \subseteq [K-1]$ \AND $\F{d(i)}{\Gamma}\in \cQ_{i}$}
		\State Worker node $W_{i}$ decodes subfile $\F{d(i)}{\Gamma}$ from the sub-message $X_{\{i\} \cup \Gamma}$ using its cache contents $\cZ_i$
		\EndFor
		\ForAll{$\Gamma \subseteq [K]$ \AND $K\in \Gamma$ \AND $\F{d(i)}{\Gamma}\in \cQ_{i}$}
		\State Worker node $W_{i}$ decodes subfile $\F{d(i)}{\Gamma}$ from the sub-message $X_{(\Gamma\setminus\{K\}) \cup \{i,d(i)\}}$ using its cache contents $\cZ_i$ and other subfiles already decoded by $W_{i}$
		\EndFor
		\Else
		\ForAll{$\Gamma \subseteq [K-1]$ \AND $\F{d(K)}{\Gamma}\in \cQ_{K}$}
		\State Worker node $W_K$ decodes subfile $\F{d(K)}{\Gamma}$ from the sub-messages $\bigoplus_{j \in [K-1]\setminus \Gamma} X_{\{j\} \cup \Gamma}$ using its cache contents~$\cZ_K$
		\EndFor
		\EndIf
		\State $\cQ_i(t) \gets \lb \F{d(i)}{\Gamma}: \Gamma  \subseteq [K] \setminus \{d(i)\}, |\Gamma|=S-1\rb \setminus \cZ_i(t)$
		\EndFor
	\end{algorithmic}
\end{algorithm}

\vspace{-2.3mm}
\begin{algorithm}[H] 
	\caption{$\updateCaches$}
	\label{AchvScheme1_5}
	\begin{algorithmic}[1]
		\State {\bfseries Input:} $\displaystyle K, S, t, \left\{d(i) : i \in [K]\right\}, \left\{\cZ_i(t) : i \in [K]\right\}$, $\left\{\cQ_i(t) : i \in [K]\right\}$
		\State {\bfseries Output:} $\left\{\cZ_i(t+1) : i \in [K]\right\}$
		\For{$i \gets 1$ {\bfseries to} $K$}
		\Comment{Updating caches of worker nodes before iteration $t+1$}
		\State $j \gets d^{-1}(i)$
		\State $\cA \gets \left\{\F{i}{\Gamma}: j \in \Gamma,\: \Gamma \subseteq [K]\setminus\{i\},\: |\Gamma|=S-1\right\}$.
		\State $\cS \gets \left\{\F{d(i)}{\Gamma}: i \in \Gamma,\: \Gamma \subseteq [K]\setminus\{d(i)\},\: |\Gamma|=S-1\right\}$
		\State $\proc{i}(t+1) \gets \left\{\F{d(i)}{\Gamma}: \Gamma \subseteq [K]\setminus\{d(i)\},\: |\Gamma|=S-1\right\}$
		\State $\ex{i}{}(t+1) \gets \left( \ex{i}{}(t) \setminus \cS \right) \cup \cA$
		\State $\cZ_i(t+1) \gets \proc{i}(t+1) \cup \ex{i}{}(t+1)$
		\EndFor
		\For{$i \gets 1$ {\bfseries to} $K$}
		\Comment{Relabeling subscripts of a set of subfiles of each worker node}
		\State $j \gets d^{-1}(i)$
		\ForAll{$\F{i}{\Gamma}: j \in \Gamma,\: \Gamma \subseteq [K] \setminus \{i\},\: |\Gamma|=S-1$}
		\State $\Lambda \gets \left(\Gamma \setminus \left\{j\right\}\right) \cup \left\{i\right\}$
		\State Replace $\Gamma$ in $\F{i}{\Gamma}$ by $\Lambda$
		\EndFor
		\EndFor
		\For{$i \gets 1$ {\bfseries to} $K$}
		\Comment{Relabeling superscripts of all subfiles of each worker node}
		\State $j \gets d^{-1}(i)$
		\ForAll{$\F{i}{\Gamma}: \Gamma \subseteq [K] \setminus \{i\},\: |\Gamma|=S-1$}
		\State Replace $i$ in $\F{i}{\Gamma}$ by $j$
		\EndFor
		\EndFor
	\end{algorithmic}
\end{algorithm}

\begin{algorithm}
	\caption{\decomposeGraph}
	\label{algo_decomposeGraph}
	\begin{algorithmic}[1]
		\State {\bfseries Input:} $\displaystyle N, K, S, \{u(i), i\in [K]\}, \{d(i), i\in [K]\}$ 
		\State {\bfseries Output:} $N/K$ perfect matchings
		\State Construct a file transition graph matrix $G$ that has size $K \times K$, where $G(i,j)$ is the number of files where $u(i) = d(j)$, otherwise, $G(i,j) = \infty$, for $i,j \in [K]$.
		\State $H \gets G$.
		\For{$p \gets 1$ {\bfseries to} $N/K$}
		\State $[v, cost] = \Hungarian(H)$.
		\Comment $v$ denotes the matching vector of size $K \times 1$, where $W_i$ is matched with $W_{v(i)}$ for $i \in [K]$.
		\For{$q \gets 1$ {\bfseries to} $K$}
		\If{$H(q,v(q)) = 1$}
		\State $H(q,v(q)) \gets \infty$
		\ElsIf{$H(q,v(q)) > 1$}
		\State $H(q,v(q)) \gets H(q,v(q)) - 1$
		\EndIf
		\EndFor 
		\EndFor
	\end{algorithmic}
\end{algorithm}

\bibliographystyle{IEEEtran}
\bibliography{codedDS_ref}

\end{document}

%% file: mySymbs.tex
\newcommand{\F}{\mathbb{F}}

\newcommand{\cA}{\mathcal{A}}

\newcommand{\cD}{\mathcal{D}}
\newcommand{\cE}{\mathcal{E}}

\newcommand{\cG}{\mathcal{G}}
\newcommand{\cH}{\mathcal{H}}

\newcommand{\cJ}{\mathcal{J}}

\newcommand{\cN}{\mathcal{N}}

\newcommand{\cP}{\mathcal{P}}
\newcommand{\cQ}{\mathcal{Q}}

\newcommand{\cS}{\mathcal{S}}
\newcommand{\cT}{\mathcal{T}}
\newcommand{\cU}{\mathcal{U}}
\newcommand{\cV}{\mathcal{V}}
\newcommand{\cW}{\mathcal{W}}
\newcommand{\cX}{\mathcal{X}}

\newcommand{\cZ}{\mathcal{Z}}

\newtheorem{lm}{Lemma}

\newtheorem{prop}{Proposition}

\newtheorem{ex}{Example}